\documentclass[plain,biber]{nowfnt} 

\title{Refinement Types}
\subtitle{A Tutorial Introduction}

\maintitleauthorlist{
Ranjit Jhala \\
University of California, San Diego, USA \\
jhala@cs.ucsd.edu

\and 

Niki Vazou \\
IMDEA Software Institute, Madrid, Spain \\
niki.vazou@imdea.org
}

\issuesetup
{%
 copyrightowner={A.~Heezemans and M.~Casey},
 volume        = xx,
 issue         = xx,
 pubyear       = 2020,
 isbn          = xxx-x-xxxxx-xxx-x,
 eisbn         = xxx-x-xxxxx-xxx-x,
 doi           = 10.1561/XXXXXXXXX,
 firstpage     = 1, 
 lastpage      = 18
 }

\usepackage{mwe}

\usepackage{booktabs}
\usepackage{hyperref}
\usepackage{amsmath,amssymb,latexsym,amstext}
\usepackage{pifont}
\usepackage{cleveref}
\usepackage{array}
\usepackage{url}
\usepackage{comment}
\usepackage{listings}
\usepackage{showexpl}
\usepackage{enumitem}
\usepackage{booktabs}
\usepackage{stmaryrd}
\usepackage[dvipsnames]{xcolor}
\usepackage{mathtools}
\usepackage{apptools}
\usepackage{chngcntr}
\usepackage{scalerel}
\usepackage{xspace}
\usepackage{mathpartir}
\usepackage{graphicx}
\usepackage{caption}
\usepackage{csvsimple}
\usepackage{pdfpages}
\usepackage{tikz}
\usepackage{anyfontsize}
\usepackage{listings}

\def\withcolor{}

\ifdefined\withcolor
	\definecolor{haskellblue}{rgb}{0.0, 0.0, 1.0}
	\definecolor{gray_ulisses}{gray}{0.55}
	\definecolor{castanho_ulisses}{rgb}{0.0,0.4,0.0}
	\definecolor{preto_ulisses}{rgb}{0.41,0.20,0.04}
	\definecolor{green_ulisses}{rgb}{0.8,0.0,0.8}
\else
	\definecolor{haskellblue}{gray}{0.1}
	\definecolor{haskellred}{gray}{0.1}
	\definecolor{gray_ulisses}{gray}{0.1}
	\definecolor{castanho_ulisses}{gray}{0.1}
	\definecolor{preto_ulisses}{gray}{0.1}
	\definecolor{green_ulisses}{gray}{0.1}
\fi

\def\incodesize{\small}
\def\codesize{\small}

\lstdefinelanguage{ocaml} {
 columns=[c]fixed,
 keywordstyle=\bfseries,
 upquote=true,
 commentstyle=,
 breaklines=true,
 showstringspaces=false,
 stringstyle=\color{blue},
 literate={'"'}{\textquotesingle "\textquotesingle}3
}

\lstdefinelanguage{HaskellUlisses} {
	basicstyle=\ttfamily\codesize,
	sensitive=true,
	mathescape=true,
	morecomment=[l][\color{gray_ulisses}\ttfamily\codesize]{--},
	morestring=[b]",
	literate={'"'}{\textquotesingle "\textquotesingle}3,
	stringstyle=\color{haskellred},
	showstringspaces=false,
	numberstyle=\codesize,
	numberblanklines=true,
	showspaces=false,
	breaklines=true,
	showtabs=false,
	emph=
	{[1]
		FilePath,IOError,acos,acosh,all,any,appendFile,approxRational,asTypeOf,asin,
	},
	emphstyle={[1]\color{haskellblue}},
	emph=
	{[2]
		bool,char,int,string,array,nat,pos,int8,list,olist,incList,decList,pair,
		incPair,Bool,Char,Double,Either,Float,IO,Integer,Int,Maybe,Ordering,
		Rational,Ratio,ReadS,ShowS,String,Word8,Nat,NonZero,Nat64,Text,
		ByteString,ByteStringSZ,ByteStringN,Ptr,ForeignPtr,CSize
    InPacket,Tree,Prop,TreeEq,TreeLt,Vec,
    NullTerm,IncrList,DecrList,UniqList,BST,MinHeap,MaxHeap,
    PtrN,ByteStringN,ByteStringEq,VO,ByteStringsEq,ByteStringNE,
		List,Even
	},
	emphstyle={[2]\color{castanho_ulisses}},
	emph=
	{[3]
		case,class,data,deriving,do,else,if,switch,return,import,in,infixl,infixr,instance,val,let,rec,
		requires,ensures,assume,val,def,
		module,measure,predicate,of,primitive,then,refinement,type,where,lazy
	},
	emphstyle={[3]\color{preto_ulisses}\textbf},
	emph=
	{[4]
		quot,rem,div,mod,elem,notElem,seq
	},
	emphstyle={[4]\color{castanho_ulisses}\textbf},
	emph=
	{[5]
		PS,Tip,Node,EQ,true,false,GT,Just,LT,Left,Nothing,Right,Show,Eq,Ord,Num,
		Cons,Nil,OCons,ONil
	},
	emphstyle={[5]\color{green_ulisses}}
}

\lstnewenvironment{fscode}
{\lstset{language=HaskellUlisses,basicstyle=\ttfamily\small}}
{}

\lstnewenvironment{code}
{\lstset{language=HaskellUlisses}}
{}

\lstnewenvironment{codebox}
{\lstset{language=HaskellUlisses,frame=tlbr,basicstyle=\ttfamily\codesize}}
{}

\lstMakeShortInline[language=HaskellUlisses,basicstyle=\ttfamily\incodesize]@

\usepackage{commands}

\usepackage{color}
\usepackage{fancyvrb}
\usepackage{longtable}
\usepackage[libertine]{newtxmath}

\DefineVerbatimEnvironment{Highlighting}{Verbatim}{commandchars=\\\{\}}

\usepackage[english]{babel}

\newcommand{\cmark}{\ding{51}}
\newcommand{\xmark}{\ding{55}}

\usepackage{inconsolata}

\usepackage{pgf, tikz}
\usetikzlibrary{arrows, automata}    

\articledatabox{\nowfntstandardcitation}

\begin{document}

\makeabstracttitle


\chapter{Introduction}\label{sec:intro}

The type systems of modern languages like C\#, Haskell, Java, Ocaml, 
Rust and Scala are \emph{the} most widely used method for establishing 
guarantees about the correct behavior of software. 
In essence, types allow the programmer to describe \emph{legal} 
sets of values for various operations, thereby eliminating, at 
compile-time, the possibility of a large swathe of unexpected 
and undesirable run-time errors.
Unfortunately, well-typed programs \emph{do} go wrong.

\begin{enumerate}

  \item \emphbf{Divisions by zero} 
  The fact that a divisor is an @int@ does not preclude
  the possibility of a run-time divide-by-zero, or that 
  a given arithmetic operation will over- or under-flow;

  \item \emphbf{Buffer overflows} 
  The fact that an @array@ or @string@ index is an @int@ does 
  not eliminate the possibility of a segmentation fault, 
  or worse, leaking data from an out-of-bounds access;

  \item \emphbf{Mismatched dimensions}
  Moving up a level, the fact that a product operator 
  is given two @matrix@ values does not prevent errors
  arising from the matrices having incompatible dimensions;

  \item \emphbf{Logic bugs}
  Classical type systems can ensure that each @date@ structure 
  contains suitable (\eg @int@ valued) fields holding the 
  @day@, @month@ and @year@, but cannot guarantee that the 
  @day@ is valid for the given @month@ and @year@; 

  \item \emphbf{Correctness errors}
  Finally, at the extreme end, a type system can ensure that 
  a sorting routine produces a list, and that a compilation 
  routine produces a sequence of machine instructions, but 
  cannot guarantee that the list was, in fact, an ordered 
  permutation of the input, or that the machine instructions 
  faithfully implemented the program source.

\end{enumerate}

\mypara{Refining Types with Predicates}
Refinement types allow us to enrich a language's type 
system with \emph{predicates} that circumscribe the set 
of values described by the type. For example, while an 
@int@ can be any integer value, we can write the refined
type
\begin{code}
  type nat = int[v|0 <= v]
\end{code}
that describes only non-negative integers.
By combining types and predicates the programmer can write 
precise \emph{contracts} that describe legal inputs and 
outputs of functions. For example, the author of an array 
library could specify that 
\begin{code}
  val size : x:array($\tvar$) $\Rightarrow$ nat[v|v = length(x)]
  val get  : x:array($\tvar$) $\Rightarrow$ nat[v|v < length(x)] $\Rightarrow$ $\tvar$ 
\end{code}
which say that (1)~a call @size(arr)@ \emph{ensures} 
the returned integer equals to the number of elements in @arr@, 
and (2)~the call @get(arr, i)@ \emph{requires} the index @i@ be 
within the bounds of @arr@.
Given these specifications, the refinement type checker can 
guarantee, at \emph{compile time}, that all operations respect 
their contracts, to ensure, \eg that all array accesses are safe at run-time.

\mypara{Language-Integrated Verification}
Refinements provide a tunable knob whereby developers can inform 
the type system about what invariants and correctness properties
they care about, \ie are important for the particular domain of 
their code.
They could begin with basic safety requirements, \eg to eliminate 
divisions by zero and buffer overflows, or ensure they don't attempt 
to access values from empty stack or collections, and then, incrementally,
dial the specifications up to include, \eg invariants about custom 
data types like dates, or ordered heaps, and, if they desire, 
ultimately go all the way to specifying and verifying the 
correctness of various routines at compile-time.
Crucially, (refinement) types eliminate the barrier between 
implementation and proof, by enabling verification within 
the same language, library and tool ecosystem.
This tight integration is essential to create a virtuous cycle 
of feedback across the phases. 
The \emph{implementation} dictates what properties are important, 
and provides hints on how to do do the verification.
Dually, the \emph{verification} provides guidance on how 
the code can be restructured, \eg to make the abstractions 
and invariants explicit enough to enable formal proof.

\section{A Brief History}

Refinement types can be thought of as a type-based
formulation of assertions from classical program 
logic \citep{Turing49,Floyd67,Hoare69}.
The idea of refining types with logical constraints 
goes back at least to \citep{Cartwright76} who described 
a means of refining LISP datatypes with constraints to 
aid in program verification.
The \textsc{ADA} programming language has a notion 
of \emph{range} types which allow the to define 
contiguous subsets of integers \citep{ada80}. 
\cite{Nordstrom83} and \cite{Constable83} introduced 
the notion of logical-refinements-as-subsets of values, 
and \cite{Constable86} turned this notion into 
a pillar of the \textsc{Nuprl} proof assistant.

\cite{FreemanPfenningDONTCITE91} introduced the
name ``refinement types'' in a paper that describes 
a syntactic mechanism to define subsets of algebraic data.
Inspired by the early work on \textsc{Nuprl},
the \textsc{PVS} proof assistant embraced the idea of 
types as subsets, and \cite{Rushby98} introduced 
the notion of \emph{predicate subtyping} which
forms the basis of the subtyping relation that 
remains the workhorse of modern refinement type 
systems.
\cite{Zenger97} and \cite{pfenningxi98}
describe a means of \emph{indexing} types 
with (symbolic) integers after which 
constraints can be used to specify 
function contracts that can be verified 
by linear programming, to, \eg perform
array bounds or list or matrix dimension 
checking at compile time, and \cite{Dunfield}
shows how combine indices with datasort
refinements to facilitate the verification 
of data structure invariants.

The \textsc{SAGE} system \citep{Gronski06} described 
how refinement like specifications could be verified 
in a \emph{hybrid} manner: partly at compile time
using SMT solvers, and partly at run-time via dynamic 
contract checks \citep{flanagan06}.
Several groups picked up the gauntlet of moving all 
the checks to compile time, leading to \textsc{F7} 
\citep{GordonTOPLAS2011} and then \fstar \citep{fstar} 
dialects of ML which has been used to formally verify 
the implementation of cryptographic routines used in 
widely used web-browsers \citep{haclstar}.
\cite{LiquidPLDI08} introduced the notion of \emph{liquid}
types which make refinements easier to use by delegating 
the task of synthesizing refinements to abstract interpretation.

The last decade has seen refinements spread over to languages 
outside the ML family.
\cite{LiquidPOPL10} and \cite{Chugh2012} show how to verify 
\textsc{C} and \textsc{JavaScript} programs by refining 
a low-level language of locations \citep{AliasTypes}.
\cite{RefinedRacket} show how refinements can be integrated 
within \textsc{Racket}'s occurrence based type system \citep{Tob08}.
\cite{rruby} integrate refinements in \textsc{Ruby}'s type system 
using just-in-time type checking. 
Finally, \cite{kuncak-stainless} present a refinement-type based 
verifier for higher-order \textsc{Scala} programs.

\section{Goals \& Outline}

Refinement types can be the vector that brings formal 
verification into mainstream software development. 
This happy outcome hinges upon the design and 
implementation of refinement type systems that 
can be retrofitted to existing languages, or 
co-designed with new ones.
Our primary goal is to catalyze the development 
of such systems by distilling the ideas developed
in the sprawling literature on the topic into 
a coherent and unified tutorial that explains 
the key ingredients of modern refinement type 
systems, by showing how to implement a refinement 
type checker.

\mypara{A Nanopass Approach}
Inspired by the \emph{nanopass framework} for teaching 
compilation pioneered by \cite{nanopass}, we will show 
how to implement refinement types via a progression of 
languages that incrementally add features to the language 
or type system.
\begin{itemize} 
\item \emphbf{$\langone$} (\S~\ref{sec:lang:one}):
  We start with the simply typed 
  $\lambda$-calculus, which will illustrate the foundations,
  namely, refinements, functions, and function \emph{application};

\item \textbf{$\langtwo$} (\S~\ref{sec:lang:two}): 
  Next, we will add branch conditions, and show 
  how refinement type checkers do \emph{path-sensitive} 
  reasoning;

\item \textbf{$\langthree$} (\S~\ref{sec:lang:three}): 
  Types are palatable only when we have to write 
  the interesting ones down: hence, next, we will see how to automatically 
  \emph{infer} the refinements to make using refinements pleasant; 
  
\item \textbf{$\langfour$} (\S~\ref{sec:lang:four}): 
  After adding inference to our arsenal, we will 
  be able to add type polymorphism which, will unlock various 
  forms of \emph{context-sensitive} reasoning;

\item \textbf{$\langfive$} (\S~\ref{sec:lang:five}): 
  Once we have polymorphic types, we can add polymorphic 
  \emph{data types} like lists and trees, and see how to 
  specify and verify properties of those structures;

\item \textbf{$\langsix$} (\S~\ref{sec:lang:six}): 
  Type polymorphism allows us to reuse functions and 
  data with different kinds of values. We will see why we often 
  need to reuse functions and data across different kinds of invariants,
  and to support this, we will develop a form of \emph{refinement polymorphism};

\item \textbf{$\langseven$} (\S~\ref{sec:lang:seven}): 
  All of the above methods allow us to verify safety properties, 
  \ie assertions about values of code. Next, we will see how refinements let 
  us verify \emph{termination};

\item \textbf{$\langeight$} (\S~\ref{sec:lang:eight}): 
  Finally, we will see how to write 
  propositions over arbitrary user defined functions and write 
  proofs of those propositions as well-typed programs, effectively 
  converting the host language into a theorem prover.

\end{itemize}

\begin{figure}[t]
\begin{center}
  \begin{tikzpicture}[
    > = stealth, 
    shorten > = 1pt, 
    auto,
    node distance = 2.2cm, 
    semithick 
]

\tikzstyle{every state}=[
    draw = black,
    thick,
    shape = rectangle,
    minimum size = 8mm
]


\node[state, fill=GreenYellow]  (l1)                     {\S~\ref{sec:lang:one}:   $\langone$};
\node[state, fill=GreenYellow]  (l2) [right of=l1]       {\S~\ref{sec:lang:two}:   $\langtwo$};
\node[state, fill=yellow] (l3) [right of=l2]       {\S~\ref{sec:lang:three}: $\langthree$};
\node[state, fill=yellow] (l4) [above right of=l3] {\S~\ref{sec:lang:four}:  $\langfour$};
\node[state, fill=pink]   (l5) [below right of=l3] {\S~\ref{sec:lang:five}:  $\langfive$};
\node[state, fill=yellow] (l6) [below right of=l4] {\S~\ref{sec:lang:six}:   $\langsix$};
\node[state, fill=Dandelion] (l7) [right of=l5]       {\S~\ref{sec:lang:seven}: $\langseven$};
\node[state, fill=Dandelion] (l8) [right of=l7]       {\S~\ref{sec:lang:eight}: $\langeight$};

\path[->] (l1) edge node {} (l2);
\path[->] (l2) edge node {} (l3);
\path[->] (l3) edge node {} (l4);
\path[->] (l3) edge node {} (l5);
\path[->] (l4) edge node {} (l6);
\path[->] (l5) edge node {} (l6);
\path[->] (l5) edge node {} (l7);
\path[->] (l7) edge node {} (l8);
\end{tikzpicture}
\end{center}
\caption{Chapter dependencies}
\label{fig:dependencies}
\end{figure}
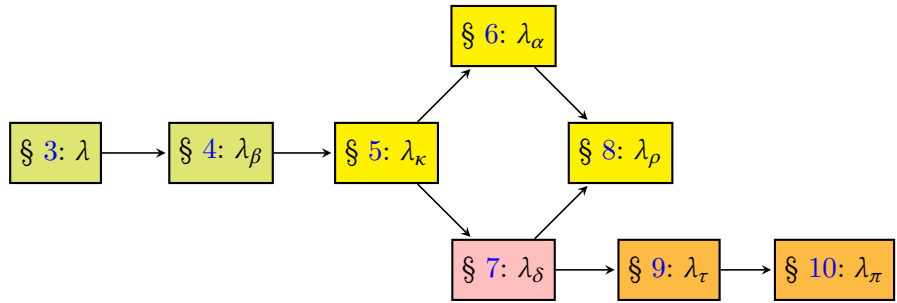

\mypara{Dependencies}
The ideal reader would, of course, devote several hours 
of thoughtful contemplation to each of the eight sub-languages.
However, life is short, and you may be interested in particular
aspects of refinement typing. If so, we suggest reading the chapters 
in the following order, summarized in \cref{fig:dependencies}
\begin{itemize}
  \item \S~\ref{sec:lang:one} and \S~\ref{sec:lang:two} are 
        essential, as they focus on the basics of refinement types 
        and \emph{path-sensitive reasoning};  

  \item \S~\ref{sec:lang:three}, \S~\ref{sec:lang:four} and \S~\ref{sec:lang:six} 
        explain how to support polymorphism via \emph{refinement inference};

  \item \S~\ref{sec:lang:five} explains how refinements allow 
        reasoning about invariants of \emph{algebraic data types}; 
        
  \item \S~\ref{sec:lang:seven} and \S~\ref{sec:lang:eight} will be
        of interest to readers who wish to learn how to scale refinements 
        up to \emph{proofs}.
\end{itemize}

\mypara{Implementation}
This article is accompanied by an implementation
\begin{verbatim}
   https://github.com/ranjitjhala/sprite-lang
\end{verbatim}
The @README@ that accompanies the code has directions on how to
build, modify and execute the sequence of type checkers that we 
will develop over the rest of this article.  
We welcome readers who like to get their hands dirty 
to clone the repository and follow along with the code.

\medskip \noindent And now, lets begin!
\chapter{Refinement Logic} \label{sec:smt}

Refinements are \emph{predicates} drawn from decidable logics.
Refinement type checking yields \emph{constraints} whose validity 
implies that the program is well-typed. 
Lets start with a quick overview of the logic of predicates 
and constraints that we will use in this article, and our 
rationale for choosing it.
Readers familiar with SMT solvers can skip ahead, and those 
interested in learning more should consult \citep{Nelson81} 
or \citep{KroeningBook}.

\section{Syntax}

\begin{figure}[t]
\begin{tabular}{rrcll}
\emphbf{Predicates}
  & \pred,\predb & $\bnfdef$ & $x, y, z, \ldots$            & \emph{variables} \\
  &       & $\spmid$  & $\ttrue, \tfalse$                   & \emph{booleans}  \\
  &       & $\spmid$  & $0, -1, 1, \ldots$                  & \emph{numbers}   \\
  &       & $\spmid$  & $\pred_1 \bowtie \pred_2$           & \emph{interp. ops.} \\
  &       & $\spmid$  & $\pred_1 \wedge  \pred_2$           & \emph{conjunction} \\
  &       & $\spmid$  & $\pred_1 \vee    \pred_2$           & \emph{disjunction} \\
  &       & $\spmid$  & $\neg \pred$                        & \emph{negation} \\
  &       & $\spmid$  & $\ite{\pred}{\pred_1}{\pred_2}$     & \emph{conditional} \\
  &       & $\spmid$  & $\uif{f}{\params{\pred}}$  & \emph{uninterp. Function} \\[0.05in]

\emphbf{Constraints}
  & \cstr & $\bnfdef$ & $\pred$                             & \emph{predicates} \\
  &       & $\spmid$  & $\csand{\cstr_1}{\cstr_2}$          & \emph{conjunction} \\
  &       & $\spmid$  & $\csimp{x}{\base}{\pred}{\cstr}$    & \emph{implication} \\[0.05in]
\end{tabular}
\caption{{Syntax of Predicates and Constraints}}
\label{fig:smt:pred}
\end{figure}

The syntax of predicates and constraints is summarized in \cref{fig:smt:pred}.

\subsection{Quantifier-free Predicates}\label{sec:smt:pred}

We will work with predicates $\pred$ drawn 
from the \emph{quantifier-free} fragment of 
linear arithmetic and uninterpreted functions 
(QF-UFLIA) \citep{SMTLIB2}. 
These include boolean literals ($\ttrue$, $\tfalse$), 
integer literals ($0,1,2,\ldots$), 
variables ranging over booleans and integers ($x, y, z, \ldots$), 
linear arithmetic operators ($\pred_1 + \pred_2, \ldots$),
and boolean combinations ($\pred_1 \wedge \pred_2$).
For convenience, we include a ternary choice operator 
$(\ite{\pred}{\pred_1}{\pred_2}$) which abbreviates 
$(\pred \Rightarrow x = \pred_1) \wedge (\neg \pred \Rightarrow x = \pred_2)$.
We use $\bowtie$ to represent all interpreted operators; this set can be 
extended to include those from other SMT decidable logics \eg operations 
over sets, strings or bitvectors.

\mypara{Uninterpreted Functions}
All other operations will be modeled as applications 
of uninterpreted functions $\uif{f}{\params{x}}$.
These functions earn their name from the fact that the 
\emph{only} information that the SMT solver has about 
their behavior is encoded in the axiom of \emph{congruence}
$$ \forall \params{x},\params{y}.\ \params{x} = \params{y} \Rightarrow \uif{f}{\params{x}} = \uif{f}{\params{y}}$$
We will see how this provides an extremely general 
mechanism to encode all manner of specifications,
from the sizes and heights of trees (\cref{sec:lang:five}), 
to polymorphic refinement variables (\cref{sec:lang:six}),
to arbitrary user-defined functions (\cref{sec:lang:eight}).

\begin{myexample}{Predicates} 
Examples of predicates include $0 \leq \vvar$ that we used 
to denote non-negative integers, and $0 \leq \vvar \wedge \vvar < \uif{{length}}{x}$
that we used to specify valid indices for the array $x$, 
where $\mw{length}$ is an uninterpreted function in the logic.
\end{myexample}

\subsection{Constraints}\label{sec:smt:constr}

Refinement type checking will produce \emph{verification condition} (VC) 
constraints \citep{Floyd67,Hoare71} whose syntax is 
summarized in \cref{fig:smt:pred}.
A constraint is either 
a single (quantifier-free) predicate $\pred$, 
or a \emph{conjunction} of two sub-constraints $\cstr_1 \wedge \cstr_2$, 
or an \emph{implication} of the form $\csimp{x}{\base}{\pred}{\cstr}$ which 
says that for each $x$ of type $\base$, if the condition $\pred$ holds 
then so must $\cstr$.
The type $\base$ is defined in the rest of the chapters and as you 
will notice includes polymorphic variables and functions. 
In the implementation, we syntactically map these types to SMT sorts 
using the standard techniques of monomorphization and defunctionalization. 
For brevity, we will often omit the sort $\base$ of the quantifier bound 
variable when it is clear from the context.

\begin{myexample}{Constraints} 
The constraints $\cstr$ and $\cstr'$ 
could be generated from source programs 
that use an index @i@ to access an array @x@
$$\begin{array}{rlrl}
\cstr  \ \doteq & \forall \tb{x}{\primty{array}}.\ {0 \leq \uif{length}{x}} \Rightarrow & 

\ \cstr' \ \doteq & \forall \tb{x}{\primty{array}}.\ {0 < \uif{length}{x}} \Rightarrow \\

              & \quad \forall \tb{n}{\tint}.\ {n = \uif{length}{x}} \Rightarrow & 

              & \quad \forall \tb{n}{\tint}.\ {n = \uif{length}{x}} \Rightarrow \\

              & \quad \quad \forall \tb{i}{\tint}.\ {i = n - 1} \Rightarrow & 

              & \quad \quad \forall \tb{i}{\tint}.\ {i = n - 1} \Rightarrow \\

              & \quad \quad \quad 0 \leq i \wedge i < \uif{length}{x} & 

              & \quad \quad \quad  0 \leq i \wedge i < \uif{length}{x}
\end{array}$$
The variable $x$ is assigned an \emph{uninterpreted} sort $\primty{array}$ 
that is distinct from all others in the refinement logic.
\end{myexample}

\section{Semantics}

Programs are well-typed when their VCs are valid, as defined next.

\mypara{Substitution}
We will write $\SUBST{\pred}{x}{\predb}$ (resp. $\SUBST{\cstr}{x}{\predb}$) 
to denote the (capture avoiding) substitution of all (free) 
occurrences of $x$ in $\pred$ (resp. $\cstr$) with $\predb$.

\mypara{Validity} 
Let $\Sigma$ denote an interpretation mapping each uninterpreted 
function $f$ to a corresponding function from the domain to the 
co-domain of $f$. 
We define the notion that $\Sigma$ models a constraint $\cstr$, 
written $\Sigma \models \cstr$ as follows.
$\Sigma \models \pred$ if $\pred$ has no free variables 
and $\pred$ is a tautology under $\Sigma$ (\ie $\Sigma(\pred) \equiv \ttrue$). 
$\Sigma \models \cstr_1 \wedge \cstr_2$ if
if $\Sigma \models \cstr_1$ and $\Sigma \models \cstr_2$.
Finally, $\Sigma \models \csimp{x}{\base}{\pred}{\cstr}$ 
if for every constant $v$ of sort $\base$ such that $\Sigma \models \SUBST{\pred}{x}{v}$ 
we have $\Sigma \models \SUBST{\cstr}{x}{v}$.
A constraint $\cstr$ is \emph{valid} if $\Sigma \models \cstr$ 
for all interpretations $\Sigma$.

\mypara{Checking Validity via SMT}
Satisfiability Modulo Theories (SMT) solvers 
can algorithmically determine whether 
a constraint $\cstr$ is valid by \emph{flattening} 
it into a collection of sub-formulas of 
the form $\cstr_i \doteq \forall \params{x}. \pred_i \Rightarrow \predb_i$, 
such that $\cstr$ is valid iff $\cstr_i$ is valid.
The validity of each $\cstr_i$ is determined by checking the \emph{satisfiability}
of the (quantifier free) predicate $\pred_i \wedge \neg \predb_i$: the formula is 
valid \emph{iff} no satisfying assignment exists.

\begin{myexample}{Validity}
The constraint $\cstr$ shown above is \emph{invalid},
as demonstrated by the interpretation where $\uif{length}{x} \doteq 0$, 
and then $n \doteq 0$ and $i \doteq -1$.
However, the modified version $\cstr'$ is \emph{valid} 
as every interpretation for $\mw{length}$ yields a model 
for the constraint.
\end{myexample}

\section{Decidability}

Modern SMT solvers do support quantified formulas
and SMT-based verifiers like \textsc{Dafny} \citep{dafny}
and \fstar \citep{fstar} use this support to automate verification.
However, we make a deliberate choice to restrict the predicates
to \emph{quantifier-free} formulas to ensure that the 
generated VCs remain \emph{decidable}.
Decidability is important as a matter of \emph{principle}, 
as we do not want typability to rely upon the heuristics 
implemented in different solvers. Instead we want to 
provide a precise, solver-agnostic, language-based 
characterization of when a program is well-typed.
Decidability is also crucial in \emph{practice}, 
to ensure that type checking remains predictable. 
In particular, we want to circumscribe the possible 
causes that a user need investigate when type checking 
fails (\cref{sec:outro}) and decidability ensures 
that when debugging failures, we can safely avoid 
worrying about the brittleness of solver heuristics.





\chapter{The Simply Typed $\lambda$-calculus} 
\label{sec:lang:one} 

Our first language is the simply typed
$\lambda$-calculus equipped with primitive 
arithmetic values and operations. 
This language has just enough constructs to 
orient our understanding of refinements, and 
hence, equips us with the tools needed to explore 
more advanced features.

\section{Examples}\label{sec:one:examples}

First, lets build up a mental model of refinements 
\emph{should} work by working through a few simple 
examples. 
Here are two refinement types that describe the 
subsets of \emph{non-negative} an \emph{positive} 
integers 
\begin{align}
\primty{\textbf{type}}\ \tnat & = \refb{\tint}{\reft{\vvar}{0 \leq \vvar}} \label{eq:type:nat} \\
\primty{\textbf{type}}\ \ttpos & = \refb{\tint}{\reft{\vvar}{0    < \vvar}} \label{eq:type:pos}
\end{align} 
In the above, $\vvar$ is the \emph{value variable} which
\emph{names} the value being refined (\eg you can use 
\texttt{this} or \texttt{self} if you prefer). 
The condition \(0 \leq \vvar\) is the \emph{refinement}
(from~\cref{fig:smt:pred}) 
that must be satisfied by any member of the type. 
Hence, \texttt{0,1,2,3,...} are all elements of 
the refined type @nat@, but not \texttt{-1,-2,-3,...}.

\mypara{Aliases} 
To save ourselves some typing, and 
as it is often convenient to name 
concepts, we will define refinement 
type \emph{aliases}, such as @nat@ 
which is a name for the type of 
non-negative integers described above.

\mypara{Ex1: Primitive Constants}
First, lets consider the simplest possible example 
of typing code: we should be able to ascribe the 
@nat@ type to the numeric literal @6@

\begin{code}
  val six: nat
  let six = 6;
\end{code}

\mypara{Ex2: Primitive Operations}
The next example illustrates \emph{sequences} 
of variable definitions (bindings) and how 
values can be \emph{combined} with various 
operators. 
In the below, the types of @six@ and @nine@ 
should be composed with that of @add@ to let us 
assign @fifteen@ the type @nat@:

\begin{code}
  val fifteen: nat
  let fifteen = {
    let six = 6;
    let nine = 9;
    add(six, nine)
  };
\end{code}

\mypara{Ex3: Functions}
Next, consider a function \texttt{inc} that takes an 
\texttt{nat} and returns its successor. 
We should be able to ascribe \texttt{inc} 
a \emph{dependent function} type
\(\trfun{x}{\tnat}{\refb{\tint}{{\reft{\vvar}{x < \vvar}}}}\) 
that specifies that the function \emph{ensures} 
that the output value exceeds the input \(x\):

\begin{code}
  val inc: x:nat => int[v|x < v]
  let inc = (x) => { 
    let one = 1;
    add(x, one) 
  };
\end{code}

\mypara{Ex4: Function Calls}
Suppose now  another function @inc2@ that calls
@inc@ with the predecessor of its input @y@ and returns that
result. 
We should be able to give @inc2@ the type which states that
the function \emph{requires} that the input @y@ 
is @pos@itive and \emph{ensures} its output is also 
@pos@itive:

\begin{code}
  val inc2: y:pos => pos
  let inc2 = (y) => {
    let one = 1;
    let y1  = sub(y, one); 
    inc(y1)
  };
\end{code}

\section{Types and Terms}
\label{sec:one:types}
\label{sec:one:terms}

We summarize the language of types and terms for $\langone$ in \cref{fig:one:types}. 

\begin{figure}[t!]
\begin{tabular}{rrcll}
\emphbf{Basic Types}
  & \base     & $\bnfdef$ & \tint                & \emph{integers} \\ [0.05in]

\emphbf{Refinements}
  & \refi     & $\bnfdef$ & \reft{\vvar}{\pred}    & \emph{known} \\[0.05in]

\emphbf{Types}
  & \type,\typeb     & $\bnfdef$ & \refb{\base}{\rreft{\refi}}     & \emph{refined base}  \\
  &           & $\spmid$  & \trfun{x}{\type}{\type} & \emph{dependent function} \\[0.05in]

\emphbf{Kinds}
  & \kind     & $\bnfdef$ & \basekind    & \emph{base kind}  \\
  &           & $\spmid$  & \starkind & \emph{star kind} \\[0.05in]

\emphbf{Environments}
  & \tcenv    & $\bnfdef$ & $\emptyset$                      & \emph{empty} \\
  &           & $\spmid$  & \tcenvext{x}{\type}              & \emph{variable-binding} \\[0.05in]
\end{tabular}
\caption{{Syntax of Types and Environments}}
\label{fig:one:types}
\label{fig:one:env}
\end{figure}

\mypara{Types}
A \emph{basic} type $\base$ is an atomic primitive type, 
\eg the set of all integers \(\tint\). 
A \emph{refinement} $\reft{\vvar}{\pred}$ is a pair of 
a \emph{value variable} $\vvar$ and a logical predicate 
$\pred$ drawn from the SMT logic \cref{fig:smt:pred}, 
\eg $0 \leq \vvar$.
A \emph{refined base type} $\refb{\base}{\reft{\vvar}{\pred}}$
is a basic type combined with a refinement 
\eg @nat@ shown in (\ref{eq:type:nat}).
A \emph{function type} $\trfun{x}{\typeb}{\type}$ 
comprises an input binder $x$ of type $\typeb$ and an 
output type $\type$ in which $x$ may appear (within a refinement),
\eg $\trfun{x}{\tnat}{ \refb{\tint}{\reft{\vvar}{x < \vvar}}}$, 
the type assigned to @inc@.

\mypara{Kinds}
We use a simple kind system to check whether a type is base, 
and hence, can be refined. The base kind $\basekind$ is given 
only to refined base types, while all types can have the 
kind $\starkind$.

\mypara{Terms} 
The different kinds of terms of $\langone$ are summarized 
in \cref{fig:one:terms}.
The simplest terms are \emph{constants} $\vconst$ which 
include primitive values like $0,1,2,\ldots$ as well as 
arithmetic operators like $\mw{add}$ (addition), 
$\mw{sub}$ (subtraction), and so on.
Next, we have \emph{variables} $\evar$ that are introduced 
via \emph{let-binders} $\elet{\evar}{\expr_1}{\expr_2}$ that bind the 
value of $\expr_1$ to $\evar$ before evaluating $\expr_2$ and 
\emph{function} definitions $\elam{\evar}{\expr}$ that create 
functions with a parameter $\evar$ that produce the value of 
$\expr$ as the result.
We can \emph{apply} functions $\eapp{\expr}{x}$ where 
$\expr$ is the function and $x$ its argument 
\footnote{We will explain why the syntax restricts 
arguments to variables in \S~\ref{one:decl:synth}}.
Finally, the \emph{annotation} form $\eann{\expr}{\type}$ 
lets us ascribe the type $\type$ to the term $\expr$.

\mypara{Abbreviations}
We write \base to abbreviate \refb{\base}{\reft{\vvar}{\ttrue}}.
We abbreviate \refb{\base}{\reft{\vvar}{\pred}} to 
\reft{\vvar}{\pred} when \base is clear from the context 
and to \refb{\base}{\rreft{\pred}} when \pred does not refer to the binder \vvar. 
For environment bindings \tb{x}{\refb{\base}{\reft{\vvar}{\pred}}}
we assume that the environment and refinement 
binder names are the same, \ie \tb{x}{\refb{\base}{\reft{x}{\pred}}}
and omit the refinement binder name to write \tb{x}{\refb{\base}{\rreft{\pred}}}.

\begin{figure}[t!]
\begin{tabular}{rrcll}
\emphbf{Terms}
  & \expr     & $\bnfdef$ & \vconst                        & \emph{constants} \\
  &           & $\spmid$  & \evar                          & \emph{variables} \\
  &           & $\spmid$  & \elet{\evar}{\expr}{\expr}     & \emph{let-binding} \\
  &           & $\spmid$  & \elam{\evar}{\expr}            & \emph{functions} \\
  &           & $\spmid$  & \eapp{\expr}{\evar}            & \emph{application} \\
  &           & $\spmid$  & \eann{\expr}{\type}            & \emph{type-annotation} \\[0.05in]
\end{tabular}
\caption{{Syntax of Terms}}
\label{fig:one:terms}
\end{figure}

\section{Declarative Typing}\label{one:decl}

We are now ready to look at the different judgments and 
rules that establish when a term $\expr$ \emph{has} 
type $\type$.

\subsection{Substitution}
We use the notation \SUBST{\type}{y}{z} to denote the type 
where all free occurrences of $y$ are substituted with the
variable $z$. 
That is, substitution is defined in a capture avoiding
manner:
\begin{align*}
  \SUBST{\refb{\base}{\reft{\vvar}{\pred}}}{\vvar}{z} & \doteq \refb{\base}{\reft{\vvar}{\pred}} \\ 
  \SUBST{\refb{\base}{\reft{\vvar}{\pred}}}{y}{z}     & \doteq \refb{\base}{\reft{\vvar}{\SUBST{\pred}{y}{z}}}  \\ 
  \SUBST{(\trfun{x}{\typeb}{\type})}{x}{z}            & \doteq \trfun{x}{\SUBST{\typeb}{x}{z}}{\type} \\
  \SUBST{(\trfun{x}{\typeb}{\type})}{y}{z}            & \doteq \trfun{x}{\SUBST{\typeb}{y}{z}}{\SUBST{\type}{y}{z}}
\end{align*}

\subsection{Judgments}

\begin{figure}[t!]

\judgementHead{Well-formedness}{\wf{\tcenv}{\type}{\kind}}

\begin{mathpar}
\inferrule
  {\wfr{\tcenvext{\evar}{\base}}{\pred}}
  {\wf{\tcenv}{\refb{\base}{\reft{\evar}{\pred}}}{\basekind}}
  {\ruleName{Wf-Base}}

\inferrule
  { \wf{\tcenv}{\typeb}{\kind_\typeb} \\
    \wf{\tcenvext{\evar}{\typeb}}{\type}{\kind_\type}
  }
  {\wf{\tcenv}{\trfun{\evar}{\typeb}{\type}}{\starkind}}
  {\ruleName{Wf-Fun}}

\inferrule
  {\wf{\tcenv}{\type}{\basekind}}
  {\wf{\tcenv}{\type}{\starkind}}
  {\ruleName{Wf-Kind}}
\end{mathpar}
\caption{Well-formedness of types}
\label{fig:one:wf}
\end{figure}

A \emph{context} $\tcenv$ is a sequence of variable-type 
bindings $\tb{\evar_1}{\type_1},\ldots, \tb{\evar_n}{\type_n}$. 
The type system uses contexts to define five kinds of judgments.

\mypara{Well-formedness} judgments $\wf{\tcenv}{\type}{\kind}$ 
state that in the context $\tcenv$ the type $\type$ 
is \emph{well-formed} with kind \kind, \ie, that each 
refinement is a @bool@-valued proposition over variables 
bound in the context or type. 
The rule \ruleName{Wf-Base} says that a base type 
$\refb{\base}{\reft{\evar}{\pred}}$ is well-formed 
with base kind in a context $\tcenv$ if the refinement $\pred$ is 
a well-sorted predicate in the context extended 
with $\evar$. 
Well-sortedness of predicates \wfr{\tcenv}{\pred}
is using the standard, unrefined type checking to check that 
the predicate $\pred$ is boolean under the environment $\tcenv$
with all refinements erased.
%
%
The rule \ruleName{Wf-Fun} says that a function type
$\trfun{\evar}{\typeb}{\type}$ is well-formed with star kind 
in a context $\tcenv$ if the \emph{input} type $\typeb$ 
is well-formed for some kind under $\tcenv$ and the \emph{output} 
type is well-formed for some kind under the context extended with 
the parameter $\evar$, thereby allowing refinements 
in the output to \emph{depend upon} the inputs. 
The rule \ruleName{Wf-Kind} says that any well formed type
with base kind also has star kind. 

The next two judgments formalize when the set 
of values of one type are \emph{subsumed by} 
(\ie contained within) another.

\begin{figure}[t!]

\judgementHead{Entailment}{\entl{\tcenv}{\cstr}}

\begin{mathpar}
\inferrule
  {\smtvalid{\cstr}}
  {\entl{\emptyset}{\cstr}}
  {\ruleName{Ent-Emp}}

\inferrule
  {\entl{\tcenv}{\csimp{x}{\base}{\pred}{\cstr}}}
  {\entl{\tcenvext{x}{\refb{\base}{{\reft{x}{\pred}}}}}{\cstr}}
  {\ruleName{Ent-Ext}}
\end{mathpar}

\judgementHead{Subtyping}{\issub{\tcenv}{\type_1}{\type_2}}

\begin{mathpar}
\inferrule
  {\entl{\tcenv}{\csimp{\vvar_1}{\base}{\pred_1}{\SUBST{\pred_2}{\vvar_2}{\vvar_1}}}}
  {\issub{\tcenv}{\refb{\base}{\reft{\vvar_1}{\pred_1}}}{\refb{\base}{\reft{\vvar_2}{\pred_2}}}}
  {\ruleName{Sub-Base}}

\inferrule
  {\issub{\tcenv}{\typeb_2}{\typeb_1} \\
   \issub{\tcenvext{\evar_2}{\typeb_2}}{\SUBST{\type_1}{\evar_1}{\evar_2}}{\type_2}
  }
  {\issub{\tcenv}{\trfun{\evar_1}{\typeb_1}{\type_1}}{\trfun{\evar_2}{\typeb_2}{\type_2}}}
  {\ruleName{Sub-Fun}}
\end{mathpar}

\caption{Entailment and Subtyping}
\label{fig:one:entail}
\label{fig:one:subtype}
\end{figure}

\mypara{Entailment} judgments $\entl{\tcenv}{\cstr}$ 
state that in the context $\tcenv$ the logical 
constraint $\cstr$ is \emph{valid} \ie, ``is true''. 
For example the entailment judgment
\begin{equation}
\entl
  {\tb{x}{\refb{\tint}{\rreft{0 \leq x}}};\
   \tb{y}{\refb{\tint}{\rreft{y = x + 1}}} 
  }
  {
   0 \leq y 
  }  \label{ex:entl}
\end{equation}
reduces, via the rule {\ruleName{Ent-Ext}}, to 
the \emph{verification condition}
\begin{equation}
\csimp{x}{\tint}{0 \leq x}{\csimp{y}{\tint}{y = x + 1}{0 \leq y}}
\label{ex:vc}
\end{equation}
which is deemed to be \emph{valid} by an SMT solver.

\mypara{Subtyping} judgments $\issub{\tcenv}{\type_1}{\type_2}$
state that $\type_1$ is a subtype of $\type_2$ in a typing context 
$\tcenv$ comprising a sequence of type bindings. 
The rule \ruleName{Sub-Base} reduces subtyping on (refined) basic 
types into an entailment. For example, the subtyping judgment 
\begin{equation}
  \issub
    {\tb{x}{\refb{\tint}{\rreft{0 \leq x}}}}
    {\refb{\tint}{\reft{y}{y = x + 1}}}
    {\refb{\tint}{\reft{\vvar}{0 \leq \vvar}}}
  \label{ex:sub:base}
\end{equation}
reduces to the entailment (\ref{ex:entl}), which is deemed valid 
by SMT.
The rule \ruleName{Sub-Fun} decomposes subtyping on functions into 
\emph{contra-variant} subtyping on the input types, and 
\emph{co-variant} subtyping on the output types.
For example, the subtyping judgment 
\begin{equation}
    \issub{\emptyset} 
    {\trfun{x}{\tint}{\refb{\tint}{\reft{y}{y = x + 1}}}}
    {\trfun{x}{\refb{\tint}{\rreft{0 \leq x}}}{\refb{\tint}{\reft{\vvar}{0 \leq \vvar}}}}
    \label{ex:sub:fun}
\end{equation}
holds because it reduces to following judgments on 
the respective input and output types:
\begin{align}
    {\emptyset} & \vdash
    {\refb{\tint}{\reft{x}{0 \leq x}}} 
    \lqsubt
    {\tint}
    \label{ex:sub:fun:in} \\
    {\tb{x}{\refb{\tint}{\rreft{0 \leq x}}}} & \vdash
    {\refb{\tint}{\reft{y}{y = x + 1}}} 
    \lqsubt
    {\refb{\tint}{\reft{\vvar}{0 \leq \vvar}}}
    \label{ex:sub:fun:out}
\end{align}
The former holds trivially, as $\entl{\tcenv}{\ttrue}$. 
The latter subtyping judgment is the same as (\ref{ex:sub:base}), 
and so, holds via the entailment (\ref{ex:entl}).

\mypara{Bidirectional Typing}
The next two kinds of judgments present typing in a 
\emph{bidirectional} style \citep{pierce-turner,bidir-survey}, 
where we separate the terms where types are \emph{checked} 
from those for which the types are \emph{synthesized}.

\begin{itemize}
  \item \emphbf{Synthesis} judgments $\tsyn{\tcenv}{\expr}{\type}$ 
    state that the type $\type$ can be \emph{generated} for  
    the term $\expr$ in the context $\tcenv$.
%

  \item \emphbf{Checking} judgments $\tchk{\tcenv}{\expr}{\type}$
  state that a given type $\type$ is indeed \emph{valid} 
  for a term $\expr$ in the context $\tcenv$, by \emph{pushing}
  typing goals for terms into typing goals for sub-terms.
%
\end{itemize}

Lets take a closer look at the checking and synthesis rules.

\subsection{Synthesis}\label{one:decl:synth}

Figure~\ref{fig:one:synth} shows the rules for deriving synthesis 
judgments $\tsyn{\tcenv}{\expr}{\type}$ for terms $\expr$ 
whose type can be generated from the context $\tcenv$.

\begin{figure}[t!]
\judgementHead{Type Synthesis}{\tsyn{\tcenv}{\expr}{\type}}
\begin{mathpar}

\inferrule
  {
    \tcenvget{\evar} = \type
  }
  {
    \tsyn{\tcenv}{\evar}{\type}
  }
  {\ruleName{Syn-Var}}

\inferrule
  {
    \constty{\vconst} = \type
  }
  {
    \tsyn{\tcenv}{\vconst}{\type}
  }
  {\ruleName{Syn-Con}}

\inferrule
  {
    \wf{\tcenv}{\type}{\kind} 
    \quad
    \tchk{\tcenv}{\expr}{\type} 
  }
  {
    \tsyn{\tcenv}{\eann{\expr}{\type}}{\type}
  }
  {\ruleName{Syn-Ann}}

\inferrule
  {
    \tsyn{\tcenv}{\expr}{\trfun{\evar}{\typeb}{\type}} 
    \quad
    \tchk{\tcenv}{\evarb}{\typeb}
  }
  {
    \tsyn{\tcenv}{\eapp{\expr}{\evarb}}{\SUBST{\type}{\evar}{\evarb}}
  }
  {\ruleName{Syn-App}}
\end{mathpar}
\caption{Bidirectional Typing: Synthesis}
\label{fig:one:synth}
\end{figure}

\mypara{Variables} $\evar$ synthesize the type ascribed to $\evar$ 
in the context $\tcenv$ (\rulename{Syn-Var}). For example, we can 
deduce that $\tsyn{\tcenv_0}{\evar}{\tnat}$ when 
\begin{equation}
\tcenv_0 \ \doteq\  \tb{\evar}{\tnat};\ \tb{{one}}{\rreft{{one} = 1}} 
  \label{eq:syn:env} \\
\end{equation}

\mypara{Constants} $\vconst$ synthesize their ``built-in'' 
primitive type denoted by $\constty{\vconst}$, which is 
usually the most precise reflection of the semantics of 
the constant that can be represented in the refinement 
logic (\rulename{Syn-Con}). 
For example, primitive $\tint$ values like $0$ and $1$ 
are assigned the \emph{singleton} types inhabited only 
by those values 
\begin{align*} 
\constty{0} &\ \doteq\ \refb{\tint}{\reft{\vvar}{\vvar = 0}} \\
\constty{1} &\ \doteq\ \refb{\tint}{\reft{\vvar}{\vvar = 1}} \\
\intertext{and arithmetic operators have primitive types that 
reflect their semantics}
\constty{{add}} &\ \doteq\ {\trfun{x}{\tint}{\trfun{y}{\tint}{\refb{\tint}{\reft{\vvar}{\vvar = x + y}}}}} \\
\constty{{sub}} &\ \doteq\ {\trfun{x}{\tint}{\trfun{y}{\tint}{\refb{\tint}{\reft{\vvar}{\vvar = x - y}}}}}
\end{align*} 
For exposition, we deliberately use ${add}$ and ${sub}$ 
for the primitive arithmetic operators of the language, 
to syntactically distinguish the names from $+$ and $-$ 
in the refinement logic. 

\mypara{Applications} $\eapp{\expr}{\evar}$ synthesize the 
\emph{output} type of $\expr$ after substituting the input
binder with the actual $\evar$ (\rulename{Syn-App}).
In the environment $\tcenv_0$ from (\ref{eq:syn:env}), 
the term $\eapp{\eapp{{add}}{\evar}}{{one}}$, would synthesize the type 
\begin{equation}
  \tsyn{\tcenv_0}{\eapp{\eapp{{add}}{\evar}}{{one}}}{\refb{\tint}{\reft{\vvar}{\vvar = \evar + {one}}}}
  \label{eq:syn:app}
\end{equation}

\mypara{Applications \& ANF}
%
%
The rule \rulename{Syn-App} returns the function's output type 
after substituting the input
binder with the actual argument.
We require 
that the application terms be in \emph{Administrative Normal Form} (ANF) 
so that this substitution only replaces binders with other
binders, and not arbitrary expressions.
This restriction ensures that all the intermediate refinements 
produced during type checking belong to the same (decidable) 
fragment of the refinement logic that the user-defined 
specifications are from. In particular, it prevents 
arbitrary terms from seeping into the refinements, 
which would complicate SMT-based subtyping.
\cite{Knowles09} propose an alternative dependent application 
rule that uses an \emph{existential type} to ensure that terms 
do not seep into refinements:
$$
\inferrule
  {
    \tsyn{\tcenv}{\expr_1}{\trfun{\evar}{\typeb}{\type}} \\
    \tsyn{\tcenv}{\expr_2}{\typeb}
  }
  {
    \tsyn{\tcenv}{\eapp{\expr}{\evarb}}{\existy{\evar}{\typeb}{\type}} 
  }
  {\ruleName{Syn-App-Ex}}
$$
The above rule ensures that the existentials only appear 
on the left-side of subtyping obligations, at which point they can simply 
be pulled into the environment, \ie via a subtyping rule:
$$
\inferrule 
  {
    \issub
      {\tcenvext{\evar}{\typeb}}
      {\type_1}
      {\type_2} 
  }
  {
    \issub{\tcenv}{\existy{\evar}{\typeb}{\type_1}}{\type_2} 
  }
  {\ruleName{Sub-Ex}}
$$
This existential-based rule requires an extra subtyping step, 
but has the benefit of not requiring terms to be in ANF which 
is problematic for the meta-theory, as the small-step evaluation 
does not preserve ANF structure. 
However, apart from the meta-theory, the two rules are equivalent 
and we pick ANF for ease of exposition and implementation.

For brevity and readability, we will present 
programs that do not follow the ANF structure 
and assume they are converted to ANF form before 
type checking. 

\mypara{Annotated} terms $\eann{\expr}{\type}$ synthesize the 
(annotated) type $\type$, after ensuring that the annotation 
$\type$ is well-formed in the given context and checking that 
$\expr$ indeed can be ascribed the type $\type$ (\ruleName{Syn-Ann}).

\subsection{Checking}\label{one:decl:check}

Figure~\ref{fig:one:check} shows the rules for deducing checking 
judgments $\tchk{\tcenv}{\expr}{\type}$. 
Typically, we use these judgments to verify that a $\lambda$-term 
has its given (annotated) type, and to push those top-level obligations 
inside let-binders to get localized obligations for the inner expressions. 

\begin{figure}[t!]
\judgementHead{Type Checking}{\tchk{\tcenv}{\expr}{\type}}
\begin{mathpar}
\inferrule
  {
    \tsyn{\tcenv}{\expr}{\typeb} \\
    \issub{\tcenv}{\typeb}{\type}
  }
  {
    \tchk{\tcenv}{\expr}{\type}
  }
  {\ruleName{Chk-Syn}}

\inferrule
  {
    \tchk{\tcenvext{\evar}{\type_1}}{\expr}{\type_2}
  }
  {
    \tchk{\tcenv}{\elam{\evar}{\expr}}{\trfun{\evar}{\type_1}{\type_2}}
  }
  {\ruleName{Chk-Lam}}

\inferrule
  {
    \tsyn{\tcenv}{\expr_1}{\type_1} \\
    \tchk{\tcenvext{\evar}{\type_1}}{\expr_2}{\type_2}
  }
  {
    \tchk{\tcenv}{\elet{\evar}{\expr_1}{\expr_2}}{\type_2}
  }
  {\ruleName{Chk-Let}}
\end{mathpar}
\caption{Bidirectional Typing: Checking}
\label{fig:one:check}
\end{figure}

\mypara{Functions} $\elam{\evar}{\expr}$ can be 
checked against the type $\trfun{\evar}{\typeb}{\type}$ 
in a context $\tcenv$ if their bodies $\expr$ 
can be checked against the output type $\type$ 
in the environment extended by binding the 
parameter to the input type $\typeb$ (\ruleName{Chk-Lam}).
For example, we check @inc@ (from \cref{sec:one:examples}) 
against its ascribed type
$$
\tchk
  {\emptyset}
  {\elam{\evar}{\elet{{one}}{1}{\pgadd{\evar}{{one}}}}}
  {\trfun{\evar}{\tnat}{\refb{\tint}{\reft{\vvar}{x < \vvar}}}}
$$
by checking the body against the output type in the extended context
\begin{equation}
  \tchk
    {\tb{\evar}{\tnat}}
    {\elet{{one}}{1}{\eapp{\eapp{{add}}{\evar}}{{one}}}}
    {\refb{\tint}{\reft{\vvar}{\evar < \vvar}}}
  \label{eq:chk:incr:out}
\end{equation}

\mypara{Let-bindings} $\elet{\evar}{\expr_1}{\expr_2}$ can be checked against 
the type $\type_2$ if we can check that $\expr_2$ has type $\type_2$ in the 
environment extended by binding $\evar$ to the type $\type_1$ synthesized for 
$\expr_1$ (\ruleName{Chk-Let}).
In effect, this rule \emph{pushes} the obligation $\type_2$ for the whole 
expression, into an obligation for the let-body $\expr_2$.
For example, the judgment (\ref{eq:chk:incr:out}) reduces to the 
below check that pushes the obligation inside:
\begin{equation}
\tchk
  {\tb{\evar}{\tnat};\ \tb{{one}}{\rreft{{one} = 1}}}
  {\pgadd{\evar}{{one}}}
  {\refb{\tint}{\reft{\vvar}{\evar < \vvar}}}
  \label{eq:chk:incr:out:body}
\end{equation}
That is, we must check the term $\pgadd{\evar}{{one}}$ in the environment 
extended by binding the local ${one}$ to the type synthesized from
its (constant) expression $1$.

\mypara{Subsumption} rule \rulename{Chk-Syn} lets us 
connect the checking and synthesis judgments: if the 
term $\expr$ synthesizes a type $\typeb$ which is 
subsumed by $\type$, then we can check $\expr$ 
against $\type$.
For example, the judgment (\ref{eq:chk:incr:out:body})
that checks the body of @inc@ is established by using 
\rulename{Chk-Syn} to yield the obligation
\begin{equation}
\issub{\tb{\evar}{\tnat};\ \tb{{one}}{\rreft{ {one} = 1}}}{\refb{\tint}{\reft{\vvar}{\vvar = \evar + {one}}}}{\refb{\tint}{\reft{\vvar}{\evar < \vvar}}}\label{ex:sub}
\end{equation}
which checks that the type synthesized for
${\eapp{\eapp{{add}}{\evar}}{{one}}}$
(\ref{eq:syn:app}) is a subtype of the goal 
${\refb{\tint}{\reft{\vvar}{\evar < \vvar}}}$.

Readers familiar with Floyd-Hoare program logics may 
be reminded of the \emph{consequence rule}
$$
\inferrule
  {
    P \Rightarrow P' \quad 
    \rreft{P'}\ C\ \rreft{Q'} \quad 
    Q' \Rightarrow Q
  }
  {
    \rreft{P}\ C\ \rreft{Q}
  }
$$
which allows us to strengthen the preconditions 
and weaken the post-conditions (via implication) 
for a command $C$ similar to how \rulename{Chk-Sym} 
lets us weaken the type for a term $\expr$ (via subtyping).

\section{Verification Conditions}\label{one:algo}

The typing rules show how refinement verification works 
in a \emph{declarative} fashion: lets finish our discussion 
with a concrete \emph{implementation} of a verifier. 
For readers familiar with program logics, the declarative 
rules are akin to Floyd-Hoare style rules. 
Instead, we will describe an implementation of a 
\emph{verification condition} (VC) generator that takes 
as input a program annotated with refinement types, and  
returns a VC, a constraint (\S~\ref{sec:smt:constr})
whose \emph{validity} can (a)~be determined by an SMT 
solver, and (b)~implies the typability of the program.
In particular, we will describe the implementation 
of three functions, that are algorithmic counterparts 
of the respective typing judgments.


\begin{figure}[t]
$$\begin{array}{lcl}
\toprule
\subsym & : & (\Type \times \Type) \rightarrow \Cstr \\
\midrule
\sub{\refb{\base}{\reft{\vvar_1}{\pred_1}}}{\refb{\base}{\reft{\vvar_2}{\pred_2}}}
  & \doteq
  & \csimp{\vvar_1}{\base}{\pred_1}{\SUBST{\pred_2}{\vvar_2}{\vvar_1}} \\[0.05in]

\sub{\trfun{\evar_1}{\typeb_1}{\type_1}}{\trfun{\evar_2}{\typeb_2}{\type_2}}
  & \doteq
  & \csand{\cstr_i}{ \cswith{\evar_2}{\typeb_2}{\cstr_o} } \\
\quad \mbox{where} & & \\
\quad \quad \cstr_i
  & =
  & \sub{\typeb_2}{\typeb_1} \\
\quad \quad \cstr_o
  & =
  & {\sub{\SUBST{\type_1}{\evar_1}{\evar_2}}{\type_2} } \\[0.05in]
\bottomrule
\end{array}$$
\caption{Algorithmic Subtyping for $\langone$}
\label{fig:one:algo:sub}
\end{figure}

\mypara{Implication Constraints}
We write $\cswith{\evar}{\type}{\cstr}$ for the \emph{implication} 
constraint which is defined as: 
$$
\cswith{\evar}{\type}{\cstr} \ \doteq\  
  \begin{cases}
    \csimp{\evar}{\base}{\SUBST{\pred}{\vvar}{\evar}}{\cstr}
      & \mbox{if}\ \type\ \equiv\ \refb{\base}{\reft{\vvar}{\pred}} \\
    \cstr & \mbox{otherwise}
  \end{cases}
$$

\mypara{Subtyping} $\sub{\typeb}{\type}$ summarized 
in \cref{fig:one:algo:sub}, mirrors the subtyping 
rules \cref{fig:one:subtype}. 
The function takes as input two types $\typeb$ and $\type$ 
and returns as output a constraint $\cstr$ whose validity 
implies that $\typeb$ is a subtype of $\type$:

\begin{proposition} \label{prop:sub}
  If $\sub{\typeb}{\type} = \cstr$ and $\entl{\tcenv}{\cstr}$, 
  then $\issub{\tcenv}{\typeb}{\type}$.
\end{proposition}

The two cases shown in \cref{fig:one:algo:sub} correspond 
directly to the rules \rulename{Sub-Base} and \rulename{Sub-Fun}
from \cref{fig:one:subtype}.
For refined base types, the generated constraint states 
that sub-refinement $\pred_1$ implies the super-refinement 
$\pred_2$.
For function types, we recursively invoke $\subsym$ 
to conjoin the constraint on the input type and an 
implication constraint that checks the output 
subtyping holds assuming the stronger input type.

\begin{myexample}{Subtyping Constraint}
For example, let $\typeb$ and $\type$ respectively 
be the sub- and super-type in subtyping 
judgment (\ref{ex:sub:fun}).
Then $\sub{\typeb}{\type}$ returns the VC constraint:
\begin{align*}
        & \csimp{x}{\tint}{0 \leq x}{\ttrue} \\
\wedge\ & \csimp{x}{\tint}{0 \leq x}{\csimp{y}{\tint}{y = x + 1}{0 \leq y}}
\end{align*}
whose first and second conjuncts correspond to 
the input (\ref{ex:sub:fun:in}) and 
output (\ref{ex:sub:fun:out}) 
subtyping obligations respectively.
\end{myexample}


\begin{figure}[t]
$$\begin{array}{lcl}
\toprule
\synsym & : & (\tcenv \times \Expr) \rightarrow (\Cstr \times \Type) \\
\midrule
\syn{\tcenv}{\evar}
  & \doteq & (\ttrue,\ \tcenvget{\evar}) \\[0.05in]

\syn{\tcenv}{\vconst}
  & \doteq & (\ttrue,\ \constty{\vconst}) \\[0.05in]

\syn{\tcenv}{\eapp{\expr}{\evarb}}
  & \doteq & (\csand{\cstr}{\cstr'},\ \SUBST{\type}{\evar}{\evarb}) \\
\quad \mbox{where} & & \\
\quad \quad (\cstr,\ \trfun{\evar}{\typeb}{\type})
  & =      & \syn{\tcenv}{\expr} \\
\quad \quad \cstr'
  & =      & \chk{\tcenv}{\evarb}{\typeb} \\[0.05in]

\syn{\tcenv}{\eann{\expr}{\type}}
  & \doteq & (\cstr,\ \type) \\
\quad \mbox{where} & & \\
\quad \quad \cstr
  & =      & \chk{\tcenv}{\expr}{\type}\\[0.05in]

\bottomrule
\end{array}$$
\caption{Algorithmic Synthesis for $\langone$}
\label{fig:one:algo:syn}
\end{figure}

\mypara{Synthesis} $\syn{\tcenv}{\expr}$ summarized in 
\cref{fig:one:algo:syn}, is the analogue of the synthesis 
rules from \cref{fig:one:synth}. 
The function takes as input a context $\tcenv$ 
and a term $\expr$ whose type we want to synthesize under 
$\tcenv$ and returns a pair $(\cstr,\ \type)$ such 
that the validity of $\cstr$ implies that $\expr$ 
synthesizes type $\type$:

\begin{proposition} \label{prop:syn}
  If $\syn{\tcenv}{\expr} = (\cstr,\ \type)$ and $\entl{\tcenv}{\cstr}$,
  then $\tsyn{\tcenv}{\expr}{\type}$.
\end{proposition}

The cases for variables $\evar$ and constants $\vconst$ simply 
return the types for the respective element by looking up the 
context $\tcenv$ or the primitive type respectively.
In both these cases, the constraint (VC) is simply 
$\ttrue$ as the synthesized types hold unconditionally 
as shown in \rulename{Syn-Var} and \rulename{Syn-Con} 
from \cref{fig:one:synth}.
For application terms $\eapp{\expr}{\evarb}$, we recursively 
invoke $\synsym$ to synthesize the type and VC for the function $\expr$,
invoke $\chksym$ to generate a VC that holds if the argument 
$\evarb$ is of the expected input type, and then return 
the conjunction of the VCs for the function and argument,
together with the function's output type.
For annotated terms $\eann{\expr}{\type}$ we generate the VC 
by invoking $\chksym$, mimicking \rulename{Syn-Ann}.
Consider the following environment that arises in the body 
of @inc2@:
\begin{align*}
\tcenv_2 \ \doteq  & \ \tb{{inc}}{\trfun{x}{\tnat}{\refb{\tint}{\reft{\vvar}{x < \vvar}}}};\  
                     \tb{\evarb}{\tnat};\ \tb{\evarb_1}{\rreft{\evarb_1 = \evarb - 1}} \\
\intertext{Here, $\syn{\tcenv_2}{\eapp{{inc}}{\evarb_1}}$ returns the VC and type}
\cstr     \ \doteq & \ \csimp{\vvar}{\tint}{\vvar = \evarb_1}{0 \leq \vvar}  \\ 
\type     \ \doteq & \ \refb{\tint}{\reft{\vvar}{\evarb_1 < \vvar}} 
\end{align*}
where the VC $\cstr$ checks that the input $\evarb_1$ meets 
the required precondition of ${inc}$ (\ie is a $\tnat$), and 
the synthesized $\type$ is the output of ${inc}$ with 
the arguments $\evarb_1$ substituted for the formal 
$\evar$.


\begin{figure}[t!]
$$\begin{array}{lcl}
\toprule
\chksym & : & (\tcenv \times \Expr \times \Type) \rightarrow \Cstr \\
\midrule
\chk{\tcenv}{\elam{\evar}{\expr}}{\trfun{\evar}{\typeb}{\type}}
                                & \doteq & \cswith{\evar}{\typeb}{\cstr} \\
\quad \mbox{where}              &        & \\
\quad \quad \cstr               & =      & \chk{\tcenvext{\evar}{\typeb}}{\expr}{\type} \\[0.05in]

\chk{\tcenv}{\elet{\evar}{\expr_1}{\expr_2}}{\type_2}
                                & \doteq & \csand{\cstr_1}{\cswith{\evar}{\type_1}{\cstr_2}} \\
\quad \mbox{where} & & \\
\quad \quad (\cstr_1, \type_1)  & =      & \syn{\tcenv}{\expr_1} \\
\quad \quad \cstr_2             & =      & \chk{\tcenvext{\evar}{\type_1}}{\expr_2}{\type_2} \\[0.05in]

\chk{\tcenv}{\expr}{\type}      & \doteq & \csand{\cstr}{\cstr'} \\
\quad \mbox{where} & & \\
\quad \quad (\cstr, \typeb)     & =      & \syn{\tcenv}{\expr} \\
\quad \quad \cstr'              & =      & \sub{\typeb}{\type} \\[0.05in]
\bottomrule
\end{array}$$
\caption{Algorithmic Checking for $\langone$}
\label{fig:one:algo:chk}
\end{figure}

\mypara{Checking} $\chk{\tcenv}{\expr}{\type}$ summarized 
in \cref{fig:one:algo:chk}, implements the checking judgment. 
The function takes as input a context $\tcenv$, a term 
$\expr$, and a type $\type$ we want to check for $\expr$
and returns as output a constraint $\cstr$ whose 
validity implies that $\expr$ checks against $\type$:

\begin{proposition} \label{prop:chk}
  If $\chk{\tcenv}{\expr}{\type} = \cstr$ and $\entl{\tcenv}{\cstr}$, 
  then $\tchk{\tcenv}{\expr}{\type}$.
\end{proposition}

The first case for terms $\elam{\evar}{\expr}$ implements 
\rulename{Chk-Lam} from \cref{fig:one:check}, by using 
$\chksym$ to generate the VC for the body $\expr$ 
under the environment extended with the type for the 
formal $\evar$, which is added as an antecedent to the
returned VC.
The second case for terms $\elet{\evar}{\expr_1}{\expr_2}$,
follows \rulename{Chk-Let} to synthesize a VC 
$\cstr_1$ and type $\type_1$ for $\expr_1$, 
which is used to generate a VC $\cstr_2$ for 
the body $\expr_2$. The VCs for the two terms 
are conjoined after weakening the body's with 
antecedent constraining the local binder $\evar$.
The last case implements the subsumption rule 
\rulename{Chk-Syn} by generating a VC $\cstr$ 
and type $\typeb$ for $\expr$ and then conjoining 
$\cstr$ with the VC stating $\typeb$ is subsumed 
by the (goal) super-type $\type$.

\begin{myexample}{VC for \texttt{inc}}
Lets see how $\chksym$ and $\synsym$ yield
a VC for @inc@ from \S~\ref{sec:one:examples}. 
Let
\begin{align*}
\tcenv_0 
  & \ \doteq\ \tb{\evar}{\tnat};\ \tb{{one}}{\rreft{{one} = 1}} \\
\type_0  
  & \ \doteq\ \refb{\tint}{\reft{\vvar}{\evar < \vvar}} \\
\intertext{The invocation $\chk{\tcenv_0}{\eapp{\eapp{{add}}{\evar}}{{one}}}{\type_0}$
matches the last case (\rulename{Chk-Syn}) to return the VC}
  \cstr_0 & \ \doteq\ \csimp{\vvar}{\tint}{\vvar = \evar + {one}}{\evar < \vvar} \\
\intertext{The antecedent in $\cstr_0$ comes 
from the (sub-) type synthesized by the call 
$\syn{\tcenv_0}{\pgadd{\evar}{{one}}}$ (\ref{eq:syn:app}), 
and the consequent comes from the (super-) type 
corresponding to the goal $\type_0$ (\ref{ex:sub}). Next, let}
\tcenv_1 
  & \ \doteq\ \tb{\evar}{\tnat} \\
\expr_1 
  & \ \doteq\ \elet{{one}}{1}{\eapp{\eapp{{add}}{\evar}}{{one}}} \\
\intertext{The invocation $\chk{\tcenv_1}{\expr_1}{\type_0}$
matches the second case (\ie \rulename{Chk-Let}) to return the VC}
\cstr_1 
  & \ \doteq\ \csimp{{one}}{\tint}{{one} = 1}{\cstr_0} \\
\intertext{obtained by recursively invoking $\chksym$ on the body 
to get $\cstr_0$ and then weakening it with the type 
synthesized by $\synsym$ for the binder ${one}$. 
Finally, $\chk{\emptyset}{\elam{\evar}{\expr_1}}{\trfun{\evar}{\tnat}{\type_0}}$
matches the first case (\ie \rulename{Chk-Lam}) to return the VC}
\cstr_2 
  &\ \doteq\ \csimp{\evar}{\tint}{0 \leq \evar}{\cstr_1}
\end{align*}
which is the VC for the body $\expr_1$ weakened with   
the type for the binder $\evar$. The constraint $\cstr_2$ 
is the formula
$$
  \csimp{\evar}{\tint}{0 \leq \evar}
    {
      \csimp{{one}}{\tint}{{one} = 1}
       {
         \csimp{\vvar}{\tint}{\vvar = \evar + {one}}{\evar < \vvar} 
       }
    }
$$
which is proved valid by the SMT solver, thereby verifying @inc@.
\end{myexample}

\section{Discussion} \label{one:summary}

Before we move on to our next language 
feature lets ponder some key 
lessons learned from developing refinement 
types for $\langone$.

\mypara{Primitive Types} connect the semantics 
of terms and refinements. That is, one can think 
of the semantics of the program at \emph{two} levels.
First, the concrete \emph{dynamic} (``operational'') 
semantics corresponding to how terms reduce to values.
Second, the abstract or \emph{static} (``logical'') 
semantics that describe the sets of values that a 
term can reduce to.
The types of the primitive constants 
\eg $\constty{\vconst}$ are the glue 
that connects the two semantics. 
We start by giving primitive constants 
like @3@ and @4@ very precise \emph{singleton}
types like $\refb{\tint}{\reft{\vvar}{\vvar = 3}}$
that reflect their concrete semantics in the 
refinement logic. 
Next, we give arithmetic operators 
like $\mw{add}$ similarly precise 
types like 
$\trfun{x}{\tint}{\trfun{y}{\tint}{\refb{\tint}{\reft{\vvar}{\vvar = x + y}}}}$,
and after this, the function application and 
subtyping suffice to logically (overapproximate) 
the sets of values that a term can reduce.

\mypara{Arithmetic Overflow}
Consequently, we can use different specifications 
to more precisely track machine operations. 
For example, for simplicity of exposition, we 
give $\mw{add}$ the type that says that the returned 
integer is in fact the logical (mathematical) 
addition of the two arguments. However, this may 
not hold if $\tint$ is implemented as 32- or 64- bit 
integers, which may \emph{overflow}. 
However, it is quite straightforward to 
write more restrictive specifications for 
primitives like $\mw{add}$ such that 
verification statically guarantees 
the absence of arithmetic overflows~\citep{overflow-blog-lh}.

\mypara{Assumptions for Soundness}
Our system relies on the assumption that 
the primitives of the language satisfy 
their specified types $\constty{\vconst}$.
This assumption does not always hold.
For example, returning to overflows, 
we can @inc@rease the ``maximum'' 
(fixed-width) integer by one 
(@inc(maxInt, 1)@) to get back 
a result than is \emph{smaller} 
than the input, violating the 
specification of @inc@. 
Similarly, one can break the 
system when primitive operations, 
like addition, equality, comparisons, \etc 
do not satisfy the laws of the respective 
primitive logic operators. 
It is important to note that the goal 
of our refinement type system is not 
to validate these assumptions but instead, 
to verify more sophisticated properties 
on a programming model where these 
assumptions hold. 

\mypara{Types as Program Logics} 
The development for $\langone$ already shows that 
refinement types can be viewed as a generalization
of Floyd-Hoare style program logics.
Such logics typically have \emph{monolithic} 
assertions that describe the entire state of 
the machine at a given program point.
Types allow us to \emph{decompose} those 
assertions into more fine-grained refinements 
on the values of individual terms.
Similarly, pre- and post-conditions correspond 
directly to input- and output-types for functions.
The function application rule \emph{checks} 
pre-conditions (input types) and \emph{assumes} 
the returned value satisfies the post-condition (output type).
Dually, the function definition rule \emph{assumes} the 
pre-conditions (input types) hold for the input parameters 
and \emph{checks} that the returned value satisfies the 
post-condition (output type).

\mypara{Higher-Order Contracts} 
One immediate benefit of using types instead of monolithic 
assertions, is that they naturally scale up to handling 
higher-order functions, such as @incf@ shown below.

\begin{code}
  val incf: x:nat => pos
  let incf = (x) => {
    val tmp : f:(nat => nat) => pos
    let tmp = (f) => { 
      add(f(x), 1) 
    };
    tmp(inc)
  };
\end{code}

The specification for @incf@ says that if it is given a @nat@
value @x@ then it also returns a @nat@. To implement this contract, 
the function creates a higher-order @tmp@ which increments the value 
returned by invoking its argument @f@ on @x@. Types help in two ways.
First, they let us \emph{specify} a suitable contract for @f@, namely 
it takes a @nat@ and returns one. 
Second, they let us \emph{verify} the application @tmp(inc)@ 
by using the function subtyping rule \rulename{Sub-Fun} to 
check that @inc@ (\cref{sec:one:examples}) whose type is 
$\trfun{z}{\tint}{\refb{\tint}{\reft{\vvar}{z < \vvar}}}$
is a valid input for @tmp@, hence verifying the call, 
and that @incf@ returns a @pos@.
While this example is contrived, we will see how this form 
of type-directed decomposition of the verification goals
greatly simplifies the verification of programs with 
polymorphic data and higher-order functions \citep{types-v-logic-blog-lh}.
\chapter{Branches and Recursion}
\label{sec:lang:two} 

The programs one can write in $\langone$ are dreadfully 
predictable: they compute simple \emph{arithmetic} 
expressions over their inputs. 
Next, lets study $\langtwo$ which enriches $\langone$ 
with two constructs -- conditional branching and 
recursion -- that are essential to facilitate 
\emph{computation}.
Consequently, we will see how to extend typing 
and VC generation to account for these constructs 
to enable \emph{path-sensitive} verification 
that precisely accounts for the conditions 
accumulated along the course of evaluation. 

\section{Examples} 
\label{sec:two:examples}

Let us get appetized with some examples that showcase
the new features. Lets assume that $\langtwo$ has a new 
primitive type @bool@ with two values @true@ and @false@:
\begin{code}
    type bool = true | false
\end{code}
We will see how to support user-defined algebraic 
data types and pattern-matching in $\langfive$ 
(\S~\ref{sec:lang:five}).

\begin{myexample}{Branches}
First, suppose that the two @bool@ constants of $\langtwo$ are 
given the following refinement types that connect the values 
to the truth or falsehood of refinement predicates:
\begin{align}
\constty{\vtrue}  &\ \doteq\ \refb{\tbool}{\reft{b}{b}}         \label{eq:prim:true}\\
\constty{\vfalse} &\ \doteq\ \refb{\tbool}{\reft{b}{\neg b}}    \label{eq:prim:false}
\end{align} 
That is the constants $\vtrue$ and $\vfalse$ map directly to 
the corresponding proposition in the refinement logic.
Consider the function @not@ that implements negation 
for @bool@ values. We would like to verify the specification
for @not@, @and@ and @or@, that reflects their semantics 
in their types:
\begin{code}
  val not: x:bool => bool[b|b $\Leftrightarrow$ $\neg$x]
  let not = (x) => { if (x) { false } else { true } };
  
  val and: x:bool => y:bool => bool[b|b $\Leftrightarrow$ x $\wedge$ y]
  let and = (x, y) => { if (x) { y } else { false } };
    
  val or: x:bool => y:bool => bool[b|b $\Leftrightarrow$ x $\vee$ y]
  let or = (x, y) => { if (x) { true } else { y } };
\end{code}
\end{myexample}

\begin{myexample}{Recursion}
Next, suppose that $\langtwo$ has primitive arithmetic 
comparison operators, analogous to @add@ and @sub@ from 
\S~\ref{one:decl:synth}
\begin{align*}
\constty{{leq}} &\ \doteq\ {\trfun{x}{\tint}{\trfun{y}{\tint}{\refb{\tbool}{\reft{\vvar}{\vvar \Leftrightarrow x \leq y}}}}} \\
\constty{{geq}} &\ \doteq\ {\trfun{x}{\tint}{\trfun{y}{\tint}{\refb{\tbool}{\reft{\vvar}{\vvar \Leftrightarrow x \geq y}}}}}
\end{align*} 
As we can now test @int@ values, we ought to be able 
to write recursive functions, such as @sum@ that takes 
as input a number $n$ and returns the summation $0 + 1 + \ldots + n$. 
Again, we would like to verify that the returned value is 
a @nat@ that exceeds @n@: 
\begin{code}
  val sum : n:int => nat[v|n <= v]
  let rec sum = (n) => {
    let c = leq(n, 0);
    if (c) {
      0
    } else {
      let n1 = sub(n,1);
      let t1 = sum(n1);
      n + t1
    }
  }
\end{code}
\end{myexample}

\section{Types and Terms}
\label{sec:two:types}
\label{sec:two:terms}

$\langtwo$ extends the syntax of $\langone$ 
with a few extensions summarized in \cref{fig:two:syntax}.

\mypara{Types}
The syntax of types is extended with 
the basic type $\tbool$.

\mypara{Terms}
The syntax of terms is extended in two ways.
First, $\langtwo$ introduces a \emph{conditional}
expression $\eif{\evar}{\expr_1}{\expr_2}$, that 
evaluates to $\expr_1$ when $\evar$ evaluates to 
$\ttrue$ and to $\expr_2$ otherwise.
We require that the condition be a variable 
$\evar$ instead of an expression for reasons 
similar to that of the ANF conversion required 
for \rulename{Syn-App} from \S~\ref{one:decl:synth}.
Second, we introduce a \emph{recursive binder} 
expression $\eletr{\evar}{\expr_1}{\type_1}{\expr_2}$,
where the binder $\evar$ may appear free in $\expr_1$.
Unlike plain let-binders, we require that the types 
of such recursive binders be annotated with their 
type $\type_1$.

\begin{figure}[t!]
\begin{tabular}{rrcll}
\emphbf{Basic Types}
  & \base & $\bnfdef$ & \ldots & \emph{from $\langone$ (\S~\ref{fig:one:types})}\\
  &       & $\spmid$  & \tbool & \emph{booleans} \\ [0.05in]

\emphbf{Terms}
  & \expr & $\bnfdef$ & $\ldots$                           & \emph{from} $\langone$  \\
  &       & $\spmid$  & \eif{\evar}{\expr}{\expr}          & \emph{branches}  \\
  &       & $\spmid$  & \eletr{\evar}{\expr}{\type}{\expr} & \emph{recursion} \\[0.05in]
\end{tabular}
\caption{{Syntax of Types and Terms}}
\label{fig:two:syntax}
\label{fig:two:types}
\label{fig:two:terms}
\end{figure}

\section{Declarative Typing} \label{two:decl}

Next, let us consider the rules used to determine whether a term $\expr$ 
has a given type $\type$. As before, we have four kinds of judgments. 
The rules for {entailment} and {subtyping} are exactly the same 
as for $\langone$. 
However, to support branching, path-sensitivity and recursion, we must 
extend the rules that establish the  checking and synthesis judgments.

\begin{figure}[t]
\judgementHead{Type Checking}{\tchk{\tcenv}{\expr}{\type}}
\begin{mathpar}
\inferrule
  {  
    \evarb\ \mbox{is fresh}                                        \\
    \tchk{\tcenv}{\evar}{\tbool}                                   \\\\
    \tchk{\tcenvext{\evarb}{\reft{\tint}{x}}}{\expr_1}{\type}      \\
    \tchk{\tcenvext{\evarb}{\reft{\tint}{\neg x}}}{\expr_2}{\type}
  }
  {
    \tchk{\tcenv}{\eif{\evar}{\expr_1}{\expr_2}}{\type}
  }
  {\ruleName{Chk-If}}

\inferrule
  {
    \wf{\tcenv}{\type_1}{\kind} \\
    \tchk{\tcenvext{\evar}{\type_1}}{\expr_1}{\type_1} \\
    \tchk{\tcenvext{\evar}{\type_1}}{\expr_2}{\type_2}
  }
  {
    \tchk{\tcenv}{\eletr{\evar}{\expr_1}{\type_1}{\expr_2}}{\type_2}
  }
  {\ruleName{Chk-Rec}}
\end{mathpar}
\caption{Bidirectional Checking: Other rules from $\langone$ (\cref{fig:one:check})}
\label{fig:two:check}
\end{figure}

\subsection{Checking} \label{two:decl:check}

Conditionals and recursive binders are handled 
by the checking rules summarized in \cref{fig:two:check}.

\mypara{Conditionals} $\eif{\evar}{\expr_1}{\expr_2}$ 
can be checked to have the type $\type$ in a context 
$\tcenv$ when $\evar$ is a $\tbool$ and \emph{both} 
branches $\expr_1$ and $\expr_2$ can be checked to 
have type $\type$ (\rulename{Chk-If}). 
Unlike with classical type checking, we want to 
check $\expr_1$ (resp. $\expr_2$) in a context 
that is extended with the fact that $\evar$ 
evaluated to $\vtrue$ (resp. $\vfalse$).
Without this extra information, we cannot, \eg, establish 
that the body of @not@ (\cref{sec:two:examples}) returns 
a boolean $b$ that is the logical negation of the input $x$.
The rule \rulename{Chk-If} incorporates branch condition 
by binding a \emph{fresh} variable $\evarb$ to a refinement 
that captures the value of the condition $\evar$. 
That is we check the ``then'' branch $\expr_1$ by extending 
the context with a binding $\tb{\evarb}{\reft{\tint}{x}}$
that says that $\evar$ is $\ttrue$.
On the other hand, we check the ``else'' branch $\expr_2$ by extending 
the context with $\tb{\evarb}{\reft{\tint}{\neg x}}$ which 
records the fact that $\expr_2$ is evaluated when $\evar$ 
is $\vfalse$.
The binder $\evarb$ is used only to capture the value of the condition 
and could be of any type. 

\begin{myexample}{Checking \mw{not}}
Lets see how this method of \emph{branch strengthening} 
allows us to check the implementation of @not@ from 
\cref{sec:two:examples} against its specification.
First, rule \rulename{Chk-Lam} gives us the following 
obligation where the context has only $\tb{\evar}{\tbool}$
from the specification for the input parameter $\evar$:
$$
    {\tb{\evar}{\tbool}} \vdash
    {\eif{\evar}{\vfalse}{\vtrue}} 
    \Leftarrow
    {\refb{\tbool}{\reft{b}{b \Leftrightarrow \neg \evar}}} 
$$
\rulename{Chk-If} splits the above 
into two obligations, for the then- and 
else- branch respectively:
\begin{align}
   \tb{\evar}{\tbool};\ \tb{\evarb}{\refb{\tint}{\rreft{\evar}}} & \vdash 
   {\vfalse} 
   \Leftarrow 
   {\refb{\tbool}{\reft{b}{b \Leftrightarrow \neg \evar}}} \notag \\
   \tb{\evar}{\tbool};\ \tb{\evarb}{\refb{\tint}{\rreft{\neg \evar}}} & \vdash 
   {\vtrue} 
   \Leftarrow 
   {\refb{\tbool}{\reft{b}{b \Leftrightarrow \neg \evar}}} \notag \\
\intertext{\ruleName{Chk-Syn} and \rulename{Syn-Con}, 
using the constant types for $\vtrue$ and $\vfalse$, 
reduce the above to the respective subtyping obligations:} 
   \tb{\evar}{\tbool};\ \tb{\evarb}{\refb{\tint}{\rreft{\evar}}} & \vdash 
   {\refb{\tbool}{\reft{b}{\neg b}}} 
   \lqsubt
   {\refb{\tbool}{\reft{b}{b \Leftrightarrow \neg \evar}}} \notag\\
   \tb{\evar}{\tbool};\ \tb{\evarb}{\refb{\tint}{\rreft{\neg \evar}}} & \vdash 
   {\refb{\tbool}{\reft{b}{b}}} 
   \lqsubt
   {\refb{\tbool}{\reft{b}{b \Leftrightarrow \neg \evar}}} \notag\\
\intertext{\rulename{Sub-Base} boils the above down 
to the subsumption checks:}
   \tb{\evar}{\tbool};\ \tb{\evarb}{\refb{\tint}{\rreft{\evar}}}
   & \vdash 
   \csimp{b}{\tbool}{\neg b}{b \Leftrightarrow \neg \evar} \notag\\
   \tb{\evar}{\tbool};\ \tb{\evarb}{\refb{\tint}{\rreft{\neg \evar}}} 
   & \vdash 
   \csimp{b}{\tbool}{b}{b \Leftrightarrow \neg \evar} \notag\\
\intertext{\rulename{Ent-Ext} turns these into the VCs that are proved valid by SMT:}
\csimp{\evar}{\tbool}{\ttrue}{}
\csimp{\evarb}{\tint}{\evar}{}
&
\csimp{b}{\tbool}{\neg b}{
b \Leftrightarrow \neg \evar} \label{eq:not:then:vc} \\ 
\csimp{\evar}{\tbool}{\ttrue}{}
\csimp{\evarb}{\tint}{\neg \evar}{}
&
\csimp{b}{\tbool}{\neg b}{
b \Leftrightarrow \neg \evar} \label{eq:not:else:vc} 
\end{align}
\noindent
Note that validity depends crucially on the 
hypotheses $\evar$ and $\neg \evar$ introduced by branch 
strengthening. Without those, the VCs would be invalid 
and hence $\mcode{not}$ would fail to typecheck.
\end{myexample}

\mypara{Recursive binders} $\eletr{\evar}{\expr_1}{\type_1}{\expr_2}$
have type $\type_2$ in a context $\tcenv$ if in the context where 
$\evar$ is also \emph{assumed} to have type $\type_1$, 
(1)~the recursive term $\expr_1$ can be \emph{guaranteed} 
    to have type $\type_1$ and 
(2)~the body $\expr_2$ can be checked to have type $\type_2$.
Note that $\langtwo$ requires explicit type annotations for 
recursive binders to facilitate bidirectional checking, so 
the rule \ruleName{Chk-Rec} additionally checks that the 
annotation $\type_1$ is well-formed in $\tcenv$.  
(While top-level signatures are invaluable for design and 
documentation, we will see how they may be elided via 
refinement inference in \cref{sec:lang:three}.)

\begin{myexample}{Checking \mcode{sum}}
Lets see how the assume-guarantee method allows us to verify
the implementation of \mcode{sum}. Let us introduce a few abbreviations:
\begin{align} 
\type_s  & \ \doteq\ \refb{\tint}{\reft{\vvar}{0 \leq \vvar \wedge n \leq \vvar}} \label{eq:sum:out}\\
\tcenv_s & \ \doteq\ \tb{{sum}}{\trfun{n}{\tint}{\type_s}};\ \tb{n}{\tint}  \label{eq:sum:env}\\
\expr_s  & \ \doteq\ \mbox{the body of \mcode{sum}} \label{eq:sum:body}
\end{align}
\rulename{Chk-Rec} and then \rulename{Chk-Lam} get the ball rolling 
by yielding the checking obligation:
$$
\tcenv_s \vdash \expr_s \Leftarrow \type_s
$$ 
\rulename{Chk-Let} and then \rulename{Chk-If} split the above into two obligations:
\begin{align}
\tcenv_s; \tb{c}{\rreft{c \Leftrightarrow n \leq 0}}; \tb{\evarb}{\rreft{c}} & \vdash
  0 
  \Leftarrow
  \type_s \label{eq:sum:then} \\ 
\tcenv_s; \tb{c}{\rreft{c \Leftrightarrow n \leq 0}}; \tb{\evarb}{\rreft{\neg c}} & \vdash
  \expr_2 
  \Leftarrow
  \type_s \label{eq:sum:else}
\end{align}
where $\expr_2$ is the \kw{else}-branch of the body of \mcode{sum}. 
\rulename{Chk-Syn}, \rulename{Syn-Con}, \rulename{Sub-Base} and 
\rulename{Ent-Ext} turn (\ref{eq:sum:then}) to the valid VC:
\begin{equation}
\forall n, c, y, \vvar.\ (c \Leftrightarrow n \leq 0) \Rightarrow c \Rightarrow (\vvar = 0) 
  \Rightarrow  (0 \leq \vvar \wedge n \leq \vv) \label{eq:sum:then:vc}
\end{equation}
\rulename{Chk-Let} reduces the judgment (\ref{eq:sum:else})
for the \kw{else} branch to the following subtyping obligation
that checks that the value of @n + t1@ is indeed a subtype of 
the return $\type_s$:
\begin{align}
  {\tcenv_s'} \ \vdash \ & 
  {\refb{\tint}{\reft{\vvar}{\vvar = n + t_1}}}
  \lqsubt
  \type_s 
  \label{eq:sum:else:sub}\\ 
\intertext{where the context $\tcenv_s'$ is $\tcenv$ extended 
with the $\kw{else}$-branch condition and bindings for $n_1$ 
and $t_1$:}
  \tcenv_s' \ \doteq \ & \tcenv_s; 
                         \tb{c}{\rreft{c \Leftrightarrow n \leq 0}}; 
                         \tb{y}{\rreft{\neg c}}; \notag \\ 
                       & \tb{n_1}{\rreft{n_1 = n - 1}}; 
                         \tb{t_1}{\rreft{0 \leq t_1 \wedge n_1 \leq t_1}} \notag
\end{align}
In $\tcenv_s'$, the binding for $n_1$ is the output type of \mcode{sub} 
with formals replaced with $n$ and $1$. Similarly, the binding for $t_1$
is the (assumed) output type of \mcode{sum}, with the formal $n$ replaced 
with the actual parameter $t_1$. 
\rulename{Sub-Base} and \rulename{Ent-Ext} reduce the 
above subtyping (\ref{eq:sum:else:sub}) to the VC:
\begin{align}
\forall & n, c, y, n_1, t_1, \vvar. \notag \\ 
  & (c \Leftrightarrow n \leq 0) \Rightarrow (\neg c) \Rightarrow (n_1 = n - 1) \Rightarrow (0 \leq t_1 \wedge n_1 \leq t_1) \Rightarrow (\vvar = n + t_1) \notag \\ 
  & \Rightarrow (0 \leq \vvar \wedge n \leq \vv) \label{eq:sum:else:vc}
\end{align}
which the SMT solver proves valid, guaranteeing that \mcode{sum} 
indeed meets its given specification.
\end{myexample}

\subsection{Synthesis} \label{two:decl:synth}

\begin{figure}[t!]
$$\begin{array}{lcl}
\toprule
\singtysym & : & (\Evar \times \Type) \rightarrow \Type \\
\midrule
\singty{\evar}{\refb{\base}{\reft{\vvar}{\pred}}}
  & \doteq & \refb{\base}{\reft{\vvar}{\pred \wedge \vvar = \evar}} \\[0.05in]

\singty{\evar}{\type}
  & \doteq & \type \\[0.05in]
\bottomrule
\end{array}$$
\caption{Selfification: Singleton Type Strengthening}
\label{fig:singleton}
\end{figure}

Both the new language constructs in $\langtwo$, 
\ie branches and recursive binders have checking 
judgments. 
However, to precisely use the information gleaned 
from branch conditions, to enable path-sensitive 
verification, we will need to modify the rule for 
\emph{variables}.

\begin{myexample}{Path Sensitivity}
\label{ex:abs}
The function @abs@ returns the 
absolute value of its input @x@.
\begin{code}
  val abs : x:int => nat[v|x $\leq$ v]
  let abs = (x) => {
    let c = leq(0, x);
    if (c) {
      x
    } else {
      sub(0, x)
    };
  };
\end{code}
However, note that the \kw{then} branch is simply @x@.
If we directly applied \rulename{Syn-Var} from 
\cref{fig:one:synth}, then the synthesized type 
would be @int@, the type that @x@ is bound to in 
the context, which is clearly not a subtype of @nat@!
Hence, we require another way to type 
the variable lookup @x@ in a manner that 
is precise enough to let us use the branch 
condition to prove that the result is a @nat@.
\end{myexample}

\mypara{Variables and Selfification} 
This problem can be solved by the so-called 
\emph{selfification} rule introduced by \cite{Ou2004}
where the variable $\evar$ is given a singleton type 
$\refb{\base}{\reft{\vvar}{\vvar = \evar}}$, \ie where 
the refinement says the value is equal to $\evar$.
We formalize this idea in the updated \rulename{Syn-Var}
in \cref{fig:two:synth}. In context $\tcenv$, the type 
synthesized for $\evar$ is $\singty{\evar}{\type}$, where 
$\type$ is what $\evar$ is bound to in $\tcenv$.
Figure~\ref{fig:singleton} summarizes definition 
of $\singtysym$. When invoked on a base type $\base$ 
-- for which equality is defined in the refinement 
logic -- the function strengthens the refinement 
$\pred$ with the singleton conjunct $\vvar = \evar$. 
When invoked on other (\eg function) types, 
the input is returned as is.

\begin{myexample}{Selfification in \mcode{abs}}
Lets use selfification to verify \mcode{abs}. Let
\begin{align}
\tcenv_a & \ \doteq\  \tb{x}{\tint};\ \tb{c}{\rreft{c \Leftrightarrow 0 \leq x}} \notag \\
\intertext{As we've seen before with \mcode{not} and \mcode{sum}, \rulename{Chk-Lam}, 
\rulename{Chk-Let} and \rulename{Chk-If} produce the two obligations
for the \kw{then} and \kw{else} branches:}
\tcenv_a;\ \tb{y}{\rreft{c}} & \vdash
  x 
  \Leftarrow 
  \refb{\tint}{\reft{\vvar}{0 \leq \vvar \wedge x \leq \vvar}} 
  \notag \\
\tcenv_a;\ \tb{y}{\rreft{\neg c}} & \vdash 
    ({sub}\ \mcode{0}\ x) 
    \Leftarrow 
    \refb{\tint}{\reft{\vvar}{0 \leq \vvar \wedge x \leq \vvar}} 
  \notag \\
\intertext{\rulename{Chk-Syn}, \rulename{Syn-Var} and \rulename{Syn-App} reduce 
the above to, respectively:}
\tcenv_a;\ \tb{y}{\rreft{c}} & \vdash 
  \refb{\tint}{\reft{\vvar}{\vvar = x}} 
  \lqsubt
  \refb{\tint}{\reft{\vvar}{0 \leq \vvar \wedge x \leq \vvar}} 
  \notag \\
\tcenv_a;\ \tb{y}{\rreft{\neg c}} & \vdash
  \refb{\tint}{\reft{\vvar}{\vvar = 0 - x}} 
  \lqsubt
  \refb{\tint}{\reft{\vvar}{0 \leq \vvar \wedge x \leq \vvar}} 
  \notag \\
\intertext{Finally, \rulename{Sub-Base} and \rulename{Ent-Ext} establish the above 
by verifying the validity of the VCs:}
  \forall x,c,y,\vvar.\ (c \Leftrightarrow 0 \leq x) & \Rightarrow c       \Rightarrow (\vvar = x) \Rightarrow (0 \leq \vvar \wedge x \leq \vv) \label{eq:abs:then:vc}\\ 
  \forall x,c,y,\vvar.\ (c \Leftrightarrow 0 \leq x) & \Rightarrow \neg c  \Rightarrow (\vvar = 0 - x) \Rightarrow (0 \leq \vvar \wedge x \leq \vv) \label{eq:abs:else:vc}
\intertext{Notice that the hypothesis $\vvar = x$ obtained from 
selfification is essential for the validity of the first (\kw{then})
VC; without it, \ie if we simply gave the term $\evar$ its type from 
the context $\tint$, we would get the \emph{invalid} VC}
  \forall x,c,y,\vvar.\ (c \Leftrightarrow 0 \leq x) & \Rightarrow c       \Rightarrow \ttrue \Rightarrow (0 \leq \vvar \wedge x \leq \vv) \notag 
\end{align}
which would make us foolishly reject \mcode{abs}.
\end{myexample}

\begin{figure}[t]
\judgementHead{Type Synthesis}{\tsyn{\tcenv}{\expr}{\type}}
\begin{mathpar}
\inferrule
  {
    \tcenvget{\evar} = \type        \\
  }
  {
    \tsyn{\tcenv}{\evar}{\singty{\evar}{\type}}
  }
  {\ruleName{Syn-Var}}
\end{mathpar}
\caption{Bidirectional Synthesis: Other rules from $\langone$ (\cref{fig:one:check}) 
\RJ{Niki we already ensure self is just id for function types, so why not simplify the rule to the above?}
}
\label{fig:two:synth}
\end{figure}

\section{Verification Conditions}\label{two:algo}

Next, lets implement the checking and synthesis rules as a pair 
of VC generation functions $\chksym$ and $\synsym$.


\mypara{Checking} Recall that $\chk{\tcenv}{\expr}{\type}$ 
returns the VC $\cstr$ whose validity implies that 
$\tchk{\tcenv}{\expr}{\type}$ holds (\cref{prop:chk}).
Figure~\ref{fig:two:algo:chk} summarizes the new cases 
of $\chksym$ for $\langtwo$.

For conditional expressions $\eif{\evar}{\expr_1}{\expr_2}$ the VC is 
the conjunction of the VCs $\cstr_1$ and $\cstr_2$ that are generated 
by invoking $\chksym$ on the branches $\expr_1$ and $\expr_2$ respectively,
and then conditioning the result to track whether the branch condition 
was true ($\rreft{\evar}$) or false ($\rreft{\neg \evar}$).
For example, $\chk{\emptyset}{\expr_{n}}{\type_{n}}$ where 
$\expr_{n}$ and $\type_{n}$ are respectively the implementation
and specification of \mcode{not} (from \cref{sec:two:examples}) 
returns the VC constraint which is the conjunction of the two 
VCs (\ref{eq:not:then:vc}, \ref{eq:not:else:vc}):
$$ 
\forall \evar, \evarb, b.\ (\evar \Rightarrow \neg b \Rightarrow (b \Leftrightarrow \neg \evar)) 
                    \ \wedge\  (\neg \evar \Rightarrow b \Rightarrow (b \Leftrightarrow \neg \evar))
$$

For recursive binders $\eletr{\evar}{\expr_1}{\type_1}{\expr_2}$ the VC 
is the conjunction of the VCs obtained for $\expr_1$ and $\expr_2$, both 
generated using the environment $\tcenv$ extended by binding $\evar$ to 
its specified type $\type_1$.
For example, $\chk{\tcenv_s}{\expr_s}{\type_s}$, where  $\tcenv_s$, 
$\expr_s$ and $\type_s$ are the environment, body and 
output type of \mcode{sum} as shown in (\ref{eq:sum:env}), (\ref{eq:sum:body}), 
(\ref{eq:sum:out}), yields the following VC, which is 
the conjunction of the two VCs (\ref{eq:sum:then:vc}, \ref{eq:sum:else:vc}) 
\begin{align*}
\forall n &, c, y, n_1, t_1, \vvar.\ (c \Leftrightarrow n \leq 0) \Rightarrow \\
        & (c  \Rightarrow \vvar = 0 \Rightarrow (0 \leq \vvar \wedge n \leq v)) \ \wedge \\
        & (\neg c \Rightarrow n_1 = n - 1 \Rightarrow (0 \leq t_1 \wedge n_1 \leq t_1) \Rightarrow \vvar = n + t_1 \Rightarrow 
          (0 \leq \vvar \wedge n \leq \vv))
\end{align*}


\mypara{Synthesis} The synthesis function $\syn{\tcenv}{\expr}$ returns 
a pair of a VC $\cstr$ and type $\type$ such that the validity of $\cstr$ 
implies that $\tsyn{\tcenv}{\expr}{\type}$ holds (\cref{prop:syn}).
Figure~\ref{fig:two:algo:syn} summarizes the updated case for variable lookup 
using selfification.
Here, the generated VC is trivial \ie $\ttrue$, but the synthesized type 
is $\singty{\evar}{\type_\evar}$ where $\type_\evar$ is the type that $\evar$
is bound to in the context $\tcenv$.
For example, using the updated selfified version of $\synsym$, the invocation
of $\chk{\emptyset}{\expr_a}{\type_a}$ --- where $\expr_a$ and $\type_a$ are the 
implementation and specification of $\mcode{abs}$ --- yields the following VC which 
is the conjunction of the \kw{then} and \kw{else} VCs (\ref{eq:abs:then:vc}, \ref{eq:abs:else:vc})
\begin{align*}
  \forall x,c,y,\vvar.\ (c \Leftrightarrow 0 \leq x) \Rightarrow\ & (c     \Rightarrow \vvar = x \Rightarrow (0 \leq \vvar \wedge x \leq \vv)) \ \wedge \\ 
                                                               & (\neg c \Rightarrow \vvar = 0 - x \Rightarrow (0 \leq \vvar \wedge x \leq \vv))
\end{align*}

\begin{figure}[t!]
$$\begin{array}{lcl}
\toprule
\chksym & : & (\tcenv \times \Expr \times \Type) \rightarrow \Cstr \\
\midrule
\chk{\tcenv}{\eif{\evar}{\expr_1}{\expr_2}}{\type}
                     & \doteq & \csand{\cstr_1}{\cstr_2}     \\
\quad \mbox{where}   &        &                              \\
\quad \quad \cstr_1  & =      & \cswith{\evarb}{\refb{\tint}{\rreft{\evar}}}{\chk{\tcenv}{\expr_1}{\type}} \\
\quad \quad \cstr_2  & =      & \cswith{\evarb}{\refb{\tint}{\rreft{\neg \evar}}}{\chk{\tcenv}{\expr_2}{\type}} \\
\quad \quad \evarb   & =      & \mbox{fresh binder}        \\[0.05in]

\chk{\tcenv}{\eletr{\evar}{\expr_1}{\type_1}{\expr_2}}{\type}
                     & \doteq & \csand{\cstr_1}{\cstr_2}       \\
\quad \mbox{where}   &        &                                \\
\quad \quad \cstr_1  & =      & \chk{\tcenv_1}{\expr_1}{\type} \\
\quad \quad \cstr_2  & =      & \chk{\tcenv_1}{\expr_2}{\type} \\
\quad \quad \tcenv_1 & =      & \tcenvext{\evar}{\type_1}      \\[0.05in]

\mbox{\emph{\ldots plus cases from}}\ \langone\ (Fig.~\ref{fig:one:algo:chk}) & & \\[0.05in]

\toprule
\synsym & : & (\tcenv \times \Expr) \rightarrow (\Cstr \times \Type) \\
\midrule

\syn{\tcenv}{\evar}
  & \doteq & (\ttrue,\ \singty{\evar}{\type_\evar}) \\
\quad \mbox{where}       &        &                                \\
\quad \quad \type_\evar  & =      & \tcenvget{\evar} \\[0.05in]

\mbox{\emph{\ldots plus cases from}}\ \langone\ (Fig.~\ref{fig:one:algo:syn}) & & \\[0.05in]
\bottomrule
\end{array}$$
\caption{Algorithmic Checking for $\langtwo$}
\label{fig:two:algo:syn}
\label{fig:two:algo:chk}
\end{figure}

\section{Discussion} \label{two:summary}

At this point, we have seen enough to write refinement type 
checkers for interesting languages, with functions, branching 
and recursion. Lets glance back at the mechanisms that 
make verification tick in $\langtwo$.

\mypara{Recursion via Assume-Guarantee Reasoning}
First, we account for recursion using the 
classic \emph{assume-guarantee} method where, 
to check $\eletr{\evar}{\expr_1}{\type_1}{\expr_2}$ 
we \emph{assume} that the recursive binder $\evar$ 
has the type $\type_1$, and then, \emph{guarantee} 
that fact by checking its implementation $\expr_1$ 
against $\type_1$.
As in classical Floyd-Hoare logic, this only gives 
us a so-called \emph{partial} correctness guarantee; 
we will look at verifying \emph{total} correctness 
later in \cref{sec:lang:seven}.

\mypara{Path-Sensitivity via Branch Strengthening}
We incorporate path-sensitive reasoning in conditional 
expressions $\eif{\evar}{\expr_1}{\expr_2}$ by introducing 
a fresh variable (\ie $\evarb$ in \rulename{Chk-If}) and 
binding it to a refinement that states that the condition 
$\evar$ is true (resp. false) when check the \kw{then} 
branch $\expr_1$ (resp. \kw{else} branch $\expr_2$).
We will generalize this strategy to account for user-defined 
data-types and pattern matching in \cref{sec:lang:five}.

\mypara{Occurence Typing via Selfification}
Finally, the presence of branches allows binders to have 
strong or more precise types under branches (as in @abs@).
We account for this form of path-sensitive strengthening 
by updating the variable lookup rule \rulename{Syn-Var} 
with \emph{selfification} which says the type of $\evar$ 
is a singleton whose value equals $\evar$ \citep{Ou2004}.
This method, dubbed ``occurrence typing'' by \cite{Komondoor05} 
and \cite{Tob08}, allows us to then use the rest of the refinement 
typing machinery (\eg branch strengthening) to precisely type 
each \emph{occurrence} of a variable $\evar$ under different 
branches.
\chapter{Refinement Inference}
\label{sec:lang:three}

Bidirectional typing's separate checking and synthesis modes 
ensure that the programmer need only write type signatures 
for functions, after which the refinement checker can 
synthesize the types of intermediate sub-expressions 
to produce verification conditions for the SMT solver 
to validate. 
However, as \cite{pierce-turner} observe, 
to make higher-order programming pleasant, 
we will want to spare the programmer the 
tedium of having to type local function 
definitions like those passed as arguments 
to \mcode{map} or \mcode{fold}.
Similarly, to make type polymorphism 
(\S~\ref{sec:lang:four}) usable, we 
want to avoid cluttering the code 
with explicit (refinement) type 
annotations at polymorphic 
instantiation sites.
Thus, lets study $\langthree$ which 
extends $\langtwo$ with a mechanism 
for \emph{inferring} refinements via
the following strategy. 
\begin{itemize}
\item{\emphbf{Step 1: Types to Templates}} 
  First, we generalize refinement type signatures
  to allow them to contain \emph{holes} denoting 
  unknown refinements. Type checking begins by 
  replacing these holes with \emph{Horn Variables} 
  that represent the unknown refinements.

\item{\emphbf{Step 2: VCs to Horn Constraints}}
  Second, we run the VC generation procedures as 
  described in the preceding chapters. 
  Now, however, these procedures yield 
  \emph{Horn Constraints} which are VCs 
  containing Horn (Variable) applications 
  in addition to predicates.

\item{\emphbf{Step 3: VC Validity to Horn Solving}}
  Third, instead of asking an SMT solver to determine 
  the validity of a VC, we will invoke a Horn Solver 
  that will perform a \emph{fixpoint} computation to
  determine whether there are refinements that can 
  be substituted for the Horn variables that make
  the resulting VCs valid.   
\end{itemize}

\section{Examples}
\label{sec:three:examples}

Before plunging into the formal 
details of \langthree, lets build 
up some intuition by studying how 
the three-step strategy plays out 
on an example.

\mypara{Encoding Assertions} \label{ex:assert}
Many languages have some form of \emph{assertion}
statement which allows the programmer to test, 
typically at run-time, that some condition holds 
and to halt execution otherwise.
The following \mcode{assert} function allows 
the programmer to write such \emph{assertions}
but the refined (input) type ensures that 
any client that calls @assert(cond)@ only 
typechecks if the refinement type checker 
can verify that @cond@ \emph{always} 
evaluates to @true@ at run-time.
\begin{code}
  val assert : bool[b|b] => int
  let assert = (b) => { 0 };
\end{code}
Recall the \mcode{abs} function from \S~\ref{ex:abs} 
whose type signature has been deliberately elided
\begin{code}
  let abs = (x) => {
    let c = leq(0, x);
    if (leq(0, x)) { x } else { sub(0, x) };
  };
\end{code}
Finally, consider \mcode{main} that calls 
\mcode{abs} and \mcode{assert}s that the returned 
value is non-negative:
\begin{code}
  val main : int => int 
  let main = (y) => {
    let z  = abs(y); 
    let c  = leq(0, z);
    assert(c) 
  }
\end{code}

\mypara{Recap: Verification Conditions}
Suppose that we are given a type signature 
for \mcode{abs}, for example
\begin{code}
  val abs : x:int => int[v|0 $\leq$ v]
\end{code}
Then the type checker from \cref{sec:lang:two} 
would produce the VC:
$$\begin{array}{rrllllr}
       & \forall \mw{x}, \mw{c}, \vvar.   & (\mw{c} \Leftrightarrow 0 \leq \mw{x}) & \Rightarrow \mw{c} & \Rightarrow \vvar = \mw{x}     & \Rightarrow 0 \leq \vvar                                   & \quad \quad {(a)} \label{ref:vc:1} \\
       &                                &                                        & \wedge \neg \mw{c} & \Rightarrow \vvar = 0 - \mw{x} & \Rightarrow 0 \leq \vvar                                   & \quad \quad {(b)} \label{ref:vc:2} \\
\wedge & \forall \mw{y}, \mw{z}, \mw{c}, \mw{b}.& \multicolumn{3}{l}{0 \leq \mw{z} \Rightarrow (\mw{c} \Leftrightarrow 0 \leq \mw{z}) \Rightarrow (\mw{b} \Leftrightarrow \mw{c})} & \Rightarrow \mw{b} & \quad \quad {(c)} \label{ref:vc:3} \\
\end{array}$$
The first two conjuncts of the VC arise from verifying 
that the implementation of \mcode{abs} satisfies the output 
the above signature, \ie its specified post-condition.
The conjuncts respectively state that in the \kw{then} 
(conjunct $(a)$) and \kw{else} (conjunct $(b)$) branches, 
the output value $\vvar$ must be non-negative.
The last conjunct comes from checking the call to 
@assert@ \ie verifying that the boolean value 
@c@ that @assert@ is invoked on, is indeed 
always @true@. 
Note that the third conjunct \emph{uses}
(1)~the output type of \mcode{abs} to assume 
    that @z@ is non-negative,
(2)~the output type of the primitive @leq@ 
    which we typed as (\S~\ref{sec:two:examples}) 
$$    
\constty{{leq}} \ \doteq\ {\trfun{x}{\tint}{\trfun{y}{\tint}{\refb{\tbool}{\reft{\vvar}{\vvar \Leftrightarrow x \leq y}}}}} 
$$    
to assume that @c@ is @true@ if and only if @z@ is non-negative.
The above assumptions suffice to prove that 
the \mcode{assert}'s input @b@ (which at this 
call-site is @c@) is indeed always @true@.

\subsection{Step 1: Holes and Templates}
Suppose that we wrote the 
following specification where \rhole
denotes a \emph{refinement-hole}: 
an unknown refinement that we want 
to \emph{infer}
\begin{code}
  val abs : x:int => int[$\rhole$]
\end{code}
Readers may be reminded of Haskell's 
notion of a \emph{type-hole} which 
allows programmers to partially 
specify type signatures that can 
then be automatically filled by 
type inference~\citep{padl2014}.
The key trick to filling refinement 
holes is to generalize VCs to constraints 
containing Horn applications that 
represent unknown refinements.
To do so, every type signature with 
a hole is transformed into a \emph{template}
containing (distinct) Horn variables 
that represent the unknown refinements. 

\begin{myexample}{Template for \mcode{abs}}
For example, the signature for \mcode{abs} yields the template:
\begin{equation}
  \mcode{abs} \ : \  \trfun{x}{\tint}{\refb{\tint}{\reft{\vvar}{\kva{\kvar}{x,\vvar}}}} \label{sig:abs:template}
\end{equation}
where $\kvar$ is a \emph{Horn variable} 
such that $\kva{\kvar}{z_1, z_2}$ denotes 
an (unkown) refinement (relation) over the
Horn variable's \emph{parameters} $z_1$ and $z_2$.
\end{myexample}

\subsection{Step 2: Horn Constraints}
Next, we use the templates to run exactly the 
same VC generation procedure as before. However, 
instead of producing a VC we get a \emph{Horn} 
constraint which is a VC with Horn variable 
applications appearing at various positions. 

\begin{myexample}{Constraints for \mcode{abs}}
For example, if we run $\chksym$ on the above 
code with \mcode{abs} and \mcode{main}, but using the 
template (\ref{sig:abs:template}) as the specification 
for \mcode{abs}, we get the Horn constraint:
\begin{equation}
\begin{array}{rrllllr}
       & \forall \mw{x}, \mw{c}, \vvar.   & (\mw{c} \Leftrightarrow 0 \leq \mw{x}) & \Rightarrow \mw{c} & \Rightarrow \vvar = \mw{x}     & \Rightarrow \kva{\kvar}{\mw{x}, {\vvar}}                                    & \quad \quad {(a')} \\
       &                                &                                        & \wedge \neg \mw{c} & \Rightarrow \vvar = 0 - \mw{x} & \Rightarrow \kva{\kvar}{\mw{x}, {\vvar}}                                      & \quad \quad {(b')} \\
\wedge & \forall \mw{y}, \mw{z}, \mw{c}, \mw{b}.& \multicolumn{3}{l}{ \kva{\kvar}{\mw{y}, \mw{z}} \Rightarrow (\mw{c} \Leftrightarrow 0 \leq \mw{z}) \Rightarrow (\mw{b} \Leftrightarrow \mw{c})} & \Rightarrow \mw{b} & \quad \quad {(c')} \label{eq:abs:hc}
\end{array}
\end{equation}
Notice that this constraint is mostly identical 
to the VC shown above, with three conjuncts $(a)$, 
$(b)$ and $(c)$, except that instead of:
(1)~the consequent $0 \leq \vvar$ that appears 
    in the conjuncts $(a)$ and $(b)$ stipulating 
    that the output value $\vvar$ is non-negative,
    we have the Horn application $\kva{\kvar}{\mw{x},\vvar}$ 
    representing the output $\vvar$ is related to 
    the input $\mw{x}$ via an (as yet not known) 
    refinement $\kvar$, and
(2)~the assumption $0 \leq \mw{z}$ that appears 
    as a hypothesis in $(c)$ stating that $\mw{z}$
    is non-negative, we have the Horn application
    $\kva{\kvar}{\mw{y}, \mw{z}}$ that says 
    that the value of $\mw{z}$ is related 
    to that of (the argument) $\mw{y}$ by 
    the as yet unknown refinement $\kvar$.
\end{myexample}

\subsection{Step 3: Horn Solving}
At this point, we cannot ask an SMT solver
to simply check the validity of a VC, as the 
constraints contain unknown Horn relations.
Instead we invoke a \emph{Horn solver} 
to determine whether \emph{there exist} 
a satisfying assignment for the Horn variables.
A Horn \emph{assignment} is a mapping of Horn 
variables to refinement predicates over the 
Horn variables' parameters.
An assignment \emph{satisfies} a Horn 
constraint if the result of substituting 
the horn variables with their assignments 
yields a \emph{valid} (Horn-variable free) 
formula.

\begin{myexample}{Solution for \mcode{abs}}
For example $\soln_1$, $\soln_2$ and $\soln_3$ 
are three possible assignments for $\kvar$:
\begin{align}
  \kva{\soln_1(\kvar)}{z_1, z_2} & \ \doteq\ z_1 \leq z_2  \label{eq:sol:1} \\
  \kva{\soln_2(\kvar)}{z_1, z_2} & \ \doteq\ 0   <    z_2  \label{eq:sol:2} \\
  \kva{\soln_3(\kvar)}{z_1, z_2} & \ \doteq\ 0   \leq z_2  \label{eq:sol:3} 
\end{align}
The assignment $\soln_1$ (\ref{eq:sol:1}) 
does not satisfy the Horn constraint (\ref{eq:abs:hc}) 
as substitution yields the following VC 
whose last conjunct is invalid:
\[
\begin{array}{rrllllr}
       & \forall \mw{x}, \mw{c}, \vvar.   & (\mw{c} \Leftrightarrow 0 \leq \mw{x}) & \Rightarrow \mw{c} & \Rightarrow \vvar = \mw{x}     & \Rightarrow {\mw{x} \leq {\vvar}}                                          & \quad \quad (\mbox{\cmark}) \\
       &                                &                                        & \wedge \neg \mw{c} & \Rightarrow \vvar = 0 - \mw{x} & \Rightarrow {\mw{x} \leq {\vvar}}                                          & \quad \quad (\mbox{\cmark}) \\
\wedge & \forall \mw{y}, \mw{z}, \mw{c}, \mw{b}.& \multicolumn{3}{l}{ {\mw{y} \leq \mw{z}} \Rightarrow (\mw{c} \Leftrightarrow 0 \leq \mw{z}) \Rightarrow (\mw{b} \Leftrightarrow \mw{c})} & \Rightarrow \mw{b} & \quad \quad (\mbox{\xmark}) \\ 
\end{array} 
\]
Assignment $\soln_2$ (\ref{eq:sol:2}) also 
fails to satisfy the constraint (\ref{eq:abs:hc})
as substituting it yields the following VC 
whose first conjunct is invalid:
$$
\begin{array}{rrllllr}
       & \forall \mw{x}, \mw{c}, \vvar.   & (\mw{c} \Leftrightarrow 0 \leq \mw{x}) & \Rightarrow \mw{c} & \Rightarrow \vvar = \mw{x}     & \Rightarrow {0 < {\vvar}}                                          & \quad \quad (\mbox{\xmark}) \\
       &                                &                                        & \wedge \neg \mw{c} & \Rightarrow \vvar = 0 - \mw{x} & \Rightarrow {0 < {\vvar}}                                          & \quad \quad (\mbox{\cmark}) \\
\wedge & \forall \mw{y}, \mw{z}, \mw{c}, \mw{b}.& \multicolumn{3}{l}{ {0 < \mw{z}} \Rightarrow (\mw{c} \Leftrightarrow 0 \leq \mw{z}) \Rightarrow (\mw{b} \Leftrightarrow \mw{c})} & \Rightarrow \mw{b} & \quad \quad (\mbox{\cmark}) \\ 
\end{array} 
$$
However, assignment $\soln_3$ (\ref{eq:sol:3}) 
\emph{does} satisfy the Horn constraint (\ref{eq:abs:hc})
as substitution produces the \emph{valid} VC
$$
\begin{array}{rrllllr}
       & \forall \mw{x}, \mw{c}, \vvar.   & (\mw{c} \Leftrightarrow 0 \leq \mw{x}) & \Rightarrow \mw{c} & \Rightarrow \vvar = \mw{x}     & \Rightarrow {0 \leq {\vvar}}                                          & \quad \quad (\mbox{\cmark}) \\
       &                                &                                        & \wedge \neg \mw{c} & \Rightarrow \vvar = 0 - \mw{x} & \Rightarrow {0 \leq {\vvar}}                                          & \quad \quad (\mbox{\cmark}) \\
\wedge & \forall \mw{y}, \mw{z}, \mw{c}, \mw{b}.& \multicolumn{3}{l}{ {0 \leq \mw{z}} \Rightarrow (\mw{c} \Leftrightarrow 0 \leq \mw{z}) \Rightarrow (\mw{b} \Leftrightarrow \mw{c})} & \Rightarrow \mw{b} & \quad \quad (\mbox{\cmark}) \\ 
\end{array} 
$$
Thus, we can fill the refinement-holes with the 
satisfying assignment to infer signatures that 
yield a well-typed program. For example, plugging 
$\soln_3$ into the template for \mcode{abs} yields 
the ``hand-written'' signature 
\begin{equation}
  \mcode{abs} \ : \  \trfun{x}{\tint}{\refb{\tint}{\reft{\vvar}{0 \leq \vvar}}}  \label{sig:abs:concrete}
\end{equation}
that let us verify \mcode{main}.
\end{myexample}

\section{Types and Terms}
\label{sec:three:types}
\label{sec:three:terms}

Next, we formalize our three-step strategy in $\langthree$ whose 
syntax is summarized in \cref{fig:three:syntax}.

\mypara{Predicates}
We extend the grammar of refinement predicates 
(\cref{fig:smt:pred}) to include Horn applications
of the form $\kvapp{\kvar}{\evar}$ where 
$\params{\evar}$ abbreviates a sequence of 
variables $\evar_1, \ldots, \evar_n$.
A Horn application $\kvapp{\kvar}{\evar}$ denotes
an \emph{unknown} predicate (or relation) over the 
variables \params{\evar}.

\mypara{Refinements}
Thus, $\langthree$ has two kinds of refinements.
The first are \emph{known} refinements (or just, refinements) 
$\reft{\vvar}{\pred}$, made up of predicates $\pred$ as before.
The second are refinement \emph{holes} (or just, holes) 
$\ureft$ which can appear in type annotations, and which 
denote an unknown refinement that the programmer has chosen 
to elide.

\mypara{Holes vs. Horn applications}
In $\langthree$, Horn-applications do not appear 
in the \emph{external} surface syntax, \ie in 
type annotations. Instead, the programmer elides 
refinements in annotations using holes.
During type checking we will replace all holes 
with Horn applications. That is, dually, holes 
do not appear in the \emph{internal} typing 
derivations.

\begin{figure}[t!]
\begin{tabular}{rrcll}
\emphbf{Predicates}
  & \pred & $\bnfdef$ & $\ldots$             & \emph{from \cref{fig:smt:pred}} \\
  &       & $\spmid$  & \kvapp{\kvar}{\evar} & \emph{horn application}\\[0.05in]

\emphbf{Refinements}
  & \refi & $\bnfdef$ & \reft{\vvar}{\pred}    & \emph{known} \\
  &       & $\spmid$  & \ureft               & \emph{hole} \\[0.05in]

\end{tabular}
\caption{{Syntax of Predicates and Refinements}}
\label{fig:three:syntax}
\end{figure}

\section{Declarative Typing} \label{three:decl}

Next, lets see how a term $\expr$ that may be annotated 
with refinement holes $\ureft$ can be verified 
to have a type $\type$.
The vast majority of the rules, in particular, 
the rules for well-formedness, subtyping and 
entailment, remain unchanged from $\langtwo$. 
However, we will introduce a new \emph{instantiation}
judgment that stipulates how holes can be filled by 
refinements. We will then use the instantiation 
judgment to eliminate holes in the two rules 
that pertain to type annotations.

\subsection{Instantiation} \label{three:decl:templates}

The instantiation judgment $\isfresh{\typeb}{\type}$ states 
that a type $\typeb$ can be \emph{instantiated to} a type 
$\type$ by replacing the refinement holes in $\typeb$ with 
suitable concrete refinements.
This intuition is formalized by the three rules summarized 
in \cref{fig:three:isfresh}:
\rulename{Ins-Hole}, which describes how a single hole is instantiated, 
\rulename{Ins-Conc}, which states the concrete refinements are left unmodified, and
\rulename{Ins-Fun}, which describes the component-wise instantiation of function types.

Lets focus on two important aspects of the instantiation judgment.
First, the rules ensure that if $\isfresh{\typeb}{\type}$ 
then there are no holes left in $\type$.
Second, the judgment is \emph{declarative}, it does not 
tell us \emph{how} to find suitable concrete refinements. 
Instead the rules tell us \emph{what} valid concrete 
refinements should look like for the program to be 
well-typed.

\begin{myexample}{Instantiating holes in \mcode{abs}}
The above rules establish that 
$$\isfresh 
  {\trfun{x}{\tint}{\refb{\tint}{\ureft}} \quad }
  {\quad \trfun{x}{\tint}{\refb{\tint}{\reft{\vvar}{ 0 \leq \vvar}}}}
$$
\ie the partial type signature for \mcode{abs} from (\ref{sig:abs:template}) can 
be instantiated to the concrete type in (\ref{sig:abs:concrete}).
\end{myexample}

\begin{figure}[t]
\judgementHead{Hole Instantiation}{\isfresh{\typeb}{\type}}
\begin{mathpar}

\inferrule
  { \quad }
  {\isfresh{\refb{\base}{\ureft}}{\refb{\base}{\reft{\vvar}{\pred}}}}
  {\ruleName{Ins-Hole}}


\inferrule
  { \quad }
  {\isfresh{\refb{\base}{\reft{\vvar}{\pred}}}{\refb{\base}{\reft{\vvar}{\pred}}}}
  {\ruleName{Ins-Conc}}


\inferrule
  { 
    \isfresh{\typeb_1}{\typeb_2} \quad
    \isfresh{\type_1}{\type_2}
  }
  { 
    \isfresh{\trfun{\evar}{\typeb_1}{\type_1}}{\trfun{\evar}{\typeb_2}{\type_2}}
  }
  {\ruleName{Ins-Fun}}

\end{mathpar}
\caption{Hole Instantiation}
\label{fig:three:isfresh}
\end{figure}

\subsection{Checking} \label{three:decl:check}

We need only alter the \emph{checking} rules 
in one place: the \rulename{Chk-Rec} judgment 
which deals with the annotations for recursive 
binders $\eletr{\evar_1}{\expr_1}{\typeb_1}{\expr_2}$.
The updated \rulename{Chk-Rec} rule is shown 
in \cref{fig:three:check}.
(All the other checking rules from $\langtwo$, 
shown in \cref{fig:two:check}, carry over to 
$\langthree$.)
Instead of using the user-specified annotation 
$\typeb_1$ which may contain holes, we use $\type_1$, 
which is an instantiation of $\typeb_1$ that is  
guaranteed to be free of holes. 

\subsection{Synthesis} \label{three:decl:synth}

Similarly, we need only modify the one \emph{synthesis} rule
that deals with type annotations, namely, rule \rulename{Syn-Ann}
which types the annotation terms $\eann{\expr}{\typeb}$.
The new rule is shown in \cref{fig:three:check}.
(All the other synthesis rules from $\langtwo$, 
summarized in \cref{fig:two:check}, apply unchanged, 
to $\langthree$.)
Again, as the annotated type $\typeb$ may have holes, 
we first instantiate it to $\type$ and then proceed,
as in $\langtwo$, pretending that the annotation was 
$\type$ all along.

\begin{figure}[t]
\judgementHead{Type Checking}{\tchk{\tcenv}{\expr}{\type}}
\begin{mathpar}
\inferrule
  { 
    \isfresh{\typeb_1}{\type_1} 
    \and
    \wf{\tcenv}{\type_1}{\kind_1} 
    \and
    \tchk{\tcenvext{\evar}{\type_1}}{\expr_1}{\type_1} 
    \and
    \tchk{\tcenvext{\evar}{\type_1}}{\expr_2}{\type_2}
  }
  {
    \tchk{\tcenv}{\eletr{\evar}{\expr_1}{\typeb_1}{\expr_2}}{\type_2}
  }
  {\ruleName{Chk-Rec}}
\end{mathpar}

\judgementHead{Type Synthesis}{\tsyn{\tcenv}{\expr}{\type}}
\begin{mathpar}
\inferrule
  {
    \isfresh{\typeb}{\type} 
    \and
    \wf{\tcenv}{\type}{\kind}
    \and
    \tchk{\tcenv}{\expr}{\type}
  }
  {
    \tsyn{\tcenv}{\eann{\expr}{\typeb}}{\type}
  }
  {\ruleName{Syn-Ann}}
\end{mathpar}
\caption{Bidirectional Checking and Synthesis: other rules from $\langtwo$ (\cref{fig:two:check})}
\label{fig:three:check}
\end{figure}

\section{Verification Conditions}\label{three:algo}


The declarative typing rules require 
an oracle to magically \emph{guess} 
refinements for the holes, and then 
\emph{verify} those guesses.
Next, lets see how Horn constraints 
let us \emph{automate} the process 
of guessing suitable instantiations. 
This approach, introduced by 
\cite{LiquidPLDI08}, has three steps.

\begin{enumerate} 
  \item \textbf{Templates:} First, we generate \emph{templates} 
        with \emph{Horn applications} $\kva{\kvar}{\evar}$
        that represent the unknown refinements;

  \item \textbf{Horn Constraints:} When the type annotations 
        contain templates, the VC generation procedure returns
        \emph{Horn constraints} that circumscribe the possible 
        concrete refinements that would make the program well-typed;

  \item \textbf{Horn Solving} Finally, we solve the Horn constraints
        to find either a suitable instantiation for the holes that 
        demonstrates the program is well-typed, or otherwise reject 
        the program as ill-typed, if no such solution can be found.
\end{enumerate}

Next, lets formalize each of these steps.

\subsection{Instantiating Holes with Templates}

\begin{figure}[t!]
$$\begin{array}{lcl}
\toprule
\freshsym & : & (\tcenv \times \Type) \rightarrow \Type     \\
\midrule
\fresh{\tcenv}{\refb{\base}{\ureft}}
                    & \doteq & \refb{\base}{\reft{\vvar}{\kva{\kvar}{\vv, \params{\evar}}}} \\[0.05in]

\quad \mbox{where}  &        &                              \\
\quad \quad \kvar   & =      & \mbox{fresh horn variable of sort}\ \base \times \params{\type} \\
\quad \quad \vvar     & =      & \mbox{fresh binder} \\
\quad \quad \params{\tb{\evar}{\type}} & = & \tcenv \\[0.05in]

\fresh{\tcenv}{\refb{\base}{\reft{\vvar}{\pred}}}
                    & \doteq & \refb{\base}{\reft{\vvar}{\pred}} \\[0.05in]

\fresh{\tcenv}{\trfun{\evar}{\typeb}{\type}}
                    & \doteq & \trfun{\evar}{\typeb'}{\type'}          \\
\quad \mbox{where}  &        &                                         \\
\quad \quad \typeb' & =      & \fresh{\tcenv}{\typeb}                  \\
\quad \quad \type'  & =      & \fresh{\tcenvext{\evar}{\typeb}}{\type} \\[0.05in]
\bottomrule
\end{array}$$
\caption{Generating fresh templates}
\label{fig:three:fresh}
\end{figure}

Recall from \cref{fig:three:syntax}, that in $\langthree$ the 
language of predicates includes Horn applications $\kvapp{\kvar}{\evar}$.

\mypara{Well-formedness} A predicate $\pred$ is 
well-formed if $\pred$ has no horn-applications, 
or is of the form $\pred' \wedge \kvapp{\kvar}{\evar}$ 
where $\pred'$ is well-formed.
This particular syntactic requirement 
on predicates ensures that the resulting 
constraints are indeed Horn constraints,
and hence can be solved to determine 
typeability.

\mypara{Templates} A \emph{template} 
is a type where all the refinements 
are well-formed predicates, \ie whose 
refinements are all of the form 
$\pred \wedge_i \kva{\kvar_i}{\evar_i}$ 
where $\pred$ has no Horn applications.

\mypara{Instantiation}
Instantiation arises in two crucial rules:
\rulename{Chk-Rec} and \rulename{Syn-Ann}.
In both cases, we require that if $\isfresh{\typeb}{\type}$ 
then the instantiated $\type$ be well-formed in the 
environment $\tcenv$.
The procedure $\freshsym$ summarized 
in \cref{fig:three:fresh} captures 
this requirement in an algorithmic
manner:
$\fresh{\tcenv}{\typeb}$ returns 
a \emph{template} $\type$ such that
for every \emph{assignment} for the 
Horn variables, the type obtained by 
applying the assignment to $\type$ 
is guaranteed to be well-formed under 
$\tcenv$. 

The first case (corresponding to \rulename{Ins-Hole})
instantiates a hole $\ureft$ with a Horn application 
$\kva{\kvar}{\vv, \params{\evar}}$ where $\vvar$ is a 
fresh symbol denoting the value being refined, and 
$\kvar$ is a fresh Horn variable denoting an (unknown) 
relation over the variables in the the environment 
$\tcenv$ and $\vvar$.
The second case (corresponding to \rulename{Ins-Conc}) 
returns the concrete refinement unmodified, and 
the third case (corresponding to \rulename{Ins-Fun}) 
recurses on the function's input and output types,
adding the input binder to the environment used to
instantiate the output, to let the output refinement 
\emph{depend upon} the input.
For example, to generate a template from the partial
type annotation for \mcode{abs}, we would invoke:
$$\fresh{\emptyset}{\trfun{x}{\tint}{\refb{\tint}{\ureft}}}$$
which would then return the template from (\ref{sig:abs:template})
$$\trfun{x}{\tint}{\refb{\tint}{\reft{\vvar}{\kva{\kvar}{x,\vvar}}}}$$

\subsection{Horn Constraints}

\begin{figure}[t]
$$\begin{array}{lcl}
\toprule
\chksym & : & (\tcenv \times \Expr \times \Type) \rightarrow \Cstr \\
\midrule
\chk{\tcenv}{\eletr{\evar}{\expr_1}{\typeb_1}{\expr_2}}{\type}
                     & \doteq & \csand{\cstr_1}{\cstr_2}       \\
\quad \mbox{where}   &        &                                \\
\quad \quad \cstr_1  & =      & \chk{\tcenv_1}{\expr_1}{\type} \\
\quad \quad \cstr_2  & =      & \chk{\tcenv_1}{\expr_2}{\type} \\
\quad \quad \tcenv_1 & =      & \tcenvext{\evar}{\type_1}      \\
\quad \quad \type_1  & =      & \fresh{\tcenv}{\typeb_1}       \\[0.05in]


\toprule
\synsym & : & (\tcenv \times \Expr) \rightarrow (\Cstr \times \Type) \\
\midrule
\syn{\tcenv}{\eann{\expr}{\typeb}}
                   & \doteq & (\cstr,\ \type)             \\
\quad \mbox{where} &        &                             \\
\quad \quad \cstr  & =      & \chk{\tcenv}{\expr}{\type}  \\
\quad \quad \type  & =      & \fresh{\tcenv}{\typeb}      \\[0.05in]

\bottomrule
\end{array}$$
\caption{Horn Verification Condition Generation for $\langthree$,  extends cases of~\cref{fig:two:algo:syn}}
\label{fig:three:algo}
\end{figure}

In \cref{fig:three:algo} we extend the VC generation procedure 
to use $\freshsym$ to generate templates from partial types. 

\mypara{Checking}
Procedure $\chksym$ is modified only for the case where it 
handles annotated terms $\eletr{\evar}{\expr_1}{\typeb_1}{\expr_2}$.
Now, we use $\freshsym$ to generate a template $\type_1$ 
for the annotation $\typeb_1$, after which we generate 
constraints as in $\langtwo$ (\cref{fig:two:check}) 
assuming the annotation was $\type_1$.

\mypara{Synthesis}
Similarly, procedure $\synsym$ is modified only for the case 
where it handles annotated terms $\eann{\expr}{\typeb}$. 
Again, we use $\freshsym$ to get a template $\type$ 
for the annotation $\typeb$, and then proceed as in 
$\langtwo$ (\cref{fig:two:synth}).

We encourage the reader to confirm that when run 
on the code with \mcode{abs} and \mcode{main} and with
the annotation $\trfun{\evar}{\tint}{\refb{\tint}{\ureft}}$ 
for \mcode{abs}, the VC generation procedure $\chksym$ 
returns the Horn constraint \cref{eq:abs:hc}. 

\section{Solving Horn Constraints}

Finally, to determine whether the program is typable, 
we need to \emph{solve} the Horn constraints produced 
by VC generation. 

\subsection{Constraint Satisfaction}

\mypara{Assignments}
Recall that a Horn \emph{assignment} $\soln$ 
is a mapping of Horn variables $\kvar$ to 
SMT predicates (relations) over the Horn 
variables' parameters.
We \emph{apply} an assignment $\soln$ 
to a predicate and constraint by \emph{replacing} 
each Horn application $\kva{\kvar}{\evarb}$ with its solution 
$\SUBST{\apply{\soln}{\kvar}}{\params{\evar}}{\params{\evarb}}$
where $\params{\evar}$ are the parameters 
of the Horn variable $\kvar$.

\mypara{Satisfaction}
An assignment $\soln$ \emph{satisfies} a Horn 
constraint $\cstr$ if applying the assignment 
to the constraint yields a \emph{valid} 
Horn-variable free formula, \ie if 
$\smtvalid{\apply{\soln}{\cstr}}$.
A Horn constraint $\cstr$ is \emph{satisfiable}
if there \emph{exists} an assignment $\soln$ 
that satisfies $\cstr$.

If the VC generation procedure yields a satisfiable
constraint, then the program is well-typed.

\begin{proposition} \label{prop:horn-sat}
  If $\chk{\tcenv}{\expr}{\type}$ is Horn satisfiable,
  then $\tchk{\tcenv}{\expr}{\type}$.
\end{proposition}

\subsection{Computing Satisfiability via Predicate Abstraction}

The \lh verifier uses its own Horn constraint 
solver that is based on predicate abstraction 
\citep{GrafSaidi97}, in particular, the Houdini 
algorithm \citep{flanagan01houdini}, extended 
with optimizations that enable precise 
\emph{local} type inference \citep{LiquidPLDI08, CosmanICFP17}.
This technique is summarized as the procedure 
$\algo{solve}$ shown in \cref{fig:fixpoint:solve}. 
$\solve{\cstr}{\quals}$ takes as input 
a Horn constraint $\cstr$  and set of 
candidate atomic predicates or 
\emph{qualifiers} $\quals$. 
The procedure returns SAT iff there 
exists an satisfying assignment for 
$\cstr$ that maps each $\kvar$ to a 
\emph{conjunction} of atomic predicates 
from $\quals$, and satisfies $\cstr$. 
The procedure has two essential elements. 

\mypara{1. Flatten}
First, we convert the Horn constraint $\cstr$ into 
a set of \emph{flat} constraints $\cstrs$ each of 
which is of the form 
$\csimps{\params{\tb{\evar}{\type}}}{\pred}{\pred'}$
where the \emph{head} $\pred'$ is either: 
(1)~a single Horn \emph{application} $\kvapp{\kvar}{\evarb}$, or,
(2)~a Horn-variable free \emph{concrete} predicate.
The subset of $\cstrs$ that have (resp. do not have) 
Horn applications in the head are gathered into the 
set $\cstrs_\kvar$ (resp. $\cstrs_\pred$).
We then invoke a the procedure \algo{fixpoint} 
to find a solution $\soln$ for the application 
constraints $\cstrs_\kvar$, and $\algo{solve}$ 
returns \textsc{Sat} iff $\soln$ also satisfies 
the concrete constraints $\cstrs_\pred$.

\mypara{2. Fixpoint}
The assignment that maps each $\kvar$ 
to the relation $\ttrue$ suffices to satisfy 
the application constraints, but of course, 
this assignment may not satisfy the 
concrete constraints. 
Instead, we start with an \emph{initial} 
solution $\soln_0$ that maps each Horn 
variable $\kvar$ to the conjunction of 
\emph{all} the candidate predicates in 
$\quals$.
\emph{Any} solution that that maps Horn 
variables to conjunctions over $\quals$ 
is trivially \emph{weaker} than $\soln_0$. 
Next, $\algo{fixpoint}$ iteratively 
\emph{weakens} the candidate solution 
$\soln$ by 
(1)~\emph{choosing} some constraint $\cstr$ not satisfied 
    by $\soln$, \ie where $\apply{\soln}{\cstr}$ is   
    is \emph{not valid}, and 
(2)~\emph{removing} qualifiers from the $\kvar$ 
    at the head $\cstr$, and 
(3)~\emph{iterating} the above process until all
    the application constraints $\cstrs_\kvar$ are 
    satisfied. 

The $\soln$ computed by \algo{fixpoint} 
is guaranteed to be the \emph{strongest} 
conjunction of candidate qualifiers that 
satisfies the application constraints 
$\cstrs_\kvar$.
Hence, if this $\soln$ \emph{also} satisfies 
the concrete constraints $\cstrs_\pred$ then 
it satisfies $\cstr$ and $\algo{solve}$ 
returns \textsc{Sat}.
Instead, if $\soln$ does not satisfy 
$\cstrs_\pred$, we can be sure there 
is no satisfying assignment for $\cstr$ 
over conjunctions of $\quals$, and so 
$\algo{solve}$ returns \textsc{Unsat} 
\citep{LiquidPLDI08}.

\begin{proposition}\label{prop:three:sat}
If $\solve{\quals}{\ \chk{\tcenv}{\expr}{\type}} = \textsc{SAT}$,
then $\tchk{\tcenv}{\expr}{\type}$.
\end{proposition}


\begin{figure}[t]
\[
\begin{array}{lcl}
\toprule
\algo{solve} & : & (\Cstr \times [\Pred]) \rightarrow \textsc{Sat} + \textsc{Unsat} \\
\midrule
\solve{\quals}{\cstr} & \doteq & \mbox{if}\ \smtvalid{\apply{\soln}{\cstrs_\pred}}\ 
                                 \mbox{then}\ \textsc{Sat}\ 
                                 \mbox{else}\ \textsc{Unsat} \\
\quad \mbox{where}    &        &  \\
\quad \quad  \mw{cs}  & =      & \flatten{\cstr} \\ 
\quad \quad  \cstrs_\kvar  & =  & \{ \cstr \ \mid\ \cstr \in \mw{cs}
                                                , \cstr \equiv \csimp{\params{\evar}}{\params{\type}}{\pred}{\kva{\kvar}{\evarb}} 
                                 \} \\
\quad \quad  \cstrs_\pred  & =  & \{ \cstr \ \mid\ \cstr \in \mw{cs}
                                                , \cstr \not \equiv \csimp{\params{\evar}}{\params{\type}}{\pred}{\kva{\kvar}{\evarb}} 
                                 \} \\
\quad \quad  \soln_0      & =  & \lambda \kvar. \wedge \{\qual \ \mid\ \qual \in \quals \} \\
\quad \quad  \soln        & =  & \fixpoint{\cstrs_\kvar}{\soln_0} \\
\bottomrule 
\end{array}
\] 
\label{fig:fixpoint:solve}
\end{figure}

\begin{figure}[t]
\[
\begin{array}{lcl}
\toprule
\flattensym & : & \Cstr \rightarrow [\Cstr] \\
\midrule 
\flatten{\cstr} 
  & \doteq & \{ \simpl{\emptyset}{\ttrue}{\cstr'} \ \mid\ \cstr' \in \splyt{\cstr} \} \\[0.05in]

\splyt{\pred}                              
  & \doteq & \{ \pred \} \\ 
\splyt{\csand{\cstr_1}{\cstr_2}}           
  & \doteq & \splyt{\cstr_1} \cup \splyt{\cstr_2} \\
\splyt{\csimp{\evar}{\type}{\pred}{\cstr}} 
  & \doteq & \{ \csimp{\evar}{\type}{\pred}{\cstr'}\ \mid\ \cstr' \in \splyt{\cstr} \} \\[0.1in]

\simpl{\params{\tb{\evar}{\type}}}{\pred}{\csimp{\evar}{\type}{\predb}{\cstr}}
  & \doteq & \simpl{\params{\tb{\evar}{\type}};\tb{\evar}{\type}}{\pred \wedge \predb}{\cstr} \\ 

\simpl{\params{\tb{\evar}{\type}}}{\pred}{\predb} 
  & \doteq & \csbind{\params{\tb{\evar}{\type}}}{\pred \Rightarrow \predb} \\[0.1in]

\toprule
\fixpointsym             & :      & ([\Cstr] \times \Soln) \rightarrow \Soln \\
\midrule
\fixpoint{\cstrs}{\soln} & \doteq & \mbox{case}\ \{ \cstr \mid \cstr \in \cstrs, \ \mbox{not}\ \smtvalid{\apply{\soln}{\cstr}} \}\ \mbox{of} \\
                         &        & \quad \emptyset \rightarrow \soln \\
                         &        & \quad \cstr : \_ \rightarrow \fixpoint{\cstrs}{\weaken{\soln}{\cstr}} \\[0.05in]

\weaken{\soln}{\csbind{\params{\tb{\evar}{\type}}}{\pred \Rightarrow \kva{\kvar}{\evarb}}}
  & \doteq & \SUBST{\soln}{\kvar}{\mw{qs'}} \\ 
\quad \mbox{where}    &        & \\
\quad \quad \mw{qs'}  & = & \{ \qual \ \mid\ \qual \in \apply{\soln}{\kvar} \ \mbox{s.t.}\ \algo{keep}(\qual) \} \\
\quad \quad \algo{keep}(\qual) & \doteq & \smtvalid{\csbind{\params{\tb{\evar}{\type}}}{\apply{\soln}{\pred} \Rightarrow \apply{\qual}{\evarb}}}\\ 
\\
\bottomrule
\end{array}
\]
\label{fig:fixpoint:refine}
\end{figure}

\section{Discussion}\label{three:discuss} 

\cite{PierceLICS03} observes that ``the more interesting 
your types get, the less fun it is to write them down.''
In this chapter, we saw how the programmer 
can elide refinement annotations and instead, 
let the type checker carry out the tedious task 
of writing them down. 
To do so, we introduced \emph{Horn variables} 
to represent the unknown refinements, converting 
the Verification Conditions into \emph{Horn constraints} 
whose satisfying assignments show the program is well-typed.

\mypara{Tools for Horn Constraint Satisfiability}
In addition to the simple algorithm based 
on predicate abstraction \citep{Jhala2018} 
shown above, there is a rich literature on 
techniques for solving Horn constraints 
that is comprehensively surveyed by 
\cite{BjornerHorn}.
Many of these ideas are based on the notion 
of iteratively refining solutions using 
refutations and are implemented in 
different tools including Spacer 
\citep{spacer}, Eldarica \citep{eldarica}, 
and even directly within some SMT solvers 
like Z3 \citep{z3PDR,z3horn}.
\cite{Jagannathan18} describes a Horn constraint 
solver that combines the refutation-guided approach 
with machine learning over sample data-values that 
either belong within or without the constrained 
relations.
Many of the above solvers are publicly available 
and compete regularly in an open competition 
\citep{ChcComp} that aims to benchmark 
and improve the solvers.

\mypara{Finding Candidate Qualifiers}
In practice we have found predicate abstraction 
to be particularly effective. Though the other 
approaches are fully automatic, \ie do not 
require qualifiers or templates, the general 
problem of inferring solutions is undecidable 
of course, and the solvers can easily diverge 
when searching for suitable predicates or relations, 
thereby making the end-to-end verification 
quite brittle \citep{jhalamcmillan06}.
In contrast, the \algo{solve} algorithm is 
parameterized by the set of qualifiers $\quals$ 
which should be thought of as a set of \emph{candidate} 
fragments from which refinements should be 
synthesized, thereby bounding the space of 
candidate refinements, and ensuring that 
the solver quickly \emph{terminates}, which 
is essential for predictable verification.
One natural source of candidates are all 
the atomic predicate fragments that appear 
inside type annotations written by the programmer.
Experience shows that this simple heuristic 
suffices to automatically infer refinements 
in practice \citep{Seidel14}.

\mypara{No Principal Types}
In some settings, inference 
can produce ideal results, 
such as the \emph{principal} 
types of \citep{Hindley69, Dam82}.
Unfortunately (with apologies to Pierce), 
the more interesting your types get the 
less principal they become. 
That is, we do not know of any reasonable 
definition of an \emph{ideal} refinement 
type for a function.
The reason is that refinements are expressive 
\emph{contracts} about how functions \emph{should} 
be used.
One can imagine many different 
and \emph{incomparable} contracts 
that \eg restrict the space of 
possible inputs to provide 
more precise guarantees 
about the \emph{outputs} 
of a function.
Thus, really the only \emph{ideal} 
specification would be an explicit 
enumeration of all the inputs and 
their respective outputs, which is 
both incomputable and entirely 
defeats the purpose of logical 
specification in the first place!

\mypara{Intra-Module Inference}
Consequently, we advocate moderation 
in the use of inference. In particular, 
because preconditions of a function are inferred 
based on the function's clients, 
it is impossible to deduce \emph{preconditions} 
describing the all inputs 
of library functions, \eg 
the {public} or {exported} 
functions of a module. 
Instead, inference is best used 
judiciously, in an \emph{intra-modular} 
fashion~\citep{Lahiri09}.
Here, the \emph{programmer} specifies 
the types for all exported functions,
and the \emph{verifier} uses those 
specifications, and the code of the 
module to infer all contracts for 
internal (private) functions. 

This recipe offers several benefits.
First, specifications on public functions 
provide useful documentation.
Second, the method allows modular analysis in 
that the verifier can analyze each module in 
complete isolation from the others.
Consequently, we have found that intra-modular 
inference drastically reduces the overhead of 
using ``interesting'' types \citep{LiquidPLDI08, Seidel14}, 
by eliminating the need to provide explicit 
annotations during polymorphic instantiation, 
as we shall see next.

\chapter{Type Polymorphism}
\label{sec:lang:four}

Next, lets look at $\langfour$ which 
adds support for \emph{type} polymorphism.

\section{Examples} \label{sec:four:examples}

The main challenge with polymorphic 
signatures is to \emph{instantiate} 
them appropriately at usage. 
Consider the following specification 
and implementation of the @max@ function: 
\begin{code}
  val max : forall 'a:Base. 'a => 'a => 'a
  let max = (x, y) => { 
    if (x < y) { y } else { x } 
  };
\end{code}
Comparison is permitted for any base type as illustrated below:
\begin{code}
  val (<) : forall 'a:Base. 'a => 'a => Bool
\end{code}
How can we verify the following \mcode{client} of \mcode{max} ?
\begin{code}
  val client: () => int[v|0 < v]
  let client = () => {
    let r = max(0, 5);
    r + 1
  };
\end{code}

\mypara{Problem: Instantiation}
Intuitively, at its usage in \mcode{client} 
the \mcode{max} function behaves 
as if it takes and returns 
non-negative numbers.
Thus, to verify \mcode{client} we must 
\emph{instantiate}, the type variable 
\mcode{'a} with the equivalent of the 
refinement:
${\refb{\tint}{\reft{\vvar}{0 \leq \vvar}}}$.
But how shall we determine appropriate 
instance refinements?
\footnote{Of course, we \emph{could} specify 
that \mcode{max} returns \emph{one of} 
its two inputs, via the type:
$\trfun{\evar}{\tint}
  {\trfun{\evarb}{\tint} 
    {\refb{\tint}{\reft{\vvar}{\vvar = \evar \vee \vvar = \evarb}}}}$
after which we would be able to verify 
the clients as in \langtwo.  
However, for the sake of exposition, 
lets assume this option is unavailable.}

\mypara{Solution: Decouple Type and Refinement Inference}
The key idea in \langfour is to \emph{decouple}
the inference of \emph{types} from those of 
\emph{refinements}.
The former, \ie type instances, can 
be determined by classical methods 
\eg Hindley-Milner style unification 
\citep{Milner82}, or its modern variants 
as seen in Haskell \citep{JonesVWW06,SchrijversJSV09},
or local inference methods with support for 
subtyping \citep{pierce-turner,sulzmann97type} 
or higher-rank polymorphism \citep{DunfieldK13}.
The latter, \ie refinement instances, 
can be obtained via Horn constraint 
solving as shown in $\langthree$.

\mypara{Phase 1: Type Elaboration}
In the first phase, $\langfour$ uses 
classical unification \citep{piercebook} 
to \emph{elaborate} the source program to
\begin{code}
    val max : forall 'a:Base. 'a => 'a => 'a
    let max = $\Lambda$ 'a:Base. (x, y) => { 
      if (x < y) { y } else { x } 
    };

    val client: () => int[v|0 < v]
    let client = () => {
      let r = (max[int[$\rhole$]])(0, 5); 
      r + 1
    };
\end{code}
Each instantiation site is elaborated to an 
explicit \emph{type application} that annotates 
the polymorphic function (\mcode{max}) with 
the instance (\mcode{int[$\rhole$]}) for each 
type variable (\mcode{'a}). 
Crucially, the instances are just the (unrefined) 
types with \emph{holes} for the as yet unknown
refinements.

\mypara{Phase 2: Refinement Inference}
In the second phase, we will generate and solve 
Horn constraints to infer suitable refinement 
instances. As in $\langthree$, we will create 
a new Horn variable $\kvar$ for each hole, so 
the above type application becomes 
${\tapp{\mcode{max}}{\refb{\tint}{\reft{\vvar}{\kva{\kvar}{\vvar}}}}}$ 
which has type
$$
  \tfun{\refb{\tint}{\reft{\vvar}{\kva{\kvar}{\vvar}}}}
       {\tfun{\refb{\tint}{\reft{\vvar}{\kva{\kvar}{\vvar}}}}
             {\refb{\tint}{\reft{\vvar}{\kva{\kvar}{\vvar}}}}}
$$ 
Next, the VC from $\langthree$ yields the Horn constraint
\begin{equation}
\begin{array}{rll}
       & \forall \vvar.\ \vvar = 0 & \Rightarrow\ \kva{\kvar}{\vvar} \\
\wedge & \forall \vvar.\ \vvar = 5 & \Rightarrow\ \kva{\kvar}{\vvar} \\
\wedge & \forall \mw{r}.\ \kva{\kvar}{\mw{r}} & \Rightarrow\ \forall \vvar.\ \vvar = \mw{r} + 1 \ \Rightarrow\ 0 < \vvar 
\end{array}
\label{eq:client:vc}
\end{equation}
Which has the satisfying assignment $\soln$ such that 
  $\kva{\soln(\kvar)}{z} \ \doteq\ 0 \leq z \label{eq:sol:max}$
which verifies that the code is well-typed.

\begin{figure}[t!]
\begin{tabular}{rrcll}
\emphbf{Environments}
  & \tcenv & $\bnfdef$ & $\ldots$               & \emph{from $\langone$ \cref{fig:one:types}} \\
  &        & $\spmid$  & \tcenvext{\tvar}{\kind}      & \emph{type variables}                       \\ [0.05in]

\emphbf{Basic Types}
  & \base & $\bnfdef$  & \ldots                 & \emph{from $\langthree$ \cref{fig:two:syntax}} \\
  &       & $\spmid$   & \tvar                  & \emph{type variables}                            \\ [0.05in]

\emphbf{Types}
  & \type  & $\bnfdef$ & $\ldots$                  & \emph{from \cref{fig:two:syntax}} \\
  &        & $\spmid$  & \tpoly{\tvar}{\type}{}      & \emph{type polymorphism}            \\[0.05in]

\emphbf{Bare Types}
  & \btype & $\bnfdef$ & \refb{\base}{\ureft}      & \emph{bare base}           \\
  &        & $\spmid$  & \trfun{x}{\btype}{\btype} & \emph{dependent function}  \\
  &        & $\spmid$  & \tpoly{\tvar}{\btype}     & \emph{type polymorphism}   \\ [0.05in]

\emphbf{Terms}
  & \expr  & $\bnfdef$ & $\ldots$                  & \emph{from \cref{fig:two:syntax}} \\
  &        & $\spmid$    & \tabs{\tvar}{\expr}     & \emph{type abstraction}             \\
  &        & $\spmid$    & \tapp{\expr}{\btype}    & \emph{type application}             \\[0.05in]
\end{tabular}
\caption{{$\langfour$: Syntax of Types and Terms}}
\label{fig:four:syntax}
\end{figure}

\section{Types and Terms} 
\label{sec:four:types}

Lets formalize the above intuition in 
$\langfour$, which extends $\langthree$ 
with polymorphic types, extends the terms 
to include type abstraction and application 
as summarized in \cref{fig:four:syntax}.

\mypara{Types}
We extend the language of types to include 
type \emph{variables} $\tvar$ of a kind $\kind$ which can be 
quantified to get \emph{polymorphic} types 
$\tkpoly{\tvar}{\kind}{\type}$. 
The set of \emph{bare} types $\btype$ are those 
where \emph{all} refinements are holes $\ureft$. 

\mypara{Terms}
As the procedure for determining type instances 
is classical \citep{piercebook, bidir-survey}, we assume 
that the language of terms is already elaborated with 
an explicit 
\emph{type abstraction} form $\tabs{\tvar}{\expr}$ 
and a 
\emph{type application} form $\tapp{\expr}{\btype}$.
Crucially, neither form involves refinements: 
the former just has type variables, and the 
latter uses bare types where all the refinements 
are holes.

\section{Declarative Typing} \label{sec:four:decl}

To account for type polymorphism, we must 
extend well-formedness and subtyping to 
quantified types, and then add rules for 
checking type abstraction terms and 
synthesizing types for the type 
application terms.

\begin{figure}[t]
\judgementHead{Well-formedness}{\wf{\tcenv}{\type}{\kind}}

\begin{mathpar}
\inferrule
  {\tb{\tvar}{\basekind}\in\tcenv \and \wfr{\tcenvext{\evar}{\tvar}}{\pred}}
  {\wf{\tcenv}{\refb{\tvar}{\reft{\evar}{\pred}}}{\basekind}}
  {\ruleName{Wf-Var-Base}}

\inferrule
  {\tb{\tvar}{\kind}\in\tcenv}
  {\wf{\tcenv}{\refb{\tvar}{\reft{\evar}{\rtrue}}}{\kind}}
  {\ruleName{Wf-Var}}

\inferrule
  {\wf{\tcenvext{\tvar}{\kind}}{\type}{\kind_\type}}
  {\wf{\tcenv}{\tkabs{\tvar}{\kind}{\type}}{\starkind}}
  {\ruleName{Wf-All}}
\end{mathpar}
\caption{$\langfour$: Rules for Well-formedness}
\label{fig:four:decl:wf}
\end{figure}

\mypara{Well-formedness}
We associate each type variable with a \emph{kind}, 
and use the well-formedness rules in \cref{fig:four:decl:wf} 
to ensure that only base-kinded type variables are refined. 
The rule \rulename{Wf-Var-Base} checks well-formedness of 
refined type variables of base kind by checking 
that the type variable is bound in the typing environment 
and that the refinement predicate is well formed. 
The rule \ruleName{Wf-Var} ensures that type variables 
of non-base types are trivially refined with true to 
avoid unsoundness.
The rule \ruleName{Wf-All} ensures that polymorphic types 
are well-formed with a star kind, if their body is well-formed 
in an environment extended with the bound variable. 

\begin{myexample}{Refining Non-base Variables is Unsound}
Consider the following polymorphic function 
that returns a \tfalse refined \tint.
\begin{code}
  val dead : forall a:Base. a[v|false] => int[v|false]
  let dead = (x) => { 0 };
\end{code}
For the type of @dead@ to be well-formed, 
the kind of @a@ has to be base as @a@ is 
refined with a non-trivial predicate. 
The precondition ensures that the function 
@dead@ type checks.
However, the precondition also effectively 
prohibits the function from being called at
run-time.
Suppose that we allowed a call to @dead@ 
with a non-base argument, \eg @id@
\begin{code}
  val unsound : int[v|false]
  let unsound = { 
    let id = (x) => {x}; 
    deadcode id 
  };
\end{code}
If the above call type checked, 
our system would \emph{unsoundly} 
prove that @0@ has the type @int[v|false]@.
Fortunately, we can ensure that 
the above definition is rejected 
by restricting how refined type 
variables (like @a@) are instantiated.  
(We could simply prohibit refinements 
 on type variables, but this would 
 preclude many useful specifications \S~\ref{sec:lang:five}.)
\end{myexample}

\begin{figure}[t!]
$$\begin{array}{rcll}
  \SUBSTMEET{(\tpoly{\tvar}{\btype})}{\tvar}{\type_\tvar} & \doteq & 
      \tpoly{\tvar}{\btype}
      & \\
  \SUBSTMEET{(\tpoly{\tvar'}{\btype})}{\tvar}{\type_\tvar} & \doteq & 
     \tpoly{\tvar'}{(\SUBSTMEET{\btype}{\tvar}{\type_\tvar})}
     ,& \tvar' \not = \tvar \\
  \SUBSTMEET{(\trfun{x}{\type_x}{\type})}{\tvar}{\type_\tvar} & \doteq & 
     \trfun{x}{(\SUBSTMEET{\type_x}{\tvar}{\type_\tvar})}{(\SUBSTMEET{\type}{\tvar}{\type_\tvar})}
     & \\
  \SUBSTMEET{(\refb{\tvar}{\rreft{\refi}})}{\tvar}{\refb{\base}{\rreft{\refi_\tvar}}} & \doteq & 
     \refb{\base}{\rreft{\refi \wedge \refi_\tvar }}
     & \\
  \SUBSTMEET{(\refb{\base}{\rreft{\refi}})}{\tvar}{\type_\tvar} & \doteq & 
     \refb{\base}{\rreft{\refi}}
     ,& \base \not = \tvar \\
\end{array}$$
\caption{Type Variable Instantiation}
\label{fig:four:tvsubst}
\end{figure}

\mypara{Type Variable Instantiation}
We use two functions to instantiate 
(substitute) variables in types. 
The function \SUBST{\type}{\tvar}{\base} 
\emph{substitutes} the type variable \tvar with 
the basic type \base in a standard way. 
The function \SUBSTMEET{\type}{\tvar}{\type_\tvar}, 
on the other hand, \emph{instantiates} the type variable 
\tvar with the type $\type_\tvar$ by strengthening 
refinement predicates, as defined in \cref{fig:four:tvsubst}.
Note that definition of \SUBSTMEET{\typeb}{\tvar}{\type} is partial: 
it is not defined when $\typeb \equiv \refb{\tvar}{\rreft{\refi}}$
for $\refi$ different than $\ttrue$ and $\type$ is not a base type, 
to prevent unsoundness as described above.

\begin{figure}[t]
\judgementHead{Subtyping}{\issub{\tcenv}{\type_1}{\type_2}}
\begin{mathpar}
\inferrule
  {
    \issub{\tcenvext{\tvar_1}{\kind}}{\type_1}{\SUBST{\type_2}{\tvar_2}{\tvar_1}}
  }
  {
    \issub{\tcenv}{\tpoly{\tvar_1}{\type_1}}{\tpoly{\tvar_2}{\type_2}}
  }
  {\ruleName{Sub-All}}
\end{mathpar}
\caption{$\langfour$: Rules for Subtyping}
\label{fig:four:decl:sub}
\end{figure}

\mypara{Subtyping}
The rule \rulename{Sub-All} shown in \cref{fig:four:decl:sub} 
formalizes subtyping for quantified types by renaming the 
type variables and checking that the types being quantified 
over belong to the subtyping relation, in an environment 
extended with the type variable.
For example, the rule derives
\[ \issub
     {\emptyset}
     {\tpoly{\tvar}{\trfun{\evar}{\tvar}{\refb{\tvar}{\reft{\vvar}{\vvar = \evar}}}}}
     {\tpoly{\tvarb}{\trfun{\evar}{\tvarb}{\tvarb}}}
\]
because, after substituting $\tvarb$ with $\tvar$ the above reduces to 
\[ \issub
     {\tvar}
     {\trfun{\evar}{\tvar}{\refb{\tvar}{\reft{\vvar}{\vvar = \evar}}}}
     {\trfun{\evar}{\tvar}{\tvar}}
\]
which follows from the rules \rulename{Sub-Base} and \rulename{Sub-Fun}.

\mypara{Checking}
The rule \rulename{Chk-TLam} in \cref{fig:four:decl} checks type-abstraction terms 
$\tabs{\tvar}{\expr}$ against quantified types $\tpoly{\tvar}{\type}$ 
by checking the inner expression $\expr$ against the $\type$ 
in an environment containing $\tvar$, and checking the 
well-formedness of the polymorphic type.

\begin{myexample}{Implementation of \mcode{max}}
The specification and implementation of 
the \mcode{max} function from \cref{sec:four:examples} 
are elaborated to:
\begin{align*}
  \type_{\mcode{max}} & \doteq\ \tkpoly{\tvar}{\basekind}{\tfun{\tvar}{\tfun{\tvar}{\tvar}}} \\ 
  \expr_{\mcode{max}} & \doteq\ \tkabs{\tvar}{\basekind}{\elam{\evar, \evarb}{\eif{\evar < \evarb}{\evarb}{\evar}}} \\
\intertext{The checking rules for $\lambda$- and branching terms establish that} 
  \tb{\tvar}{\basekind}           & \vdash\ \elam{\evar, \evarb}{\eif{\evar < \evarb}{\evarb}{\evar}} \Leftarrow \tfun{\tvar}{\tfun{\tvar}{\tvar}}
\end{align*}
after which the rule \rulename{Chk-TLam} lets us conclude 
$\tchk{\emptyset}{\expr_{\mcode{max}}}{\type_{\mcode{max}}}$.
\end{myexample}

\mypara{Synthesis} 
The rule \rulename{Syn-TApp} in \cref{fig:four:decl} 
synthesizes the type for a type-application term 
$\tapp{\expr}{\btype}$. 
Recall that the elaboration process inserts the 
application annotations using one of many standard 
approaches, but the applied type is \emph{bare} 
in that every refinement is a \emph{hole} $\ureft$ 
as a standard elaborator is unaware of refinements.
Instead, similar to the \rulename{Chk-Rec} and 
\rulename{Syn-Ann} from \cref{fig:three:check} 
which handle type annotations with holes,
the rule \rulename{Syn-TApp} first guesses
a suitable instantiation $\isfresh{\btype}{\type}$
such that $\type$ is well-formed in the given 
context.
The rule then \emph{substitutes} the 
concrete $\type$ for the type variable 
$\tvar$ quantified over in the signature 
for $\expr$.
To prevent unsoundness, this substitution 
is partial: type synthesis fails in cases 
that it is not defined. 

\begin{myexample}{Uses of \mcode{max}}
The function application term 
$(\mcode{max}\ 0\ 5)$ inside \mcode{client} from \cref{sec:four:examples}
is elaborated to $(\tapp{\mcode{max}}{\refb{\tint}{\ureft}} \ 0\ 5)$.
In the context $\tcenv$ with the signature for $\mcode{max}$, 
\begin{align*}
  \tcenv \doteq\ & \tb{\mcode{max}}{\type_{\mcode{max}}} \\
\intertext{we have}
  \tcenv \vdash\ & \mcode{max} \Rightarrow \tkpoly{\tvar}{\basekind}{\tfun{\tvar}{\tfun{\tvar}{\tvar}}} 
                          \quad \mbox{and} \quad \isfresh{\refb{\tint}{\ureft}}{\tnat} 
                          \quad \mbox{and} \quad \wf{\tcenv}{\tnat}{\basekind} \\
\intertext{and so, using \rulename{Syn-TApp} we conclude that} 
  \tcenv \vdash\ & \tapp{\mcode{max}}{\refb{\tint}{\ureft}} \Rightarrow \tfun{\tnat}{\tfun{\tnat}{\tnat}} \\
\intertext{after which the application rule \rulename{Syn-App} \cref{fig:one:check} yields}
  \tcenv \vdash\ & \eapp{\eapp{\tapp{\mcode{max}}{\refb{\tint}{\ureft}}}{0}}{5} \Rightarrow \tnat
\end{align*}
\end{myexample}

\begin{figure}[t]
\judgementHead{Type Checking}{\tchk{\tcenv}{\expr}{\type}}
\begin{mathpar}
\inferrule
  {
    \tchk{\tcenvext{\tvar}{\kind}}{\expr}{\type} \and 
    \wf{\tcenv}{\tpoly{\tvar}{\type}}{\starkind}
  }
  {
    \tchk{\tcenv}{\tabs{\tvar}{\expr}}{\tpoly{\tvar}{\type}}
  }
  {\ruleName{Chk-TLam}}
\end{mathpar}

\judgementHead{Type Synthesis}{\tsyn{\tcenv}{\expr}{\type}}
\begin{mathpar}
\inferrule
  {
    \tsyn{\tcenv}{\expr}{\tpoly{\tvar}{\typeb}}
    \and
    \isfresh{\btype}{\type} 
    \and 
    \wf{\tcenv}{\type}{\kind}
  }
  {
    \tsyn{\tcenv}{\tapp{\expr}{\btype}}{\SUBSTMEET{\typeb}{\tvar}{\type}}
  }
  {\ruleName{Syn-TApp}}
\end{mathpar}
\caption{$\langfour$: Rules for Checking and Synthesis}
\label{fig:four:decl}
\end{figure}

\section{Verification Conditions} \label{sec:four:algo}

\begin{figure}[t]
$$\begin{array}{lcl}
\toprule
\subsym & : & (\Type \times \Type) \rightarrow \Cstr \\
\midrule
\sub{\tpoly{\tvar_1}{\type_1}}{\tpoly{\tvar_2}{\type_2}} 
                     & \doteq & \sub{\type_1}{\SUBST{\type_2}{\tvar_2}{\tvar_1}} \\[0.05in]

\toprule
\chksym & : & (\tcenv \times \Expr \times \Type) \rightarrow \Cstr \\
\midrule
\chk{\tcenv}{\tabs{\tvar}{\expr}}{\tpoly{\tvar}{\type}}
                     & \doteq & \chk{\tcenvextv{\tvar}}{\expr}{\type} \\[0.05in]


\toprule
\synsym & : & (\tcenv \times \Expr) \rightarrow (\Cstr \times \Type) \\
\midrule
\syn{\tcenv}{\tapp{\expr}{\btype}}  
                   & \doteq & (\cstr, {\SUBST{\typeb}{\tvar}{\type}}) \\
\quad \mbox{where} &        &                                         \\
\quad \quad (\cstr,\ \tpoly{\tvar}{\typeb})  
                   & =      & \syn{\tcenv}{\expr}                     \\
\quad \quad \type  & =      & \fresh{\tcenv}{\ \btype}                 \\[0.05in]

\bottomrule
\end{array}$$
\caption{Verification Conditions for $\langfour$, extends cases of~\cref{fig:three:algo}}
\label{fig:four:algo}
\end{figure}

Figure~\ref{fig:four:algo} summarizes how we extend the Horn Verification Condition 
generation algorithm to account for type polymorphism. In essence, we add new cases 
to the procedures $\subsym$, $\chksym$ and $\synsym$ that respectively generate 
Horn Constraint for the subtyping, checking and synthesis modes to account for the 
new derivation rules shown in \cref{fig:four:decl}. 

\mypara{Subtyping}
The subtyping constraint for two polymorphic types is generated 
by recursing on the underlying types, after unifying the type 
variables.

\mypara{Checking}
Similarly, to check a type abstraction, we recursively 
invoke $\chksym$ on the inner expression using a suitably extended 
context. We also check well-formedness of the provided type 
to ensure the kind given to the abstracted type variable is
correct. 

\mypara{Synthesis}
The heavy lifting is done by $\synsym$, which
synthesizes a type and a constraint for a type 
application term $\tapp{\expr}{\btype}$.
However, we treat this analogous to synthesizing
the type of a type-annotation. 
Instead of ``guessing'' a type as in the declarative
\rulename{Syn-TApp} (\cref{fig:four:decl}), we use
$\btype$ to generate a fresh \emph{template} for the
instantiated $\type$ and then substitute the template
$\type$ for the type variable $\tvar$ to get the
template for the $\tapp{\expr}{\btype}$.

\begin{derivation}{\mcode{client}}
Let us see how the above works on \mcode{client} 
from \cref{sec:four:examples}. Let
\begin{align*}
  \tcenv   & \doteq\ \tb{\mcode{max}}{\tpoly{\tvar}{\tfun{\tvar}{\tfun{\tvar}{\tvar}}}} \\
  \expr_0  & \doteq\ \tapp{\mcode{max}}{\refb{\tint}{\ureft}} \\
  \expr_1  & \doteq\ \eapp{\eapp{\expr_0}{0}}{5} \\
  \expr_2  & \doteq\ \elet{\mcode{r}}{\expr_1}{\mcode{r} + 1} 
\end{align*}
Now, it it easy to check that as it simply returns the type of \mcode{max} in $\tcenv$,
\begin{align*}
  \syn{\tcenv}{\mcode{max}} 
    & \doteq\ (\ttrue, \tpoly{\tvar}{\tfun{\tvar}{\tfun{\tvar}{\tvar}}})  \\ 
\intertext{Thus, the instance of \mcode{max} synthesizes the type}
  \syn{\tcenv}{\expr_0}
    & \doteq\ (\ttrue, \tfun{\refb{\tint}{\reft{\vvar}{\kva{\kvar}{\vvar}}}} 
                            {\tfun{\refb{\tint}{\reft{\vvar}{\kva{\kvar}{\vvar}}}}{ \refb{\tint}{\reft{\vvar}{\kva{\kvar}{\vvar}}}}}) \\    
\intertext{by substituting all occurrences of $\tvar$ 
  with the \emph{fresh} template $\refb{\tint}{\reft{\vvar}{\kva{\kvar}{\vvar}}}$ 
  generated from $\refb{\tint}{\ureft}$. Consequently, the subsequent 
  applications in $\expr_1$ synthesize the constraint and type}
  \syn{\tcenv}{\expr_1} & \doteq\ (\cstr_1 \wedge \cstr_2,\ \refb{\tint}{\reft{\vvar}{\kva{\kvar}{\vvar}}}) \\
\intertext{where the synthesized type is the (instance) function's \emph{output}
  and $\cstr_1$ and $\cstr_2$ are constrain $0$ and $5$ to be subtypes of 
  the function's \emph{input}}
  \cstr_1 & \doteq\ \forall \vvar.\ \vvar = 0 \Rightarrow\ \kva{\kvar}{\vvar} \\
  \cstr_2 & \doteq\ \forall \vvar.\ \vvar = 5 \Rightarrow\ \kva{\kvar}{\vvar} \\
\intertext{The result, and hence, output type is bound to $\mcode{r}$ and so writing} 
  \tcenv' & \doteq\ \tcenvext{\mcode{r}}{ \refb{\tint}{\reft{\vvar}{\kva{\kvar}{\vvar}}}} \\
\intertext{then we get $\chk{\tcenv'}{\mcode{r} + 1}{\refb{\tint}{\reft{\vvar}{0 < \vvar}}} \doteq\ \cstr_3$ where}
\cstr_3  & \doteq\ \forall \mw{r}.\ \kva{\kvar}{\mw{r}} \ \Rightarrow\ \forall \vvar.\ \vvar = \mw{r} + 1 \ \Rightarrow\ 0 < \vvar 
\end{align*}
Hence, checking the body $\expr_2$ of \mcode{client} against its output type yields 
\[  \chk{\tcenv}{\expr_2}{\refb{\tint}{\reft{\vvar}{0 < \vvar}}} \ \doteq\ \cstr_1 \wedge \cstr_2 \wedge \cstr_3 \]
which is exactly the constraint (\ref{eq:client:vc}).
\end{derivation}

\section{Discussion}

Thus, the mechanism for refinement inference 
introduced for $\langthree$ \cref{sec:lang:three} makes 
using polymorphic functions very pleasant. 
The main problem here is to figure out how to \emph{instantiate}
a polymorphic signature at a particular instantiation site.
The key idea is to \emph{first} use classical methods to instantiate 
the \emph{unrefined} (``bare'') part of the type, leaving \emph{holes} 
for the unknown refinements, after which the Horn constraint based 
method from $\langthree$ can be applied to infer suitable refinements.

\mypara{Other approaches to Polymorphic Instantation}
There are several other possible ways to account for type polymorphism.

\begin{itemize}
  \item \emphbf{Annotations} 
    One approach is to have the programmer \emph{explicitly 
    specify} the instance refinements. However, this is 
    most unpalatable as polymorphism is ubiquitous in 
    modern code \citep{pierce-turner} and placing explicit 
    annotations would get tedious quickly.

  \item \emphbf{Defaults}
    Another approach would be to \emph{default} to some refinement 
    such as $\ttrue$ as done in Refined Racket \citep{RefinedRacket} 
    or the Stainless verifier \citep{kuncak-stainless}. 
    Sadly, this method is rather conservative, as it 
    precludes the verification of \mcode{client} as the 
    type checker has no information about the value 
    returned by \mcode{max} other than it is \emph{some} @int@.

  \item \emphbf{Unification}
    A third approach, deployed by the \fstar system \citep{fstar}, 
    is to try to \emph{unify} the types of the inputs 
    or outputs to determine suitable instance refinements.
    Unfortunately, the interaction with the refinement 
    logic makes unification brittle: for example, 
    in \mcode{client} it is unclear how to unify the types 
    of the two inputs @0@ and @5@ to obtain the 
    instance type ${\refb{\tint}{\reft{\vvar}{0 \leq \vvar}}}$.
\end{itemize}

\mypara{Polymorphism and HOFs}
Finally, support for type polymorphism is essential for 
being able to easily use Higher-Order functions. For example,
consider the \mcode{fold} function below that accumulates some value 
over the integers between $0$ and $n$:
\begin{code}
  val fold : ('a => int => 'a) => 'a => int => 'a
  let fold = (f, acc, n) => {
    let rec loop = (i, acc) => {
      if (i < n) {
        loop(i+1, f(acc, i))
      } else {
        acc
      }
    };
    loop(0, acc)
  }
\end{code}
We can use \mcode{fold} to sum the @int@egers from @0@ to @m@ as:
\begin{code}
  val sumTo: m:nat => nat
  let sumTo = (m) => {
    let add = (x, y) => {x + y};
    fold(add, 0, 0, m)
  }
\end{code}
Readers familiar with the classical Floyd-Hoare proof rule for loops 
might notice its similarity to the type signature of \mcode{fold}:
\[
  \tpoly{\tvar}
    {\tfun{(\tfun{\tvar}{\tfun{\tint}{\tvar}})}{\tfun{\tvar}{\tfun{\tint}{\tvar}}}}
\]
The type variable $\tvar$ is analagous to the loop \emph{invariant};
the accumulation function's type $\tfun{\tvar}{\tfun{\tint}{\tvar}}$ 
says that it \emph{preserves} the invariant, \ie if the input 
accumulated value satisfies the invariant then so does the output; 
the initial value of the accumulator must satisfy the invariant $\tvar$; 
and hence, ``by induction'', the final value, regardless of how many 
accumulation steps is guaranteed to satisfy the invariant $\tvar$.
Hence, the VC generation mechanism of $\langfour$
let the checker infer that within \mcode{sumTo}, 
the type parameter $\tvar$ is instantiated to $\tnat$,
\ie that $\tnat$ is an invariant of the accumulator. 
Consequently, the value returned by \mcode{fold}, 
and hence, \mcode{sumTo} must also be a $\tnat$.

This ability to automatically infer refinements 
in the presence of polymorphism will prove especially 
useful with user-defined \emph{data types}, 
as we shall see next. 

\chapter{Data Types} \label{sec:lang:five}

There is only so much one can do with @int@ 
and @bool@ values: programs get much more 
interesting once we start adding 
\emph{data types}. 
Next, lets look at $\langfive$ which extends 
$\langfour$ with support for precisely specifying 
and verifying properties of (algebraic) 
\emph{user-defined data types}.

\section{Examples} \label{sec:five:examples}

As usual, lets begin with a bird's eye view 
of the different kinds of specifications we 
might write for data types. 

\subsection{Properties of Data}

The simplest, but perhaps most ubiquitous and useful 
examples, pertain to properties of the data 
stored within polymorphic \emph{containers} like 
the @list@ type defined as:
\begin{code}
  type list('a) =
    | Nil
    | Cons('a, list('a))
\end{code}
The type declaration introduces the type @list('a)@ 
which has two constructors:
\begin{code}
  val Nil  : list('a)
  val Cons : 'a => list('a) => list('a)  
\end{code}

Lets use @Nil@ and @Cons@ to write a function @range@ that 
returns the sequence of @int@ values between 
a lower bound @lo@ and upper bound @hi@. 
\begin{code}
  val range: lo:int => hi:int =>  
             list(int[v|lo <= v && v < hi]) 
  let range = (lo, hi) => {
    if (lo < hi) {
      let rest = range(lo + 1, hi);
      Cons(lo, rest)
    } else {
      Nil
    }
  }
\end{code}
The signature for @range@ specifies that \emph{every} 
element in the output @list@ is in the interval between 
@lo@ and @hi@.

\subsection{Relationships between Data}
\label{subsec:ordered-lists}

The previous example showed how we can 
capture properties of \emph{individual} 
datum by refining the type parameter of 
@list@. 
What if we want to relate the values of 
\emph{multiple} data across a structure?
For example, here's a data type definition 
that specifies an \emph{ordered list} of 
non-decreasing values:
\begin{code}
  type olist('a) =
    | ONil
    | OCons (x:'a, xs:olist('a[v|x<=v]))
\end{code}
The type definition endows the constructors with refined signatures
\begin{code}
  val ONil : olist('a)
  val OCons: x:'a => olist('a[v|x<=v]) => olist('a)  
\end{code}
The signature for @OCons@ says that
the head @x@ must be smaller than \emph{each}
element of the tail. Thus, the type checker will  
\emph{accept} the term 
\begin{code}
  let okList = OCons(0, OCons(1, OCons(2, ONil)));  
\end{code}
but will \emph{reject} the term 
\begin{code}
  let badList = OCons(0, OCons(2, OCons(1, ONil)));
\end{code}
That is, the constructor's signature ensures 
that \emph{illegal} values do not \emph{inhabit} the type. 
More interestingly, we can specify and verify 
the function in \cref{fig:insert} that @insert@s 
a value @x@ into an ordered list @ys@.
We can use \mcode{insert} to implement an 
\emph{insertion-sort} function and verify 
that it always returns an ordered list:
\begin{code}
  val isort : list('a) => olist('a) 
  let rec isort = (xs) => {
    switch (xs){
      | Nil         => ONil
      | Cons (h, t) => insert(h, isort(t))
    }
  };
\end{code}

\begin{figure}[t!]
\begin{code}
  val insert : 'a => olist('a) => olist('a)            
  let rec insert = (x, ys) => {
    switch (ys) {
      | ONil => 
          let tl = ONil;
          OCons(x, tl)
      | OCons(y, ys') => 
          if (x <= y) {
            let tl = OCons(y,ys');
            OCons(x,tl)
          } else {
            let tl = insert(x,ys');
            OCons(y,tl)
          }
    }
  };
\end{code}
\caption{A function to $\mcode{insert}$ a value $\mcode{x}$ into an ordered list $\mcode{ys}$. }
\label{fig:insert}
\end{figure}

\subsection{Properties of Structure}

A third class of useful specifications 
are \emph{aggregate} properties of the 
entire structure, for example, 
the \emph{height} of a tree, or
the \emph{multi-set} of elements 
of a list.
Next, lets see how these can be specified 
by refining the \emph{output} types of the 
constructors with \emph{ghost} functions 
that specify the aggregate properties via 
two steps.

\mypara{1. Defining Measures}
To specify the \emph{length} of a list,
we introduce a function such that $\mlen{xs}$
represents the length of the list $\mw{xs}$.
\begin{code}
  measure len: list('a) => nat
\end{code}
To ensure decidable VC validity checking 
we ensure that @len@ is \emph{uninterpreted} 
in the refinement logic, \ie the SMT solver only 
knows that \mcode{len} satisfies the \emph{axiom of congruence}
\[
\forall \mw{xs}, \mw{ys}.\ \mw{xs} = \mw{ys} \ \Rightarrow\ \mlen{xs} = \mlen{ys}  
\]

\mypara{2. Refining Constructors}
We use measures to specify the structure's 
properties, by appropriately refining the type  
of the constructors' output. 
\begin{code}
  type list('a) =
   | Nil                    => [v|len(v) = 0]
   | Cons(x:'a,xs:list('a)) => [v|len(v) = 1+len(xs)]
\end{code}
In the definition above, the output for @Nil@ says
that it constructs a list of length $0$; the output 
for @Cons@ says that it constructs a list whose 
length is one greater than the tail.

\mypara{3. Using Measures}
We can now use @len@ in refinements in various ways.
First, to specify pre-conditions on \emph{partial} 
functions. For example, refinement checking ensures 
that due to the precondition --- which will be checked
at uses of @head@ --- the @assert(false)@ never fails 
at run-time:
\begin{code}
  val head : list('a)[v|0 < len(v)] => 'a
  let head = (xs) => { 
    switch(xs){
      | Cons(h, t) => h
      | Nil        => assert(false) 
    }
  };
\end{code}
Second, to specify post-conditions \eg on the result 
of list concatenation
\begin{code}
  val append: xs:list('a) => ys:list('a) => 
              list('a)[v|len(v) = len(xs)+len(ys)]
  let rec append = (xs, ys) => {
    switch (xs) {
      | Nil        => ys
      | Cons(h, t) => let rest = append(t, ys);
                      Cons(h, rest)
    }
  };
\end{code}

\mypara{Measures are Ghost Code}
Measures only exist at the level of the \emph{specification}:
they cannot be used in the implementation\footnote{Systems like 
\lh allow the programmer to specify the measure as 
a function that satisfies certain syntactic constraints, 
and then automatically \emph{generate} the constructor's 
refined types. However, that is merely a convenience: 
conceptually, a measure exists only for specification.}.
However, it is easy to connect measures to run-time 
values, via functions like
\begin{code}
  val length: xs:list('a) => int[v|v = len(xs)]
  let rec length = (xs) => {
    switch(xs){
      | Nil        => 0
      | Cons(h, t) => 1 + length(t)
    }
  }
\end{code}
We can now use the result of @length@ to determine 
whether it is safe to compute the @head@ of a @list@ 
\begin{code}
  val safeHead : 'a => list('a) => 'a
  let safeHead = (default, xs) => {
    let nonEmpty = 0 < length(xs);
    if (nonEmpty) { head(xs) } else { default }
  };
\end{code}
Refinement typing establishes that when @nonEmpty@ is @true@, 
we indeed have @0 < len(xs)@ thereby verifying the call @head(xs)@. 

\section{Types and Terms}\label{sec:five:types}

From the examples in \cref{sec:five:examples} 
one might get the impression that $\langfive$ 
must have multiple extensions over $\langfour$.
In fact, all three flavors of specifications --- 
reasoning about individual data, about 
relationships between data, and reasoning 
about structure --- are supported by a 
single pillar: \emph{refined data constructors}.
Thus, the only mechanisms we need are a way to 
``apply'' the refined type when \emph{constructing} 
new data and to ``unapply'' the type when 
\emph{destructing} the data by pattern matching.
Next, lets see how these two ideas are formalized
in $\langfive$, whose syntactic additions are summarized 
in Figure \ref{fig:five:syntax}.

\mypara{Datatypes}
First, we assume there is a set of 
\emph{type constructors} $\tcon$ (\eg $\mcode{list}$) 
and \emph{data constructors} $\dcon$ (\eg $\tnil$, $\tcons$).
The \emph{polarity} $\polar$ captures the position that 
type variables appear in definitions of data types 
and can be positive ($\ppos$), negative ($\pneg$), 
both ($\pboth$), or neither ($\pnone$).  
A \emph{datatype} $\delta$ is a triple 
$\tdef{\tcon}{\dtydefs{\tvar}{\kind}{\polar}}{\ddefs{\dcon}{\type}}$ 
comprising a type constructor $\tcon$, 
the list of type variables 
over which the datatype is parameterized
together with their kind 
and polarity $\dtydefs{\tvar}{\kind}{\polar}$,
and a set of data constructors and their 
refinement types $\ddefs{\dcon}{\type}$.

\begin{myexample}{Ordered Lists}
Suppose we wrote the following 
ordered list type, refined with a \mcode{len} 
measure tracking the list's size
\begin{code}
  type ol('a) =
   | ON                       => [v|len(v)=0]
   | OC('a,xs:ol('a[v|x<=v])) => [v|len(v)=1+len(xs)]
\end{code}
We would represent the above as 
$\delta_{\mcode{OL}} \doteq \langle \mcode{OL},\ \{\dtydef{\tvar}{\basekind}{\ppos}\},\ \{ \tb{\mcode{ON}}{\type_N}; \tb{\mcode{OC}}{\type_C} \} \rangle$
The type variable $\tvar$ appears in one positive position 
and since $\tvar$ elements are compared it is of base kind.  
The types of the ``nil'' and ``cons'' constructors are respectively:
\begin{align}
  \type_N  & \doteq\ \tkpoly{\tvar}{\basekind}{\refb{\olista}{\reft{\vvar}{\tlen{\vvar} = 0}}} \notag \\
  \type_C  & \doteq\ \tkpoly{\tvar}{\basekind}{\trfun{\evar}{\tvar}{\trfun{\mw{xs}}{\olist{\refb{\tvar}{\reft{\vvar}{\evar \leq \vvar}}}}{\refb{\olista}{\reft{\vvar}{\tlen{\vvar} = 1 + \tlen{\mw{xs}}}}}}} \label{eq:OC:type}
\end{align}
That is, the type $\type_N$ says that ``nil'' 
returns an ordered list of length $0$;
and the type $\type_C$ says that ``cons'' takes 
as input a head $\mw{x}$ of type 
$\tvar$ and a tail $\mw{xs}$ \emph{each} 
of whose elements is an $\tvar$ larger 
than $\mw{x}$, and returns an ordered 
list whose size is one more than that 
of the tail $\mw{xs}$.
\end{myexample}

\mypara{Environments}
We extend the environments to include all the 
data type definitions $\delta$. For simplicity, 
we will assume that all the data type definitions 
and measure names are \emph{global}, that is, they 
belong in the top-level environment used for type 
checking and synthesis.

\mypara{Types}
As hinted in the discussion for constructors above, 
the language of \emph{basic} types is extended to 
include a \emph{type constructor application} form 
$\tcapp{\tcon}{\params{\type}}$, where the type 
constructor $\tcon$ is applied to the type arguments 
$\params{\type}$.
Intuitively one can think of the above as the 
type obtained by instantiating the type 
parameters $\params{\tvar}$ of $\tcon$ 
with the actual type arguments $\params{\type}$.
As before, these basic types can be refined, so 
$\refb{\olist{\tnat}}{\reft{\vvar}{3 \leq \tlen{\vvar}}}$ 
would correspond to the type of ordered lists 
of non-negative $\tint$ values comprising three 
or more elements.

\mypara{Alternatives}
Each \emph{alternative} $\dcalt{\dcon}{\evar}{\expr}$ 
comprises a \emph{pattern} $\dcon(\params{\evar})$ 
and the term $\expr$ to be evaluated if the 
scrutinee matches the pattern.

\mypara{Terms}
We add the data constructors $\dcon$ to the language of terms so that 
polymorphic instantiation $\tapp{\expr}{\type}$ (from $\langfour$)
and function application $\eapp{\expr}{\evar}$ (from $\langone$) 
can be combined to \emph{construct} values of user-defined types.
To \emph{destruct} values of user-defined types, we introduce
a pattern-match form $\ecase{\evarb}{\params{\alt}}$ where 
the value bound to $\evarb$ is \emph{scrutinized} by each 
of the alternatives in $\params{\alt}$.

\begin{figure}[t!]
\begin{tabular}{rrcll}
\emphbf{Data Constructors} & \dcon & $\bnfdef$ & $\dcon_1,\dcon_2,\dcon_3,\ldots$ &   \\[0.05in]

\emphbf{Type Constructors} & \tcon & $\bnfdef$ & $\tcon_1,\tcon_2,\tcon_3,\ldots$ &   \\[0.05in]

\emphbf{Polarity}
  & $\polar$ & $\bnfdef$ & \ppos $\spmid$ \pneg $\spmid$ \pboth $\spmid$ \pnone &  \\[0.05in]

\emphbf{Datatypes}
  & $\delta$ & $\bnfdef$ & \tdef{\tcon}{\dtydefs{\tvar}{\kind}{\polar}}{\ddefs{\dcon}{\type}}  &  \\[0.05in]

\emphbf{Environments}
  & \tcenv & $\bnfdef$ & $\ldots$                       & \emph{from Fig.~\ref{fig:four:syntax}} \\
  &        & $\spmid$  & \tcenvextv{\delta}             & \emph{type definitions}  \\[0.05in]

\emphbf{Basic Types}
  & \base & $\bnfdef$  & $\ldots$                       & \emph{from Fig.~\ref{fig:four:syntax}} \\
  &       & $\spmid$   & \tcapp{\tcon}{\params{\type}}  & \emph{datatypes} \\ [0.05in]

\emphbf{Alternatives}
  & \alt  & $\bnfdef$  & \dcalt{\dcon}{\evar}{\expr}    & \emph{switch alternative} \\[0.05in]

\emphbf{Terms}
  & \expr & $\bnfdef$ & $\ldots$                        & \emph{from Fig.~\ref{fig:four:syntax}} \\
  &       & $\spmid$  & $\dcon$                         & \emph{data constructor} \\
  &       & $\spmid$  & \ecase{\evar}{\params{\alt}}    & \emph{data destructor}  \\ [0.05in]
\end{tabular}
\caption{{$\langfive$: Syntax of Types and Terms}}
\label{fig:five:syntax}
\end{figure}

\section{Declarative Typing} \label{sec:five:decl}

Next, lets see how the rules for well-formedness, subtyping, checking and synthesis 
are extended to account for constructors and destructors.

\begin{figure}[t]
\judgementHead{Well-formedness}{\wf{\tcenv}{\type}{\kind}}

\begin{mathpar}
\inferrule
  {
  \params{\kind} = \kinds{\delta}{\tcon}
  \quad
  \wf{\tcenv}{\type_i}{\kind_i}\ \mbox{for each}\ 1 \leq i \leq \mid\! \params{\kind} \!\mid
  \quad
  \wfr{\tcenvext{\evar}{\tcapp{\tcon}{\params{\type}}}}{\pred}
  }
  {\wf{\tcenv}{\refb{\tcapp{\tcon}{\params{\type}}}{\reft{\evar}{\pred}}}{\basekind}}
  {\ruleName{Wf-Data}}
\end{mathpar}
\caption{$\langfive$: Rule for Well-formedness}
\label{fig:five:wf}
\end{figure}

\begin{figure}[t]
  \judgementHead{Subtyping}{\issub{\tcenv}{\type_1}{\type_2}}
\begin{mathpar}
\inferrule
  {
    \issub{\tcenv}{\typeb_i}{\type_i} \ \mbox{for each}\ i. \polar_i \in \{\ppos, \pboth \}\\\\
    \issub{\tcenv}{\type_i}{\typeb_i} \ \mbox{for each}\ i. \polar_i \in \{\pneg, \pboth \}\\\\
    \entl{\tcenvext{\vvar_1}{\reft{\tcapp{\tcon}{\params{\typeb}}}{\pred_1}}}{\SUBST{\pred_2}{\vvar_2}{\vvar_1}}
    \and \params{\polar} = \polarities{\delta}{\tcon}
  }
  {
    \issub{\tcenv}
      {  \refb{\tcapp{\tcon}{\params{\typeb}}}{\reft{\vvar_1}{\pred_1}}  }
      {  \refb{\tcapp{\tcon}{\params{\type }}}{\reft{\vvar_2}{\pred_2}}  }
  }
  {\ruleName{Sub-Data}}
\end{mathpar}
\caption{$\langfive$: Rules for Subtyping}
\label{fig:five:sub}
\end{figure}

\subsection{Well-formedness}
The rule \ruleName{Wf-Data} shown in~\cref{fig:five:wf} formalizes well-formedness of 
datatypes. The rule checkes that the refinement of the type is well-formed,
that each of the type arguments has the proper kind, and that 
the type constructor is fully applied. 
The premises use the function $\kinds{\delta}{\tcon}$
that retrieves the kinds of $\tcon$ from the data environment $\delta$:
$$
\kinds{\delta}{\tcon} \doteq \params{\kind} \quad \text{if}\
\tdef{\tcon}{\dtydefs{\tvar}{\kind}{\polar}}{\ddefs{\dcon}{\type}} \in \delta
$$

\subsection{Subtyping}
The rule \rulename{Sub-Data} shown in \cref{fig:five:sub} 
formalizes subtyping between datatypes.
In an environment $\tcenv$, the type $\refb{\tcapp{\tcon}{\params{\typeb}}}{\reft{\vvar_1}{\pred_1}}$
is a subtype of $\refb{\tcapp{\tcon}{\params{\type }}}{\reft{\vvar_2}{\pred_2}}$,
if the base refinement $\pred_1$ entails $\pred_2$, and \emph{each} of the 
component types $\typeb_i$ is a subtype of the corresponding component 
$\type_i$.
Subtyping of the components is checked using 
\polarities{\delta}{\tcon} which retrieves the 
polarity information from the data environment $\delta$
$$
\polarities{\delta}{\tcon} = \params{\polar} \ \text{if}\
\tdef{\tcon}{\dtydefs{\tvar}{\kind}{\polar}}{\ddefs{\dcon}{\type}} \in \delta
$$
For each component with positive polarity (resp. negative) polarity 
the rule uses covariant (resp. contravariant) subtyping. 

\begin{myexample}{Subtyping in \mcode{insert}}
Consider the environment
\[ \tcenv'_\leq \ \doteq\ \tb{\tvar}{\basekind};\
                         \tb{\mw{x}}{\tvar};\ 
                          \tb{\mw{y}}{\tvar};\ 
                          \tb{\mw{ys'}}{\olist{\refb{\tvar}{\reft{\vvar}{y \leq \vvar}}}};\ 
                          \tb{\mw{ys}}{\olistp{\tvar}{\mw{ys'}}};\ 
                          \mw{x} \leq \mw{y}
                          \]
and the alias $\olistp{\tvar}{\mw{z}}$ that denotes 
ordered lists of type $\tvar$ whose size is one 
more than that of $\mw{z}$. 
%
%
As the following entailment is valid 
\begin{align}           
  \forall \mw{x}, \mw{y}, \vvar.\ & \mw{x} \leq \mw{y} \ \Rightarrow\ \mw{y} \leq \vvar \ \Rightarrow\ \mw{x} \leq \vvar 
  \notag \\
\intertext{the rule \ruleName{Sub-Data} lets us conclude}
  \tcenv'_\leq & \ \vdash\ \olist{\refb{\tvar}{\reft{\vvar}{\mw{y}\leq \vvar}}} \lqsubt\ \olist{\refb{\tvar}{\reft{\vvar}{\mw{x}\leq\vvar}}} 
  \label{eq:ins:five:c}\\
\intertext{ As the following entailment is valid}
  \forall \mw{ys},\ \mw{ys'},\ \mw{tl}.\ 
    & \tlen{\mw{ys}} = 1 + \tlen{\mw{ys'}}\ \Rightarrow\ \tlen{\mw{tl}} = 1 + \tlen{\mw{ys'}} \ \Rightarrow 
      \notag \\
    & \tlen{\vvar} = 1 + \tlen{\mw{tl}}\  \Rightarrow\ \tlen{\vvar} = 1 + \tlen{\mw{ys}} 
      \notag \\  
\intertext{the rule \ruleName{Sub-Data} lets us conclude}
  \tcenv'_\leq; \tb{\mw{tl}}{\type_\mw{y}} & \ \vdash \ \olistp{\tvar}{\mw{tl}}\ \lqsubt\ \olistp{\tvar}{\mw{ys}} 
  \label{eq:ins:seven}
\end{align}
where $\type_\mw{y} \doteq \olistp{\tvar_\mw{x}}{\mw{ys'}}$
\end{myexample}


\subsection{Checking}

\begin{figure}[t!]
\judgementHead{Type Checking}{\tchk{\tcenv}{\expr}{\type}}

\begin{mathpar}
\inferrule
  {
      \tchk{\tcenvscr{\evarb}}{\alt_i}{\type} \ \mbox{for each}\ i
  }
  {
      \tchk{\tcenv}{\ecase{\evarb}{\params{\alt}}}{\type}
  }
  {\ruleName{Chk-Swt}}
\end{mathpar}

\judgementHead{Checking Alternatives}{\tchk{\tcenvscr{\evarb}}{\alt}{\type}}
\begin{mathpar}
  \inferrule
  {
      \typeb = \dconty{\tcenv}{\dcon}{\evarb}
      \and
      \tcenv' = \unapply{\tcenv}{\evarb}{\params{\evarc}}{\typeb}
      \and
      \tchk{\tcenv'}{\expr}{\type}
  }
  {
      \tchk{\tcenvscr{\evarb}} {\dcalt{\dcon}{\evarc}{\expr}}{\type}
  }
  {\ruleName{Chk-Alt}}
\end{mathpar}
\caption{$\langfive$: Rules for Type Checking}
\label{fig:five:check}
\label{fig:five:decl}
\end{figure}

\begin{figure}[t]
$$\begin{array}{lcl}
\toprule
\unapplysym & : & (\tcenv \times \Evar \times \Evar^{*} \times \Type) \rightarrow \tcenv \\
\midrule
\unapply{\tcenv}{\evarb}{z;\params{\evarc}}{\trfun{\evar}{\typeb}{\type}}
  & \doteq & \unapply{\tcenvext{\evarc}{\typeb}}{\evarb}{\params{\evarc}}{\SUBST{\type}{\evar}{\evarc}} \\[0.05in]

\unapply{\tcenv}{\evarb}{\emptyset}{\type}
  & \doteq & \tcenvext{\evarb}{\meet{ \tcenvget{\evarb} }{\type}} \\[0.10in]

\toprule
\dcontysym & : & (\tcenv \times \dcon \times \Evar) \rightarrow \Type \\
\midrule
\dconty{\tcenv}{\dcon}{\evarb}
  & \doteq & \SUBST{\typeb}{\params{\tvar}}{\params{\type}}  \\
\quad \mbox{where} & & \\
\quad \quad \tcapp{\tcon}{\params{\type}}  & = & \tcenvget{\evarb} \\
\quad \quad \tpoly{\params{\tvar}}{\typeb} & = & \tcenvget{\dcon} \\[0.05in]
\bottomrule
\end{array}$$
\caption{Meta-functions for Type Checking Switch Alternatives} 
\label{fig:unapply}
\label{fig:dconty}
\end{figure}

\begin{figure}[t]
$$\begin{array}{lcl}
\toprule
\meetsym & : & (\Type \times \Type) \rightarrow \Type \\
\midrule
\meet{\refb{\base}{\reft{\vvar_1}{\pred_1}}}{\refb{\base}{\reft{\vvar_2}{\pred_2}}}
  & \doteq & \refb{\base}{\reft{\vvar_1}{\pred_1 \wedge \SUBST{\pred_2}{\vvar_2}{\vvar_1}}} \\[0.05in]
%
\meet{\trfun{\evar_1}{\typeb_1}{\type_1}}{\trfun{\evar_2}{\typeb_2}{\type_2}}
  & \doteq & \trfun{\evar_1}{\meet{\typeb_1}{\typeb_2}}{\meet{\type_1}{ \SUBST{\type_2}{\evar_2}{\evar_1}}} \\[0.05in]
%
\meet{\tpoly{\tvar_1}{\type_1}}{\tpoly{\tvar_2}{\type_2}}
  & \doteq & \tpoly{\tvar_1}{\meet{\type_1}{ \SUBST{\type_2}{\tvar_2}{\tvar_1} }} \\[0.05in]
\bottomrule
\end{array}$$
\caption{Conjoining Types}
\label{fig:five:meet}
\end{figure}

The rule \rulename{Chk-Swt} shown in \cref{fig:five:check} 
describes how to \emph{check} that a switch expression has 
a given type $\type$, by checking that each {alternative} of 
the switch produces a value of type $\type$.

\mypara{Checking an Alternative}
The judgment 
$\tchk{\tcenvscr{\evarb}} {\dcalt{\dcon}{\evarc}{\expr}}{\type}$
states that in the \emph{environment} $\tcenv$ 
when the \emph{scrutinee} $\evarb$ 
matches the \emph{pattern} $\dcon(\params{\evarc})$ 
the evaluated \emph{result} $\expr$ has type $\type$.
The rule \rulename{Chk-Alt} establishes this judgment 
in three steps.
\begin{enumerate}
\item We use $\dconty{\tcenv}{\dcon}{\evarb}$
      summarized in \cref{fig:dconty} to get 
      $\typeb$, the monomorphic \emph{instantiation} 
      of the polymorphic type of constructor $\dcon$. 
      In other words, $\typeb$ is the type of $\dcon$ 
      at \emph{this} particular match-instance.
\item We invoke $\unapply{\tcenv}{\evarb}{\params{\evarc}}{\typeb}$
      summarized in \cref{fig:dconty} 
      to obtain the environment $\tcenv'$ which 
      is $\tcenv$ extended with the types for 
      the pattern match bindings $\params{\evarc}$ 
      and also, with additional refinements for 
      the scrutinee $\evarb$ that are revealed 
      by matching against this particular pattern. 
\item We check that result $\expr$ has 
      the type $\type$ in environment $\tcenv'$.
\end{enumerate}

\mypara{Unapply}
At destruction sites we use 
$\unapply{\tcenv}{\evarb}{\params{\evarc}}{\typeb}$ 
summarized in \cref{fig:unapply}.
The function $\unapplysym$ can be viewed as
the dual of function application 
Given the output type of the constructed 
value ($\typeb$), we want to 
(1)~determine the types that the inputs 
      ($\params{\evarc}$) must have had, 
      and to then 
(2)~add those bindings to get the 
      environment used to check the 
      alternative's body $\expr$.
$\unapplysym$ does so by recursively 
``zipping'' together the match-binders 
$\params{\evarc}$ with the input binders 
of the constructor's (function) type $\typeb$.
If the sequence of binders is \emph{non-empty} ($\evarc; \params{\evarc}$) 
and the constructor type is $\trfun{\evar}{\typeb}{\type}$ then 
we extend $\tcenv$ with the binding $\tb{\evarc}{\typeb}$, and 
recurse on the extended environment and the remaining binders 
$\params{\evarc}$ and the ``rest'' of the constructor type, \ie 
its output $\type$ after substituting the formal $\evar$ with 
the ``actual'' $\evarc$.
Once the sequence of binders is \emph{empty} ($\emptyset$) 
then constructor type $\type$ is exactly the result of
$\dcon(\params{\evarc})$. Crucially, $\type$ can have 
extra information about the scrutinee $\evarb$ that 
holds under this particular pattern match, and so we
\emph{strengthen} the type of $\evarb$ by using $\meetsym$
to conjoin the old type $\tcenvget{\evarb}$ with the 
pattern-match result $\type$, and return the extended 
environment as the final result (that is used to check 
the alternative's body $\expr$.)

\begin{figure}[t]
\begin{align*}
\lambda \mw{x},\ \mw{ys}.\ 
& \kw{switch}\ (\mw{ys}) \\
& \quad \mid \dcapp{\mcode{OC}}{\mw{y}, \mw{ys'}} \ \rightarrow \\
& \quad \quad \kw{if}\ \mw{x} \leq \mw{y}\ \kw{then} \\ 
& \quad \quad \quad \kw{let}\ \mw{tl} = \eapp{\eapp{\tapp{\mcode{OC}}{\tvar_{\mw{x}}}}{\mw{y}}}{\mw{ys'}}\ \kw{in} \\ 
& \quad \quad \quad \eapp{\eapp{\tapp{\mcode{OC}}{\tvar}}{\mw{x}}}{\mw{tl}} \\ 
& \quad \quad \kw{else}\ \ldots \\
& \quad \mid \mcode{ON} \ \rightarrow \ldots
\end{align*}
\caption{Definition of \mcode{insert} (\cref{fig:insert}) with elaborated type applications.}
\label{fig:insert:elab}
\end{figure}

\begin{myexample}{Checking in \mcode{insert}}
Lets see how the declarative rules let 
us \emph{check} the implementation of the 
\mcode{insert} function from \cref{fig:insert}.
Suppose our goal is to check that
\mcode{insert} implements the type 
$${\trfun{\mw{x}}{\tvar}{\trfun{\mw{ys}}{\olista}{\olistp{\tvar}{\mw{ys}}}}}$$
\ie that \mcode{insert} returns an ordered list with 
one more element than the input list \mw{ys}.
The \cref{fig:insert:elab} shows a fragment of 
the definition of \mcode{insert} elaborated with 
explicit type applications at constructor application 
sites.
Rule \rulename{Chk-Lam} reduces type 
checking to the following judgment 
that checks the body of \mcode{insert} 
against the specified output type in 
the context $\tcenv$ comprising the 
bindings for the inputs $\mw{x}$ and $\mw{ys}$: 
\begin{align}
\tcenv & \ \vdash\ \ecase{\mw{ys}}{\ \dcapp{\mcode{OC}}{\mw{y}, \mw{ys'}}\rightarrow {\expr'} \ \mid\ \ldots } \Leftarrow\ \olistp{\tvar}{\mw{ys}} 
  \notag \\ 
\intertext{The rule \rulename{Chk-Alt} reduces the above into 
obligations for each alternative. For the $\mcode{OC}$ 
case, we get:}
\tcenv \ \mid\ \mw{ys} & \ \vdash\  \dcapp{\mcode{OC}}{\mw{y}, \mw{ys'}}\rightarrow {\expr'} \Leftarrow\ \olistp{\tvar}{\mw{ys}} 
  \notag \\ 
\intertext{By the definition of $\unapplysym$ the above judgment reduces to}
\tcenv' & \ \vdash\ \expr' \Leftarrow\ \olistp{\tvar}{\mw{ys}} 
  \notag \\
\intertext{where $\tcenv'$ is $\tcenv$ extended with the bindings computed by $\unapplysym$}
\tcenv' &\ \doteq\ \tcenv;\ 
                    \tb{\mw{y}}{\tvar};\ 
                    \tb{\mw{ys'}}{\olist{\refb{\tvar}{\reft{\vvar}{y \leq \vvar}}}};\
                    \tb{\mw{ys}}{\olistp{\tvar}{\mw{ys'}}} 
  \notag \\
\intertext{As $\expr'$ is $\eif{\mw{x} \leq \mw{y}}{\expr'_\leq}{\ldots}$ 
           by rule \rulename{Chk-If}, eliding the else-branch, 
           and recalling that $\expr'_\leq \doteq \elet{\mw{tl}}{\expr_\mw{y}}{\expr_\mw{x}}$ 
           (\cref{fig:insert:defs}) the above judgment reduces to} 
\tcenv';\ \mw{x} \leq \mw{y} & \ \vdash\ \elet{\mw{tl}}{\expr_\mw{y}}{\expr_\mw{x}} \Leftarrow\ \olistp{\tvar}{\mw{ys}}
  \label{eq:ins:four} \\
\intertext{In the sequel, we will show how the synthesis rules establish}
  \tcenv'_\leq & \ \vdash \ \expr_{\mw{y}} \Rightarrow \type_{\mw{y}} \label{eq:ins:five} \\
  \tcenv'_\leq; \tb{\mw{tl}}{\type_\mw{y}} & \ \vdash \ \expr_{\mw{x}} \Rightarrow \olistp{\tvar}{\mw{tl}} \label{eq:ins:six}
\intertext{which \rulename{Chk-Let} and \rulename{Chk-Syn} combine with the subtyping judgment} 
\tcenv'_\leq; \tb{\mw{tl}}{\type_\mw{y}} & \ \vdash \ \olistp{\tvar}{\mw{tl}}  \lqsubt\ \olistp{\tvar}{\mw{ys}} 
  \quad \mbox{previously derived in (\ref{eq:ins:seven})}
  \notag 
\end{align}
to yield the goal (\ref{eq:ins:four}).
\end{myexample}


\begin{figure}[t]
\[
\begin{array}{lcl}
  \type_{\mcode{ins}}       & \doteq & \trfun{\mw{x}}{\tvar}{\trfun{\mw{ys}}{\olista}{\olistp{\tvar}{\mw{ys}}}}      \\
  \olistp{\tvar}{\mw{z}} & \doteq & \olista{\reft{\vvar}{\tlen{\vvar} = 1 + \tlen{\mw{z}}}}                           \\
  \tcenv                 & \doteq & \tb{\tvar}{\basekind};\ \tb{\mcode{insert}}{\type_{\mw{ins}}};\ \tb{\mw{x}}{\tvar};\ \tb{\mw{ys}}{\olista} \\
  \tcenv'                & \doteq & \tcenv;\ \tb{\mw{y}}{\tvar};\ 
                                             \tb{\mw{ys'}}{\olist{\refb{\tvar}{\reft{\vvar}{y \leq \vvar}}}};\ 
                                             \tb{\mw{ys}}{\olistp{\tvar}{\mw{ys'}}}                  \\
  \tcenv'_\leq           & \doteq & \tcenv';\ \mw{x} \leq \mw{y}                                    \\
  \expr'                 & \doteq & \eif{\mw{x} \leq \mw{y}}{\expr^C_\leq}{\ldots}                  \\ 
  \expr'_\leq            & \doteq & \elet{\mw{tl}}{\expr_y}{\expr_x}                                \\ 
  \expr_{\mw{y}}         & \doteq & \eapp{\eapp{\tapp{\mcode{OC}}{\tvar_{\mw{x}}}}{\mw{y}}}{\mw{ys'}}  \\ 
  \expr_{\mw{x}}         & \doteq & \eapp{\eapp{\tapp{\mcode{OC}}{\tvar}}{\mw{x}}}{\mw{tl}}            \\
  \tvar_{\mw{x}}         & \doteq & \refb{\tvar}{\reft{\vvar}{\mw{x} \leq \vvar}}                       \\                      
  \type_\mw{y}           & \doteq & \olistp{\tvar_\mw{x}}{\mw{ys'}}                                 \\
\end{array}
\]
\caption{Definitions for checking and synthesizing types for \mcode{insert} \cref{fig:insert}.}
\label{fig:insert:defs}
\end{figure}

\subsection{Synthesis}

As we said at the outset, the key 
mechanism to supporting datatypes 
are refined \emph{data constructors}.
Their types are applied at construction 
sites in a manner identical to plain 
function application, where given the 
input type we synthesize the output.
Hence, to synthesize the types for constructor 
applications we add the rule \rulename{Syn-Data} 
shown in \cref{fig:five:decl}.
The rule synthesizes the type for a data 
constructor $\dcon$ by looking up its type 
in the environment $\tcenv$; after this, 
the rules for application 
\ie \rulename{Syn-App} from \cref{fig:one:synth}
suffice for constructor applications.

\begin{myexample}{Synthesis in \mcode{insert}}
Lets see how the synthesis rules let us establish the 
judgments (\ref{eq:ins:five}) and (\ref{eq:ins:six}) 
that are needed to check the \kw{then} branch of the 
implementation of \mcode{insert} from \cref{fig:insert}.

First, lets establish the type synthesized 
for the constructor application 
of \ref{eq:ins:five} (where the abbreviations $\tvar_\mw{x}$, $\type_\mw{y}$ 
are summarized in \cref{fig:insert:defs}):
\begin{align}
\tcenv'_\leq & \ \vdash\ \eapp{\eapp{\tapp{\mcode{OC}}{\tvar_{\mw{x}}}}{\mw{y}}}{\mw{ys'}}
                    \Rightarrow   \olistp{\tvar_\mw{x}}{\mw{ys'}} \notag \\
\intertext{\rulename{Syn-Data} synthesizes the polymorphic type 
  of the constructor $\mcode{OC}$ (\ref{eq:OC:type}) and 
  \rulename{Syn-TApp} synthesizes its instantiation to yield}
\tcenv'_\leq 
  & \ \vdash\ {\tapp{\mcode{OC}}{\tvar_{\mw{x}}}} \Rightarrow 
  \trfun{\mw{z}}{\tvar_{\mw{x}}}
    {
      \trfun{\mw{zs}}{\olist{\refb{\tvar_{\mw{x}}}{\reft{\vvar}{\mw{z} \leq \vvar}}}}
        {
          \olistp{\tvar_\mw{x}}{\mw{zs}}
        }
    } 
    \label{eq:ins:cons:y}
\end{align}
The rule \ruleName{Syn-App} requires that the types of 
the first and second arguments $\mw{y}$ and $\mw{ys'}$ 
must be subtypes of the constructor's respective input 
types. 
The subtyping for the first argument \mw{y} follows 
from the validity of the entailment 
\[
  \forall \mw{x}, \mw{y}, \vvar.\ \mw{x} \leq \mw{y} \ \Rightarrow\ \vvar = \mw{y} \ \Rightarrow\ \mw{x} \leq \vvar 
\]
The subtyping for the second argument was established 
in (\ref{eq:ins:five:c}). 
Hence, \ruleName{Syn-App} establishes (\ref{eq:ins:five}).

Next, lets synthesize a type for the constructor 
application (\ref{eq:ins:six}) in the environment 
$\tcenv'_\leq$ extended by binding $\type_\mw{y}$ 
synthesized in (\ref{eq:ins:five}) to $\mw{tl}$:
\begin{align}
\tcenv'_\leq; \tb{\mw{tl}}{\type_\mw{y}} 
  & \ \vdash\ \eapp{\eapp{\tapp{\mcode{OC}}{\tvar}}{\mw{x}}}{\mw{tl}} \Rightarrow \olistp{\tvar}{\mw{tl}} \notag \\
\intertext{Again via \ruleName{Syn-Data} and \ruleName{Syn-TApp} 
   the constructor's instance type is}
\tcenv'_\leq; \tb{\mw{tl}}{\type_\mw{y}} 
  & \ \vdash\ {\tapp{\mcode{OC}}{\tvar}} \Rightarrow 
    \trfun{\mw{z}}{\tvar}
    {
      \trfun{\mw{zs}}{\olist{\refb{\tvar}{\reft{\vvar}{\mw{z} \leq \vvar}}}}
        {
          \olistp{\tvar}{\mw{zs}}
        }
    } \label{eq:ins:six:a} 
\end{align}
As $\tcenvget{\mw{x}} = \tvar$ we trivially get that the first 
argument $\mw{x}$ is a subtype of the first input type $\tvar$. 
Similarly from the binding $\tb{\mw{tl}}{\type_\mw{y}}$,
the second argument $\mw{tl}$ is a subtype 
of the second input type $\olist{\refb{\tvar}{\reft{\vvar}{\mw{x} \leq \vvar}}}$, 
after substituting the parameter $\mw{z}$ for the actual $\mw{x}$. 
Hence, by \ruleName{Syn-App} and the type of the 
constructor (\ref{eq:ins:six:a}), we arrive at our 
destination (\ref{eq:ins:six}).
\end{myexample}

\begin{figure}[t]
\judgementHead{Type Synthesis}{\tsyn{\tcenv}{\expr}{\type}}
\begin{mathpar}
\inferrule
  {
    \tcenvget{\dcon} = \type
  }
  {
    \tsyn{\tcenv}{\dcon}{\type}
  }
  {\ruleName{Syn-Data}}
\end{mathpar}
\caption{$\langfive$: Rules for Type Synthesis}
\label{fig:five:syn}
\end{figure}

\section{Verification Conditions} \label{sec:five:algo}

The declarative rules for checking destructor types and synthesizing 
constructor types readily translate to an algorithm for verification 
condition generation.

\mypara{Subtyping}
The subtyping constraint for two polymorphic 
datatypes types follows the rule \rulename{Sub-Data}.
As shown in \cref{fig:five:algo:subtyping}, 
the constraint is the conjuction of the 
constraint generated by the top-level 
refinements of the data types ($\cstr_o$), 
the positive ($\cstr_p$) and negative ($\cstr_n$) 
type components. 

\begin{figure}[t!]
$$\begin{array}{lcl}
\toprule
\subsym & : & (\Type \times \Type) \rightarrow \Cstr \\
\midrule
\sub{\refb{\tcapp{\tcon}{\params{\typeb}}}{\reft{\vvar_1}{\pred_1}}}{\refb{\tcapp{\tcon}{\params{\type }}}{\reft{\vvar_2}{\pred_2}}}
  & \doteq
  & \csand{\csand{\cstr_o}{\cstr_p}}{\cstr_n} \\
\quad \mbox{where} & & \\
  \quad \quad \params{\polar} & = & \polarities{\delta}{\tcon} \\
  \quad \quad \cstr_o
  & = & \csimp{\vvar_1}{\base}{\pred_1}{\SUBST{\pred_2}{\vvar_2}{\vvar_1}}\\
\quad \quad \cstr_p
  & =
  & \bigwedge_{\forall i. \polar_i \in \{\ppos, \pboth \}} \sub{\typeb_i}{\type_i} \\
  \quad \quad \cstr_n
  & =
  & \bigwedge_{\forall i. \polar_i \in \{\pneg, \pboth \}} \sub{\type_i}{\typeb_i} \\[0.05in]
\bottomrule
\end{array}$$
\caption{Subtyping Constraints for $\langfive$, extends cases of~\cref{fig:four:algo}}
\label{fig:five:algo:subtyping}
\end{figure}

\mypara{Checking}
Recall that the $\chksym$ function takes an environment 
$\tcenv$, term $\expr$ and type $\type$ and returns a Horn 
constraint $\cstr$ whose satisfiability implies $\tchk{\tcenv}{\expr}{\type}$ 
(\ref{prop:chk}).
Figure~\ref{fig:five:algo} shows how
we extend the function to account for 
destructor (pattern-match) terms 
$\ecase{\evarb}{\params{\alt}}$ by invoking
$\chkalt{\tcenv}{\evarb}{\alt}{\type}$ which implements 
rule \ruleName{Chk-Swt} by computing the conjunction of 
the constraints for \emph{each} alternative. 
Thus, the hard work is done by
$\chkalt{\tcenv}{\evarb}{\dcalt{\dcon}{\evarc}{\expr}}{\type}$
which implements rule \ruleName{Chk-Alt}.
To do so, $\chkaltsym$ first obtains  the data 
constructor's type $\typeb$ at this instance.
Next, it uses $\unapply{\tcenv}{\evarb}{\params{\evarc}}{\typeb}$ 
to obtain the environment $\tcenv'$ extended with types 
for the pattern binders.
Finally, it recursively invokes $\chk{\tcenv'}{\expr}{\type}$
to recursively compute the constraint for the pattern-expression 
under the extended environment.

\begin{figure}[t!]
$$\begin{array}{lcl}
\toprule
\chksym & : & (\tcenv \times \Expr \times \Type) \rightarrow \Cstr \\
\midrule
\chk{\tcenv}{\ecase{\evarb}{\params{\alt}}}{\type}
    & \doteq & \bigwedge_{\alt \in \params{\alt}}\ \chkalt{\tcenv}{\evarb}{\alt}{\type} \\[0.05in]

\toprule
\chkaltsym & : & (\tcenv \times \Evar \times \Alt \times \Type) \rightarrow \Cstr \\
\midrule
\chkalt{\tcenv}{\evarb}{\dcalt{\dcon}{\evarc}{\expr}}{\type}
    & \doteq &  \cswiths{\evarc}{\typeb}{\cswith{\evarb}{\typeb'}{\cstr}} \\
\quad \mbox{where}  &   &                                        \\
\quad \quad \cstr
    & =      & \chk{\tcenv'}{\expr}{\type} \\
\quad \quad \tcenvextv{\params{\tb{\evarc}{\typeb}}; \tb{\evarb}{\typeb'}}\ \mbox{as}\ \tcenv'
    & =      & \unapply{\tcenv}{\evarb}{\params{\evarc}}{\typeb} \\
\quad \quad \typeb  
    & =      & \dconty{\tcenv}{\evarb}{\dcon}                    \\[0.05in]

\toprule
\synsym & : & (\tcenv \times \Expr) \rightarrow (\Cstr \times \Type) \\
\midrule
\syn{\tcenv}{\dcon} 
    & \doteq & (\ttrue,\ \tcenvget{\dcon} = \type) \\[0.05in]
\bottomrule
\end{array}$$
\caption{Verification Condition Generation for $\langfive$, extends cases of~\cref{fig:four:algo}}
\label{fig:five:algo}
\end{figure}

\mypara{Synthesis}
The function $\synsym$, shown in \cref{fig:five:algo}, 
implements rule \ruleName{Syn-Data} by returning the 
environment's (polymorphic) type for the constructor 
$\dcon$.
This lets us then handle constructor applications just 
like plain function applications. 

\begin{figure}[t]
\begin{align*}
\cstr \ \doteq\
  & \forall \mw{x},\ \mw{ys}. \\
  & \quad \forall \mw{y},\ \mw{ys'}.\ \tlen{\mw{ys}} = 1 + \tlen{\mw{ys'}} \ \Rightarrow \\ 
  & \quad \quad \mw{x} \leq \mw{y} \ \Rightarrow \cstr_\mw{y} \ \wedge \\ 
  & \quad \quad \quad \forall \mw{tl}.\ \tlen{\mw{tl}} = 1 + \tlen{\mw{ys'}} \ \Rightarrow\ \cstr_\mw{x} \ \wedge \\
  & \quad \quad \quad \quad \forall \vvar.\ \tlen{\vvar} = 1 + \tlen{\mw{tl}} \Rightarrow \tlen{\vvar} = 1 + \tlen{\mw{ys}} \\
  & \quad \wedge \mw{x} \not \leq \mw{y} \ \Rightarrow\ \ldots \mbox{(constraint for \kw{else} branch)} \\ 
  & \wedge \ldots \mbox{(constraint for \mcode{ON} case)}\\[0.05in]
\cstr_\mw{y} \ \doteq\ 
  & \forall{\vvar}.\ \vvar = \mw{y}    \ \Rightarrow\ \kva{\kvar_\mw{y}}{\vvar} \ \wedge \\
  & \forall{\vvar}.\ \mw{y} \leq \vvar \ \Rightarrow\ (\kva{\kvar_\mw{y}}{\vvar} \wedge \mw{y} \leq \vvar) \\[0.05in]
\cstr_\mw{x} \ \doteq\ 
  & \forall{\vvar}.\ \vvar = \mw{x}    \ \Rightarrow\ \kva{\kvar_\mw{x}}{\vvar} \ \wedge \\
  & \forall{\vvar}.\ \kva{\kvar_\mw{y}}{\vvar} \ \Rightarrow\ (\kva{\kvar_{x}}{\vvar} \wedge \mw{x} \leq \vvar)
\end{align*}
\caption{Verification Condition for definition of \mcode{insert} (\cref{fig:insert}).}
\label{fig:insert:vc}
\end{figure}

\begin{myexample}{VC for \mcode{insert}}
Figure~\ref{fig:insert:vc} shows a part 
of the VC generated by invoking \chksym on 
the environment $\tcenv$, the definition of 
\mcode{insert} from \cref{fig:insert:elab} 
-- with the type applications replaced
with holes $\refb{\tvar}{\ureft}$ -- 
and the goal type $\type_{\mcode{ins}}$.
(The abbreviations $\tcenv$ and $\type_{\mcode{ins}}$ are 
shown in \cref{fig:insert:defs}.)
The VC generation recurses into the body,  
to generate two constraints for the ``cons'' 
and ``nil'' cases respectively. 
The latter is elided for brevity.
The former adds the pattern match binders 
for \mw{y} and \mw{ys'} and adds the 
hypothesis that the size of \mw{ys} 
is one more than that of \mw{ys'}.
We then recurse into the \kw{if}-expression: 
with one sub-constraint for the \kw{then} 
branch (with the hypothesis $\mw{x} \leq \mw{y}$) 
and a second for the \kw{else} branch,
elided for brevity.
Recall that the \kw{then}-expression is
\[ \elet{\mw{tl}}{\expr_\mw{y}}{\expr_\mw{x}}  \]
Where $\expr_\mw{y}$ and $\expr_\mw{x}$ are the two 
constructor applications in \cref{fig:insert:defs}.
The \kw{then}-constraint has a sub-constraint $\cstr_\mw{y}$ 
that arises from calling $\synsym$ on $\expr_\mw{y}$ and which 
also produces the hypothesis that the size of \mw{tl} is 
one more than that of \mw{ys'}.
Similarly, we then again invoke $\synsym$ on $\expr_\mw{x}$ 
to get $\cstr_\mw{x}$ and the result type: a list $\vvar$ whose 
length is one more than $\mw{tl}$ which must then imply the 
post-condition, that the output $\vvar$'s size is in fact, 
one greater than the input \mw{ys}.

Next, lets look at the constraint $\cstr_\mw{y}$ 
shown in \cref{fig:insert:vc}, returned by $\synsym$ 
on the constructor application term
\[
  \expr_{\mw{y}} \ \doteq\ \eapp{\eapp{\tapp{\mcode{OC}}{\refb{\tvar}{\ureft} }}{\mw{y}}}{\mw{ys'}}
\]
As described in \cref{sec:four:algo}, we generate 
a fresh template $\refb{\tvar}{\reft{\vvar}{\kva{\kvar_\mw{y}}{\vvar}}}$ 
for the hole, which means that at this site, the 
constructor \mcode{OC} has the following type 
(analogous to that shown in (\ref{eq:ins:cons:y}), 
 except with $\kva{\kvar_\mw{y}}{\vvar}$ instead of 
 the to-be-inferred refinement $\mw{x} \leq \vvar$:)
\[
  \trfun{\mw{z}}{\refb{\tvar}{\reft{\vvar}{\kva{\kvar_{\mw{y}}}{\vvar}}}}
    {
      \trfun{\mw{zs}}{\olist{\refb{\tvar}{\reft{\vvar}{\kva{\kvar_{\mw{y}}}{\vvar} \wedge \mw{z} \leq \vvar}}}}
        {
          \olistp{{\refb{\tvar}{\reft{\vvar}{\kva{\kvar_{\mw{y}}}{\vvar}}}}}{\mw{zs}}
        }
    } 
\]
Thus, $\cstr_\mw{y}$ has two conjuncts, 
one each for the subtyping obligations 
for the inputs $\mw{y}$, $\mw{ys'}$. 
The first conjunct ensures that the 
first argument $\mw{y}$ is indeed a 
subtype of the constructor's first input.
The second conjunct checks that the 
second input $\mw{ys'}$ whose type, 
per $\unapplysym$ is $\olist{\refb{\tvar}{\reft{\vvar}{\mw{y} \leq \vvar}}}$ 
(as shown in $\tcenv'_\leq$ in \cref{fig:insert:defs})
is a subtype of the second input of the constructor.

Finally, lets consider the constraint $\cstr_\mw{x}$ 
shown in \cref{fig:insert:vc} that is returned by 
$\synsym$ on the constructor application term
\[
  \expr_{\mw{x}} \ \doteq\ \eapp{\eapp{\tapp{\mcode{OC}}{ \refb{\tvar}{\ureft} }}{\mw{x}}}{\mw{tl}}
\]
Again, we generate a fresh template for the hole 
$\refb{\tvar}{\reft{\vvar}{\kva{\kvar_\mw{x}}{\vvar}}}$
which means that at this site, the constructor \mcode{OC} 
has the type
\[
  \trfun{\mw{z}}{\refb{\tvar}{\reft{\vvar}{\kva{\kvar_{\mw{x}}}{\vvar}}}}
    {
      \trfun{\mw{zs}}{\olist{\refb{\tvar}{\reft{\vvar}{\kva{\kvar_{\mw{x}}}{\vvar} \wedge \mw{z} \leq \vvar}}}}
        {
          \olistp{{\refb{\tvar}{\reft{\vvar}{\kva{\kvar_{\mw{x}}}{\vvar}}}}}{\mw{zs}}
        }
    } 
\]
Hence, the constructor application yields the 
constraint $\cstr_{\mw{x}}$ with one conjunct 
for each of the input arguments \mw{x} and \mw{tl}.
As the output type of the \emph{first} 
constructor application, and so the 
type of \mw{tl} is 
${\olistp{{\refb{\tvar}{\reft{\vvar}{\kva{\kvar_{\mw{y}}}{\vvar}}}}}{\mw{ys'}}}$
the second conjunct of $\cstr_{\mw{x}}$ -- which 
represents the subtyping obligation for the 
input argument \mw{tl} -- has the antecedent 
$\kva{\kvar_{\mw{y}}}{\vvar}$.

\mypara{Solution} We encourage the reader to verify that the assignment 
\begin{align*}
  \kva{\kvar_{\mw{y}}}{\vvar} & \ \doteq\ \mw{x} \leq \vvar \\
  \kva{\kvar_{\mw{x}}}{\vvar} & \ \doteq\ \ttrue 
\end{align*}
satisfies the Horn constraint in \cref{fig:insert:vc}, by yielding, 
after substitution, a valid VC, thereby verifying that \mcode{insert} 
implements $\type_{\mcode{ins}}$.
\end{myexample}

\section{Discussion}

To sum up, in this chapter we saw how to extend the language 
to support refinements on user defined algebraic data-types.
We saw how to encode three classes of properties -- invariants 
of individual datum, properties relating multiple data, and 
properties of the structure itself -- using a single mechanism: 
a refined type for the data constructors.
This approach lets us reuse the rules for function \emph{application} 
to \emph{generalize} properties when \emph{constructing} the structure.
Dually, it lets us define a notion of \emph{un-application} to let 
us \emph{instantiate} properties when \emph{destructing} the structure.

\mypara{Refinements on Data vs. Type Constructors}
Our approach is using refinements of data constructors. 
An alternative approach would be to refine type constructors.
For example, the ordered lists of \cref{subsec:ordered-lists}
could be alternatively defined as @list int {l: isOrdered l}@, 
where @isOrdered@ is a boolean function that checks is the input 
list is ordered.
\cite{Igarashi15} compare the two alternative approaches and conclude 
that refinements of type constructors are easier for the programmer to 
\emph{specify}, but refinements on data constructors are much easier to 
(automatically) \emph{verify}.

\mypara{Types as a Decision Procedure}
The ease of verification arises from the fact that 
refined data constructors give rise to a simple 
syntax-directed approach to establish invariants 
of complex heap-allocated data structures.
Crucially, this approach does not require the 
expensive blur and materialize mechanisms of 
shape analysis~\citep{SagivRepsWilhelm02j}, 
or the undecidable (universal) quantifier-based 
reasoning used by Floyd-Hoare logic based 
deductive verifiers like \textsc{Dafny} \citep{dafny}.
Instead, \emph{subtyping} provides a syntactic 
proof system for checking entailment between 
quantfied assertions. 
Subtyping lets us prove the (quantified) assertion 
that if every element of a list is positive then 
every element is non-negative, by proving the 
(quantifier-free) assertion that if $\vvar$ 
is positive then $\vvar$ is non-negative. 
Similarly, the constructor application 
is a syntactic heuristic for generalizing 
facts about individual elements into facts 
about the whole structures: \eg if @x@ is 
a @nat@ and @xs@ is a @list(nat)@, then 
@Cons(x, xs)@ is a @list(nat)@. 
Finally, the destructor sites provide 
a syntactic heuristic for instantiating 
quantified facts about the whole structure:
\eg if @xs@ is a @list(nat)@ and 
is destructed to @Cons(x, xs')@
then \mcode{x} is a @nat@.

\mypara{Refinements for Pointer-based Structures}
While this chapter develops this ideas for a purely functional 
language, related ideas have been proposed for imperative languages. 
Notable examples include \citep{Madhu2013} which introduces the idea of 
``Natural Proofs'' which shows how separation-logic based verifiers 
need only fold and unfold recursive predicates (which encode the same 
information as our recursive types) at the equivalent of constructor 
application and destruction sites, and \citep{Bakst16} which shows 
how to add refinements to an alias type system \citep{AliasTypesRec} 
that precisely tracks locations and aliasing, yielding an expressive 
and automated way to reason about invariants of pointer-based data 
structures.

\newcommand\intk{\refb{\tint}{\reft{\vvar}{\kva{\kvar}{\vvar}}}}
\newcommand\kvarbr{\kvar_{\mcode{br}}}
\newcommand\intkbr{\refb{\tint}{\reft{\vvar}{\kva{\kvarbr}{\vvar}}}}
\newcommand\tinte{\mcode{int8}}
\newcommand\inte{\refb{\tint}{\reft{\vvar}{0 \leq \vvar < 256}}}
\newcommand\intfk{\refb{\tint}{\reft{\vvar}{\uf{\kvar}(\vvar)}}}

\chapter{Refinement Polymorphism}
\label{sec:lang:six}

Modern programming languages allow code and data to 
abstract over the concrete types that
they work with. For example, it is commonplace to 
define polymorphic arrays or hash-tables that can, 
\eg hold @int@ or @string@ or other values,
and to write type-agnostic functions that operate 
over these structures. 
In \S~\ref{sec:lang:four} we saw how polymorphic types 
could be instantiated at different use-sites
to precisely track invariants in a context-sensitive
manner.

However, we also often write monomorphic 
functions and data that nevertheless abstract 
over the refinements satisfied by the 
data they manipulate. 
This abstraction is often implicit, 
\ie not captured in the function's 
type specification, which makes it 
hard to establish the invariants 
at usage sites. 
Hence, next, lets develop a means 
for specifying signatures that 
abstract over the refinements, 
see how to automatically verify 
these signatures, and learn how 
to instantiate them automatically 
at client sites, all while keeping 
verification efficiently decidable.

\section{Examples} \label{sec:six:examples}

Lets crack our knuckles with some illustrative examples.

\subsection{Problem: Picking the Right Specification}

Consider @maxI(x,y)@ which returns the larger of @x@ and @y@. 
\begin{code}
  val maxI : int => int => int
  let maxI = (x, y) => {
    if (x < y) { y } else { x }
  };
\end{code}
Suppose that we want to verify that @maxI(a, b)@
must be non-negative if @a@ and @b@ are non-negative.
\begin{code}
  val bigger: nat => nat => nat 
  let bigger = (a, b) => { maxI(a, b) };
\end{code}
Unfortunately, the above @bigger@ will \emph{fail} 
as the specified type is silent about what properties 
hold of the @int@ returned by @maxI@! 

\mypara{Premature Specialization}
We could try to type @maxI@ as 
\begin{code}
  val maxI : nat => nat => nat
\end{code}
This signature would let us verify @bigger@ 
but would prematurely restrict @maxI@ to 
non-negative values. 
For example, if we had a refinement for 
valid 8-bit @int@s
\begin{code}
  type int8 = int[v|0 <= v && v < 256]
\end{code}
we would end up \emph{rejecting} the perfectly correct
\begin{code}
  val brighter : int8 => int8 => int8
  let brighter = (c1, c2) => {
    maxI(c1, c2)
  }
\end{code}
Instead, we might specify that the output 
\begin{code}
  val maxI: x:int => y:int => int[v|v=x || v=y] 
\end{code}
This would let us verify @bigger@ and @brighter@. 
This approach suffices when there are just two
@int@ inputs but does not scale up to unbounded 
collections, as in @maxL@ which computes the 
largest element of a \emph{list}
\begin{code}
  let rec maxL = (default, xs) => {
    fold_right(maxI, default, xs)
  };
\end{code}
Since the list is unbounded we have no way to say
the output is \emph{one of} the elements of @xs@. 
We could use an \emph{existential quantifier} but 
that would lead to verification conditions that 
are outside the boundaries of decidable SMT based 
validity checking.
%
%
Now suppose we use @maxL@ with different kinds of values, 
for example:
\begin{code}
  val biggerL : list(nat) => nat
  let biggerL = (xs) => { maxL(0, xs) }

  val brighterL : list(int8) => int8
  let brighterL = (xs) => { maxL(0, xs) }
\end{code}
How can we verify that @maxList@ returns 
a @nat@ or @int8@ values when invoked with 
lists of @nat@ and @int8@ values respectively?

\subsection{Solution: Abstracting Over Refinements}

If we take a step back, we might notice that the fact 
that @maxI@ returns \emph{one of} its two inputs @x@ 
and @y@ can be rephrased as follows: if there is some 
property $p$ that \emph{both} inputs satisfy, then the 
output must also satisfy $p$. 
That is, we can make the refinement $p$ itself be 
a \emph{parameter} in the specification of @maxI@
\begin{code}
  val maxI: int[v|p v] => int[v|p v] => int[v|p v] 
\end{code}
The above specification says that \emph{for any} 
predicate @p@ over @int@ values, @maxI@ takes two 
inputs @x@ and @y@ which respectively satisfy @p x@ 
and @p y@ and returns an output @v@ that satisfies @p v@.

It is only useful to parameterize
over refinements if there is some convenient 
way to \emph{instantiate} the parameters.
We will extend the machinery for 
inferring refinements for holes $\ureft$ to 
automatically instantiate the refinement 
parameters at the usage-sites --- \eg in @bigger@ 
and @brighter@ --- where @p@ will be 
instantiated to the concrete refinements 
\begin{align}
  p_{\mcode{nat}}  & \doteq \ \lambda \vvar.\ 0 \leq \vvar                     \label{eq:p:nat} \\
  p_{\mcode{int8}} & \doteq \ \lambda \vvar.\ 0 \leq \vvar \wedge \vvar < 256  \label{eq:p:int8}
\end{align}
thereby verifying those two client functions.

The above method scales up to unbounded lists.
We can specify 
\begin{code}
 val maxL: int[v|p v] => list(int[v|p v]) => int[v|p v]
\end{code} 
which says that when @maxL@ is given a \emph{default} @int@ 
and a @list(int)@ all of which satisfy the predicate @p@, 
the output is also an @int@ that satisfies @p@.
Once again, we can verify the clients @biggerL@ and 
@brighterL@ by instantiating @p@ as $p_{\mcode{nat}}$ and 
$p_{\mcode{int8}}$ at the respective call-sites. 

\mypara{Preserving Decidability}
At first glance, it may appear that these abstract 
predicate variables @p@ have taken us into the realm 
of higher-order logics, and that we must leave decidable 
SMT based checking at the door. 
Fortunately, that is not the case. 
We will see how to 
encode abstract refinements @p@ as \emph{uninterpreted 
functions} in the refinement logic, which will allow us 
to continue with SMT based Horn VC checking. 

\subsection{Abstracting Refinements over Data Types}

Refinement parameters are a natural fit
for \emph{datatype} definitions.

\mypara{Dependent Pairs}
For example, we can specify a @pair@ datatype where 
there is some relationship between the first and 
second component, by parameterizing the datatype 
definition with a refinement parameter @p@:
\begin{code}
  type pair('a, 'b)(p : 'a => 'b => bool) =
    | MkPair(x:'a, y:'b[v|p x v])
\end{code}
The definition is parameterized by a binary 
predicate @p@ that relates the two elements
of the @pair@.
Hence, we can define the set of @int@ pairs 
where the second component exceeds the first 
as:
\begin{code}
  type incPair = pair(int,int)((x, y) => x < y)
\end{code}
The refinement parameter is \emph{asserted}
when the pair is \emph{constructed}. 
Hence, @okPair@ will verify but @badPair@
will be rejected:
\begin{code}
  val okPair: n:int => incPair
  let okPair = (n) => { MkPair(n, n+1) };
  
  val badPair: x:int => incPair
  let badPair = (n) => { MkPair(n, n-1) };
\end{code}
Dually, the refinement is \emph{assumed} 
when the pair is \emph{destructed}, and so 
the below verifies:
\begin{code}
  val chkPair : incPair => nat
  let chkPair = (p) => { 
    switch (p) {
      MkPair(x1, x2) => x2 - x1
    }
  };
\end{code}

\mypara{Abstracting Refinements over Lists}
The combination of recursion and refinement
parameters allows us to compactly specify 
various interesting properties of collections 
without baking them into the datatype's
definition.
Recall the definition of ordered lists 
of non-decreasing values from \cref{sec:five:examples}
\begin{code}
  type olist('a) =
    | ONil
    | OCons (x:'a, xs:olist('a[v|x<=v]))
\end{code}
Order is established by the signature for @OCons@
that requires that every element of the tail \mcode{xs} 
is greater than the head \mcode{x}.
Refinement parameters let us abstract the particular 
relation: we can specify a generic @list@ as
\begin{code}
  type list('a)(p:'a =>'a => bool) =
    | Nil
    | Cons(x:'a, xs:list('a[v|p x v])((y,z) => p y z))
\end{code}
The definition is parameterized by a binary predicate @p@
that relates \emph{every} element of the tail with the head 
@x@ at each application of the constructor @Cons@.
That is, the specification says that if 
\[
  \tcons(x_1, \tcons(x_2, \ldots \tcons(x_n, \tnil))) : \tlist{\tvar}(p) 
\]
then for each $1 \leq i < j \leq n$ we have $p(x_i, x_j)$.

\mypara{Ordered Lists}
We can now specify non-decreasing lists by instantiating the 
refinement parameter @p@ appropriately:
\begin{code}
  type incList('a) = list('a)((x1, x2) => x1<=x2)
  
  val checkInc : (int) => incList(int)
  let checkInc = (x) => { 
    Cons(x, Cons(x+1, Cons(x+2, Nil)))
  };
\end{code}
Similarly, we can define non-increasing lists 
\begin{code}
  type decList('a)  = list('a)((x1,x2) => x1>=x2)
  
  val checkDec : (int) => decList(int)
  let checkDec = (x) => { 
    Cons(x+2, Cons(x+1, Cons(x, Nil)))
  };
\end{code}
or \emph{duplicate-free} lists, simply by changing the relation:
\begin{code}
  type uniqList('a) = list('a)((x1,x2) => x1!=x2)

  val checkUnique : (int) => uniqList(int)
  let checkUnique = (x) => { 
    Cons(x+2, Cons(x, Cons(x+1, Nil)))
  };
\end{code}

We can omit the refinement instantiation to write just 
@list('a)@ to denote the trivially (un)refined instance
@list('a)((x1,x2) => true)@.
Now, the machinery developed for reasoning about datatypes 
in \cref{sec:five:examples} lets us verify properties
of code manipulating (abstractly) refined lists,
\eg that the following insertion-sort produces 
ordered lists 
\begin{code}
  val isort : list('a) => incList('a)
  let isort = (xs) => { 
    foldr(insert, Nil, xs);
  };
\end{code}

\section{Types and Terms} \label{sec:six:types}

Figure~\ref{fig:six:syntax} summarizes how we extend 
the language of {types} to include abstraction  
over refinements, and the language of {terms} to 
include instantiation of refinement variables.

\mypara{Refinement Variables} We write $\rvar$ 
to denote refinement variables that abstract 
over concrete or specific refinements. 
Refinement variables $\rvar$ are of the 
form $\tb{\kvar}{\tfun{\params{\base}}{\tbool}}$, 
\ie represent a $\tbool$ valued \emph{predicates} 
over some (non-empty) sequence of base types.

\mypara{Abstracting over Refinements} We abstract 
over refinements either in function signatures of 
the form $\rpoly{\rvar}{\type}$ or in data type 
definitions $\trdef{\tcon}{\dtydef{\tvar}{\kind}{\polar}}{\drefdef{\rvar}{\polar}}{\ddefs{\dcon}{\type}}$ 
which are now parameterized by a (possibly empty) 
sequence of refinement variables and their respective polarities 
in addition to 
type variables from Figure~\ref{fig:five:syntax}.
For example, we might assign @maxI@ the type
\begin{equation}
\type_{\mcode{maxI}} \ \doteq\ \rpoly{\tb{\kvar}{\tfun{\tint}{\tbool}}}{\tfun{\intk}{\tfun{\intk}{\intk}}}
\label{eq:maxI:type}
\end{equation}
which captures the intuition that when given two $\intk$ inputs, 
the function is guaranteed to return an $\intk$ output, for any 
refinement $\kvar$ on $\tint$ values. In the sequel, we will 
elide the refinement variables' sort when it is clear from 
the context.

\mypara{Implicit Refinement Application} We instantiate refinements 
either in \emph{refinement application} terms of the 
form $\rapp{\expr}$ or in \emph{type-constructor application} 
types of the form $\trcapp{\tcon}{\params{\type}}{\params{\aref}}$,
where $\aref$ is a \emph{concrete refinement} $\ardef{x}{\base}{\pred}$
which is a boolean valued predicate over a set of variables $\params{\tb{x}{\base}}$.
We leave the refinement instances \emph{implicit} 
for terms (denoted by $\star$) as, like type instances, 
these are ubiquitous, and hence, should be automatically 
synthesized.
For example, @brighter@ is defined via the following 
term in which the refinement variable in the signature 
of @maxI@ is implicitly instantiated at the usage site: 
\begin{equation}
\kw{let}\ \mcode{brighter}\ =\ \elam{x, y}{\ \eapp{\eapp{\rapp{\mcode{maxI}}}{x}}{y}}
\label{eq:brighter}
\end{equation}

\mypara{Explicit Refinement Application}
However, we allow \emph{explicit} refinement instances
for data types as we often want to specify signatures
where the constructor is refined with a particular
concrete refinement. For example, the definitions 
of @incPair@, @incList@, and @decList@ from 
\cref{sec:six:examples} are, respectively:
\begin{align}
  \kw{type}\ \mcode{incPair} & = \tpair{\tint}{\tint}(\elam{x_1, x_2}{x_1 < x_2}) \label{eq:incPair} \\
  \kw{type}\ \mcode{incList} & = \tlist{\tint}(\elam{x_1, x_2}{x_1 \leq x_2}) \label{eq:incList} \\
  \kw{type}\ \mcode{decList} & = \tlist{\tint}(\elam{x_1, x_2}{x_1 \geq x_2}) \label{eq:decList}
\end{align}

\begin{figure}[t!]
\begin{tabular}{rrcll}
\emphbf{Ref. Var.}
  & \rvar & $\bnfdef$ & \tb{\kvar}{\tfun{\params{\base}}{\tbool}} & \\[0.05in]

\emphbf{Abs. Refine.}
  & \aref & $\bnfdef$ & \ardef{x}{\base}{\pred}                          & \\[0.05in]

\emphbf{Datatypes}
  & $\delta$ & $\bnfdef$ & \trdef{\tcon}{\dtydef{\tvar}{\kind}{\polar}}{\drefdef{\rvar}{\polar}}{\ddefs{\dcon}{\type}}  &  \\[0.05in]

\emphbf{Basic Types}
  & \base & $\bnfdef$  & $\ldots$                                        & \emph{from Fig.~\ref{fig:five:syntax}} \\
  &       & $\spmid$   & \trcapp{\tcon}{\params{\type}}{\params{\aref}}  & \emph{datatypes} \\ [0.05in]

\emphbf{Types}
  & \type & $\bnfdef$ & \ldots                                    & \emph{from $\langfive$ Fig.~\ref{fig:five:syntax}} \\
  &       & $\spmid$  & \rpoly{\rvar}{\type}                      & \emph{ref. polymorphism} \\[0.05in]

\emphbf{Terms}
  & \expr  & $\bnfdef$ & $\ldots$               & \emph{from $\langfive$ Fig.~\ref{fig:five:syntax}} \\
  &        & $\spmid$    & \rapp{\expr}         & \emph{refinement application}               \\[0.05in]
\end{tabular}
\caption{{$\langfive$: Syntax of Types and Terms}}
\label{fig:six:syntax}
\end{figure}

\section{Declarative Typing} \label{sec:six:decl}

Next, lets look at how the declarative typing rules are extended to 
accomodate abstract refinements.

\subsection{Well-formedness}

\begin{figure}[t!]
\judgementHead{Well-formedness}{\wf{\tcenv}{\type}{\kind}}
\begin{mathpar}
\inferrule
  {\wf{\tcenvext{\kvar}{\tfun{\params{\base}}{\tbool}}}{\type}{\kind}}
  {\wf{\tcenv}{\rpoly{\tb{\kvar}{\tfun{\params{\base}}{\tbool}}}{\type}}{\starkind}}
  {\ruleName{Wf-RAbs}}
\end{mathpar}
\caption{$\langsix$: Rules for Well-formedness}
\label{fig:six:wf}
\end{figure}

Rule \ruleName{Wf-RAbs} shown in \cref{fig:six:wf} 
states that a refinement polymorphic type is well-formed 
with kind $\starkind$, if the underlying (quantified) 
type is also well-formed.

\subsection{Subtyping}

\begin{figure}[t!]
\judgementHead{Abs. Refinement Implication}{\issub{\tcenv}{\aref_1}{\aref_2}}
\begin{mathpar}
\inferrule
  {
    \entl{\tcenvextv{\params{\tb{x_1}{\base}}}}{\pred_1 \Rightarrow \SUBST{\pred_2}{\params{x_2}}{\params{x_1}}}  
  }
  {
    \issub{\tcenv}{\ardef{x_1}{\base}{\pred_1}}{\ardef{x_2}{\base}{\pred_2}}
  }
  {\ruleName{Sub-CRef}}
\end{mathpar}

\judgementHead{Subtyping}{\issub{\tcenv}{\type_1}{\type_2}}
\begin{mathpar}
\inferrule
  { 
    \issub{\tcenv}{\typeb_i}{\type_i} \ \mbox{for each}\ i. \polar_i \in \{\ppos, \pboth \}\\\\
    \issub{\tcenv}{\type_i}{\typeb_i} \ \mbox{for each}\ i. \polar_i \in \{\pneg, \pboth \}\\\\
    \issub{\tcenv}{\arefb_i}{\aref_i} \ \mbox{for each}\ i. \polar_{ri} \in \{\ppos, \pboth \}\\\\
    \issub{\tcenv}{\aref_i}{\arefb_i} \ \mbox{for each}\ i. \polar_{ri} \in \{\pneg, \pboth \}\\\\
    \entl{\tcenvext{\vvar_1}{\reft{\tcapp{\tcon}{\params{\typeb}}}{\pred_1}}}{\SUBST{\pred_2}{\vvar_2}{\vvar_1}}
    \and (\params{\polar},\params{\polar_r}) = \polarities{\delta}{\tcon} \\\
  }
  {
    \issub{\tcenv}
      {  \refb{\trcapp{\tcon}{\params{\typeb}}{\params{\arefb}}}{\reft{\vvar_1}{\pred_1}}  }
      {  \refb{\trcapp{\tcon}{\params{\type }}{\params{\aref }}}{\reft{\vvar_2}{\pred_2 }}  }
  }
  {\ruleName{Sub-Con}}
\end{mathpar}
\caption{$\langsix$: Rules for Subtyping}
\label{fig:six:sub}
\end{figure}

Figure~\ref{fig:six:sub} summarizes the new rules for subtyping.
The rule \ruleName{Sub-CRef} shows the rule for checking 
subsumption between two concrete refinements $\ardef{x_1}{\base}{\pred_1}$ 
and $\ardef{x_2}{\base}{\pred_2}$, by checking that $\pred_1$ is subsumed
by $\pred_2$ after suitably renaming the bound variables.
The second rule \ruleName{Sub-Con} shows how to extend the rule for checking 
subtyping between two refined instances
$\refb{\trcapp{\tcon}{\params{\typeb}}{\params{\arefb}}}{\reft{\vvar_1}{\pred_1}}$
and
$\refb{\trcapp{\tcon}{\params{\type}}{\params{\aref}}}{\reft{\vvar_2}{\pred_2}}$
of a type constructor $\tcon$, by checking the subsumption holds between 
the corresponding concrete refinements $\arefb$ and $\aref$ of the two 
instances according to their polarity using \ruleName{Sub-CRef}, and checking the subsumption of the 
base refinements $\pred_1$ and $\pred_2$ and type components as before.

\begin{myexample}{Subtyping in \mcode{incPair}}
As an example, let 
\begin{align*}
  \tcenv  & \defeq \ \tb{a}{\tint} \\
  \aref   & \defeq \ \elam{x,y}{x = a \wedge y = a + 1} \\
  \arefb  & \defeq \ \elam{x,y}{x < y} \\ 
\intertext{and consider the subtyping obligation}
  \tcenv  & \vdash \ {\tpair{\tint}{\tint}(\aref)}\ \lqsubt\ {\tpair{\tint}{\tint}(\arefb)}\\
\intertext{that could arise in the course of checking that}
  \tcenv & \vdash \ \mcode{MkPair}(a, a+1) \Leftarrow \mcode{incPair} \\
\intertext{where \mcode{incPair} is defined in \ref{eq:incPair}. 
Rule \rulename{Sub-Con} reduces the subtyping obligation 
to the following concrete refinement subsumption}
\tcenv & \vdash \ \aref \ \lqsubt\ \arefb \\
\intertext{and then \rulename{Sub-CRef} reduces the above to the entailment}
\tcenv, \tb{x,y}{\tint} & \vdash \ (x = a \wedge y = a + 1) \ \Rightarrow\ (x < y)
\end{align*}
that is readily validated by the SMT solver.
\end{myexample}

\subsection{Checking}

\begin{figure}[t!]
\judgementHead{Type Checking}{\tchk{\tcenv}{\expr}{\type}}
\begin{mathpar}
\inferrule
  {
    \rvar = \tb{\kvar}{\tfun{\params{\base}}{\tbool}}
    \quad
    \aref = \elam{\params{\evar}}{\uf{\kvar}(\params{\evar})} 
    \\
    \tchk{\tcenvext{\uf{\kvar}}{\tfun{\params{\base}}{\tbool}}}
      {\rvinst{\expr}{\rvar}{\aref}}
      {\rvinst{\typeb}{\rvar}{\aref}} 
  }
  {
    \tchk{\tcenv}{\expr}{\rpoly{\rvar}{\typeb}}
  }
  {\ruleName{Chk-RAbs}}
\end{mathpar}
\caption{$\langsix$: Rules for Type Checking}
\label{fig:six:check}
\end{figure}

The rule \ruleName{Chk-RAbs} checks that a term $\expr$ implements the 
abstractly refined signature $\rpoly{\rvar}{\typeb}$.
The key idea is to use the refinement variable 
$\rvar = \tb{\kvar}{\tfun{\params{\base}}{\tbool}}$ to
(1)~generate a fresh \emph{uninterpreted function symbol} $\uf{\kvar}$, 
(2)~substitute all occurrences of $\kvar$ with $\uf{\kvar}$ 
    in the term $\expr$ and the type $\typeb$, to respectively
    obtain $\SUBST{\expr}{\kvar}{\uf{\kvar}}$ and $\SUBST{\typeb}{\kvar}{\uf{\kvar}}$
    and then
(3)~perform the check on the substituted type and term, in an environment 
    extended with a binding for the uninterpreted function $\uf{\kvar}$.

\mypara{Refinement Instantiation}
We replace all occurences of the refinement variable $\rvar \doteq \tb{\kvar}{\cdot}$
with an uninterpreted function $\uf{\kvar}$ via an operation $\rvinst{\typeb}{\rvar}{\aref}$
shown in \cref{fig:refinst} which \emph{instantiates} (or \emph{substitutes}) 
a concrete refinement $\aref$ for a refinement variable $\rvar$ 
in a signature $\typeb$. 
The operation traverses $\typeb$ to replace 
all occurences of $\kvapp{\kvar}{\evar}$ with $\SUBST{\pred}{\params{\evarb}}{\params{\evar}}$ 
when $\rvar \doteq \tb{\kvar}{\cdot}$ and $\aref \doteq \elam{\params{\evarb}}{\pred}$,
\ie we replace the \emph{parameters} of the refinement ($\params{\evarb}$) 
with the \emph{arguments} ($\params{\evar}$) in the concrete refinement $\pred$.

\begin{myexample}{Checking \mcode{maxI}}
Recall the implementation and specification of @maxI@ (\S~\ref{sec:six:examples}) 
\begin{align}
  \expr_{\mcode{maxI}} & \doteq\ \elam{x, y}{\eif{x < y}{y}{x}} 
  \label{eq:maxI:impl} \\
  \type_{\mcode{maxI}} & \doteq\ \rpoly{\kvar}{\tfun{\intk}{\tfun{\intk}{\intk}}} 
  \label{eq:maxI:spec}
\end{align}
Lets consider the goal of verifying that the above implementation 
checks against its specification (\ref{eq:maxI:type})
\begin{equation}
  \emptyset \vdash\ \expr_{\mcode{maxI}} \Leftarrow \type_{\mcode{maxI}} \\
  \label{eq:maxI:check}
\end{equation}
The rule \rulename{Chk-RAbs} reduces the above to
\begin{align*}
  \emptyset & \vdash\ \expr_{\mcode{maxI}} \Leftarrow {\tfun{\intfk}{\tfun{\intfk}{\intfk}}} \\
\intertext{which rules \rulename{Chk-Lam}, \rulename{Chk-If} and \rulename{Syn-Var} 
  reduce to two goals}
  \tcenv & \vdash\ \refb{\tint}{\reft{\vvar}{\vvar = x}} \ \lqsubt\ \intfk \\ 
  \tcenv & \vdash\ \refb{\tint}{\reft{\vvar}{\vvar = y}} \ \lqsubt\ \intfk \\ 
\intertext{where $\tcenv$ has bindings for the refinement and value parameters}
  \tcenv & \doteq \ \tb{\uf{\kvar}}{\tfun{\tint}{\tbool}}, \tb{x}{\intfk}, \tb{y}{\intfk} \\
\intertext{The two goals above check that the specified output type 
is indeed returned in each branch. Both the above reduce to the entailment}
  \tcenv, \tb{\vvar}{\tint} & \vdash \ \vvar = y\ \Rightarrow \ \uf{\kvar}(\vvar) 
\end{align*}
that is easily confirmed by the SMT solver.
\end{myexample}

\begin{figure}[t]
$$\begin{array}{lcl}
\toprule
\dcontysym & : & (\tcenv \times \dcon \times \Evar) \rightarrow \Type \\
\midrule
\dconty{\tcenv}{\dcon}{\evarb}
  & \doteq & \rvinst{\SUBST{\typeb}{\params{\tvar}}{\params{\type}}}{\params{\rvar}}{\params{\aref}}  \\[0.05in]
\quad \mbox{where} & & \\
\quad \quad \trcapp{\tcon}{\params{\type}}{\params{\aref}}         & = & \tcenvget{\evarb} \\
\quad \quad \tpoly{\params{\tvar}}{\rpoly{\params{\rvar}}{\typeb}} & = & \tcenvget{\dcon} \\[0.05in]
\bottomrule
\end{array}$$
\caption{Meta-functions for Checking Alternatives; $\unapplysym$ as in Figure~\ref{fig:unapply}}
\label{fig:drconty}
\end{figure}

\begin{figure}[t]
$$\begin{array}{rrcl}
\emphbf{\mbox{Predicates}}
  & \rvinst{\kvapp{\kvar}{\evar}}{\rvar}{\aref}
      & \doteq & \SUBST{\pred}{\params{\evarb}}{\params{\evar}}
                  \ \mbox{if} \ \rvar = \tb{\kvar}{\cdot},
                              \ \aref = \elam{\params{\evarb}}{\pred}  \\
  & \rvinst{\pred_1 \bowtie \pred_2}{\rvar}{\aref}
      & \doteq & \rvinst{\pred_1}{\rvar}{\aref} \bowtie \rvinst{\pred_2}{\rvar}{\aref} \\
  & \rvinst{\pred}{\rvar}{\aref}
     & \doteq & p \\[0.10in]
\emphbf{Abstr. Refs}
  & \rvinst{(\ardef{x}{\base}{\pred})}{\rvar}{\aref}
     & \doteq & \ardef{x}{\base}{(\rvinst{\pred}{\rvar}{\aref})} \\[0.10in]
\emphbf{Base Types}
  & \rvinst{\trcapp{\tcon}{\params{\type}}{\params{\arefb}}}{\rvar}{\aref}
     & \doteq & \trcapp{\tcon}{\params{\rvinst{\type}{\rvar}{\aref}}}{\params{\rvinst{\arefb}{\rvar}{\aref}}} \\
  & \rvinst{\base}{\rvar}{\aref}
     & \doteq & \base \\[0.10in]
\emphbf{Types}
  & \rvinst{\refb{\base}{\reft{\vvar}{\pred}}}{\rvar}{\aref}
     & \doteq & \refb{\rvinst{\base}{\rvar}{\aref}}{  \reft{\vvar}{\rvinst{\pred}{\rvar}{\aref}} } \\
  & \rvinst{(\trfun{x}{\typeb}{\type})}{\rvar}{\aref}
     & \doteq & \trfun{x}{ \rvinst{\typeb}{\rvar}{\aref}}{\rvinst{\type}{\rvar}{\aref}} \\[0.10in]
\end{array}$$
\caption{Instantiating a refinement variable $\rvar$ with a concrete refinement $\aref$.}
\label{fig:refinst}
\end{figure}

\mypara{Checking Case Alternatives}
We continue to check case alternatives using 
\rulename{Chk-Alt} from \cref{fig:five:check}. 
However, we must extend the definition of 
$\dconty{\tcenv}{\dcon}{\evarb}$ --- which 
defines the monomorphic instantiation of 
the polymorphic type of the data constructor 
$\dcon$ corresponding the case-scrutinee $\evarb$ --- 
to also account for refinement polymorphism.
This extension is summarized by the definition 
in \cref{fig:drconty}, where we obtain both the 
monomorphic types  $\params{\type}$ and additionally 
the concrete refinements $\params{\aref}$ from the 
environment $\tcenv$ signature of $\evarb$, and use 
those to respectively instantiate the type ($\params{\tvar}$) 
and refinement ($\params{\rvar}$) variables in the signature 
of the data constructor $\dcon$, where the latter is done 
via the refinement instantiation mechanism described above.

\begin{myexample}{Checking \mcode{chkPair}}
Recall @chkPair@ from \S~\ref{sec:six:examples} 
\begin{align*}
  \expr     & \doteq \ \elam{p}{\ecase{p}{\dcapp{\mcode{MkPair}}{x, y} \rightarrow y - x}} \\
\intertext{Lets see how the rules establish that}
  \emptyset & \vdash \ \expr \Leftarrow \tfun{\mcode{incPair}}{\tnat} \\
\intertext{First, \rulename{Chk-Lam} reduces the above to}
  \tb{p}{\mcode{incPair}} & \vdash\ \ecase{p}{ \dcapp{\mcode{MkPair}}{x, y} \rightarrow y - x } \Leftarrow \tnat \\
\intertext{Via \rulename{Chk-Alt} (\cref{fig:five:check}) we extend 
the environment with the pattern-binders derived from 
$\typeb \doteq \ \dconty{\tb{p}{\mcode{incPair}}}{\mcode{MkPair}}{p}$
\ie the type of $\mcode{MkPair}$ \emph{at} this instance. As}
  \mcode{MkPair} & :: \ 
  \tpoly{\tvar,\tvarb}{
    \rpoly{\kvar}{
      \trfun{a}{\tvar}{
        \trfun{b}{\refb{\tvarb}{\rreft{\kvar(a, b)}}}{
          \tcapp{\mcode{pair}}{\tvar, \tvarb}(\elam{a,b}{\kvar(a,b)})
        }
      }
    }
  }
  \\
\intertext{as $\tb{p}{\mcode{incPair}}$, we get}
\typeb & \doteq \ 
  \trfun{a}{\tint}{
    \trfun{b}{\refb{\tint}{\rreft{a < b}}}{
      \tcapp{\mcode{pair}}{\tint, \tint}(\elam{a, b}{a < b})
    }
  } 
\\
\intertext{Thus, $\unapplysym$ (\cref{fig:unapply}) 
extends the environment with pattern binders}
\tcenv' & \doteq \ \tcenv,\ \tb{x}{\tint},\ \tb{y}{\refb{\tint}{\rreft{x < y}}} \\ 
\intertext{to obtain the goal for the case-alternative}
\tcenv' & \vdash \ y - x \Leftarrow \tnat 
\end{align*}
which, \rulename{Syn-Var} and \rulename{Syn-App} 
reduce to the entailment
$$\tcenv, \tb{x, y, \vvar}{\tint} \vdash\ (x < y) \Rightarrow (\vvar = y - x) \rightarrow (0 \leq v)$$
that is readily verified by the SMT solver.
\end{myexample}

\subsection{Synthesis}

\begin{figure}[t!]
\judgementHead{Type Synthesis}{\tsyn{\tcenv}{\expr}{\type}}
\begin{mathpar}
\inferrule
  {
    \rvar  = \tb{\kvar}{\tfun{\params{\base}, \base}{\tbool}} 
    \quad
    \tsyn{\tcenv}{\expr}{\rpoly{\rvar}{\typeb}}
    \\
    \isfresh{\refb{\base}{\ureft}}{\refb{\base}{\reft{x}{\pred}}}
    \quad
    \wf{\tcenv; \params{\tb{x}{\base}}}{\refb{\base}{\reft{x}{\pred}}}{\basekind}
  }
  {
    \tsyn{\tcenv}{\rapp{\expr}}{\rvinst{\typeb}{\rvar}{\lambda \params{\tb{x}{\base}},\tb{x}{\base}.{\pred}}}
  }
  {\ruleName{Syn-RApp}}
\end{mathpar}
\caption{$\langsix$: Rules for Type Synthesis}
\label{fig:six:synth}
\end{figure}

To make refinement polymorphism ergonomic, we need a way 
to automatically synthesize appropriate concrete refinements
for each term $\rapp{\expr}$ corresponding to \emph{uses} 
of terms $\expr$ whose types abstract over the refinements.
Otherwise, the programmer would have to bear the burden 
of writing concrete refinements at the ubiquitous usage 
sites, and worse, the resulting code would be very difficult 
to read! 
Thus, refinement synthesis is analogous to how we usually 
want to relieve the programmer of the burden of providing 
(monomorphic) instances at uses of a polymorphic signature.
Next, lets see how this instantiation can be achieved 
by the refinement synthesis machinery introduced in 
$\langthree$ (\S~\ref{sec:lang:three}).

\mypara{Instantiating Refinement Variables}
The rule \rulename{Syn-RApp} shown in \cref{fig:six:synth} 
shows how to synthesize types at refinement application sites 
$\rapp{\expr}$.
The rule says if we synthesize for $\expr$ 
a type $\rpoly{\rvar}{\typeb}$ which uses 
an abstract refinement variable 
$\rvar \equiv \tb{\kvar}{\tfun{\params{\base}, \base}{\tbool}}$ 
then we can \emph{instantiate} $\rvar$ in $\typeb$ 
with any \emph{well-formed} concrete refinement 
$\elam{\params{\tb{x}{\base}},\tb{x}{\base}}{\pred}$
whose sort is compatible with that of $\rvar$, 
\ie that is parameterized by the sorts compatible 
with the inputs of $\kvar$.

Observe that \rulename{Syn-RApp} mimics
\rulename{Chk-Rec} from \cref{fig:three:check} 
which uses the hole instantiation judgment 
$\isfresh{\typeb}{\type}$ from \cref{fig:three:isfresh} 
to declaratively \emph{guess} a suitable 
way to replace the holes $\ureft$ in $\typeb$ 
with concrete refinements.
In essence, \rulename{Syn-RApp} views the refinement 
variable $\rvar \equiv \tb{\kvar}{\tfun{\params{\base}, \base}{\tbool}}$ 
which denotes a relation over values of types 
$\params{\base},\base$ as an \emph{unknown} 
refinement over the base type $\refb{\base}{\ureft}$ 
and requires that we fill the hole with any 
concrete refinement $\pred$ that is well-formed 
under an environment extended with (\ie can refer to) 
binders $\params{\evar}$ for the other parameters 
$\params{\base}$. 

\begin{myexample}{Synthesis in \mcode{brighter}}
Lets use \rulename{Syn-RApp} to verify @brighter@ from \S~\ref{eq:brighter}. Let
\begin{align*}
  \tinte & \doteq\ \inte \\ 
  \tcenv & \doteq \ \tb{\mcode{maxI}}{\type_{\mcode{maxI}}}, \tb{x}{\tinte}, \tb{y}{\tinte} 
\intertext{where $\type_{\mcode{maxI}}$ is from (\ref{eq:maxI:type}). Using \rulename{Syn-Var} we have}
  \tcenv & \vdash \ \mcode{maxI} \Rightarrow \type_\mcode{maxI} 
\intertext{Hence, as $\isfresh{\refb{\tint}{\ureft}}{\tinte}$ 
  and $\wf{\tcenv}{\tinte}{\Base}$ \rulename{Syn-RApp} lets us 
  instantiate the refinement variable $\kvar$ quantifying 
  the signature of $\type_\mcode{maxI}$ with the concrete 
  refinement $\elam{x}{0 \leq x < 256}$ to obtain}
  \tcenv & \doteq\ \rapp{\mcode{maxI}} \Rightarrow \tfun{\tinte}{\tfun{\tinte}{\tinte}} 
\intertext{Now, the function application rule \rulename{Syn-App} lets us establish}
  \tcenv & \doteq\ {\eapp{\eapp{\rapp{\mcode{maxI}}}{x}}{y}} \Rightarrow \tinte 
\end{align*}
which, via rules \rulename{Chk-Syn} and \rulename{Chk-Lam} 
establishes that the implementation of \mcode{brighter} 
$\elam{x, y}{\eapp{\eapp{\rapp{\mcode{maxI}}}{x}}{y}}$ 
indeed checks against its specified type 
$\tfun{\tinte}{\tfun{\tinte}{\tinte}}$.
\end{myexample}

\section{Verification Conditions} \label{sec:six:algo}

The declarative rules simply guess suitable concrete 
refinements at each instantiation site. Next, lets see how 
to implement those rules via the method of Horn constraints 
introduced in \S~\ref{sec:lang:three}, as summarized 
in \cref{fig:six:algo}.

\mypara{Checking}
To extend the $\chksym$ function to implement \rulename{Chk-RAbs}
we add a case for checking a term $\expr$ against a type of form 
$\rpoly{\rvar}{\typeb}$ under environment $\tcenv$.
To this end, we substitute the refinement variable $\rvar$ with 
a concrete refinement $\aref$ corresponding to an uninterpreted function 
application $\elam{\params{\evar}}{\uf{\kvar}(\params{\evar})}$
and then return the Horn constraint $\cstr$ corresponding to 
the VC for checking the (substituted) term $\expr'$ against 
the (substituted) type $\typeb'$.

\begin{myexample}{VC for \mcode{maxI}}
Lets see the VC 
$\chk{\emptyset}{\expr_\mcode{maxI}}{\type_\mcode{maxI}}$
generated to verify that the implementation (\ref{eq:maxI:impl})
of @maxI@ adheres to its specification (\ref{eq:maxI:spec}).
The VC is obtained by substituting the refinement 
variable $\kvar$ with the uninterpreted function 
symbol $\uf{\kvar}$ and then recursively invoking 
$\chksym$ on the substituted body and signature, 
which produces the VC:
$$ \forall \uf{\kvar}, \mw{x}, \mw{y}, \vvar.\  
   \uf{\kvar}(\mw{x}) \Rightarrow \uf{\kvar}(\mw{y}) \Rightarrow 
                                           ((\vvar = \mw{x} \Rightarrow \uf{\kvar}(\vvar) ) \wedge
                                           (\vvar = \mw{y} \Rightarrow \uf{\kvar}(\vvar) ))
$$
which mirrors the entailments for the checking judgment (\ref{eq:maxI:check}).
\end{myexample}

\mypara{Synthesis}
Finally, in \cref{fig:six:algo} we extend $\synsym$ 
to account for refinement instantiation terms 
$\rapp{\expr}$ via an algorithmic implementation 
of \rulename{Syn-RApp}. 
First, we recursively invoke $\synsym$ to 
As in the declarative rule, we recursively 
invoke $\synsym$ to synthesize the 
refinement-polymorphic signature 
$\rpoly{\rvar}{\typeb}$ for $\expr$.
Next, as in \cref{fig:three:algo} we implement the declarative
$\isfresh{\refb{\base}{\ureft}}{\refb{\base}{\reft{x}{\pred}}}$
by using $\fresh{\tcenv; \params{\tb{x}{\base}}}{\refb{\base}{\ureft}}$
to obtain a new \emph{template} $\refb{\base}{\reft{x}{\pred}}$ 
where $\pred$ contains refinement (Horn) variables for the unknown 
concrete refinements to be used for instantiation.  
We use $\pred$ to build a concrete refinement that is substituted 
for $\rvar$ in $\typeb$ to return the synthesized type for $\rapp{\expr}$, 
together with the VC $\cstr$ obtained for $\expr$.

\begin{myexample}{Checking of \mcode{brighter}}
Lets see how the above procedure computes a Horn constraint (VC) 
whose satisfiability implies that \mcode{brighter} (\ref{eq:brighter}) 
implements the specification $\tfun{\tinte}{\tfun{\tinte}{\tinte}}$.
Let 
\begin{align*}
  \tcenv  & \doteq \ \tb{\mcode{maxI}}{\type_{\mcode{maxI}}} \\
  \expr_\mcode{br} & \doteq\ \elam{x, y}{\ \eapp{\eapp{\rapp{\mcode{maxI}}}{x}}{y}} \\
  \type_\mcode{br} & \doteq\ \tfun{\tinte}{\tfun{\tinte}{\tinte}} 
\intertext{respectively be the environment, implementation and specification 
  for \mcode{brighter}, where $\type_{\mcode{maxI}}$ is from (\ref{eq:maxI:type}). 
  Let
}
  \tcenv' & \doteq \ \tcenv,\ \tb{x}{\tinte},\ \tb{y}{\tinte} 
\intertext{By consulting $\tcenv'$, $\syn{\tcenv'}{\mcode{maxI}}$ returns $(\ttrue,\ \type_\mcode{maxI})$ where}
   \type_\mcode{maxI} & \doteq \ \rpoly{\kvar}{\tfun{\intk}{\tfun{\intk}{\intk}}})
\intertext{Hence, we invoke $\fresh{\tcenv, \refb{\tint}{\ureft}}$ 
to obtain the template $\intkbr$ with a new Horn variable $\kvar_\mcode{br}$ 
denoting the unknown concrete refinement at this instantiation site. 
Upon substituting the corresponding concrete refinement for the 
refinement variable we get the template (eliding the trivial Horn constraint $\ttrue$)}
\syn{\tcenv'}{\rapp{\mcode{maxI}}} & \doteq \tfun{\intkbr}{\tfun{\intkbr}{\intkbr}} 
\end{align*}
Then, via the usual cases for function application, we get the VC 
$$
\begin{array}{rcll}
\chk{\tcenv}{\expr_\mcode{br}}{\type_\mcode{br}} 
& \doteq &  \multicolumn{2}{l}{\forall x. (0 \leq x < 256) \Rightarrow} \\ 
&        &  \multicolumn{2}{l}{\quad \forall y. (0 \leq y < 256) \Rightarrow} \\ 
& .      & \quad \quad & \forall \vvar. (\vvar = x) \Rightarrow \kva{\kvarbr}{\vvar} \\
&        & \quad \quad \wedge & \forall \vvar. (\vvar = y) \Rightarrow \kva{\kvarbr}{\vvar} \\
&        & \quad \quad \wedge & \forall \vvar. \kva{\kvarbr}{\vvar} \Rightarrow (0 \leq v < 256) 
\end{array}
$$ 
The first two conjuncts check that the arguments $x$ and $y$ respectively satisfy 
the input types (preconditions) of $\rapp{\mcode{maxI}}$, and the last conjunct 
stipulates that the output type (postcondition) of the call is a valid $\tinte$ 
value, all assuming the values $x$ and $y$ inhabit $\tinte$ as specified by the 
input type of $\mcode{brighter}$. 
The above VC can be satisfied by assigning $\kvarbr \doteq \elam{a}{0 \leq a < 256}$, 
and hence we verify that \mcode{brighter} correctly implements its specification.
\end{myexample}

\begin{figure}[t]
$$\begin{array}{lcl}
\toprule
\chksym & : & (\tcenv \times \Expr \times \Type) \rightarrow \Cstr \\
\midrule
\chk{\tcenv}{\expr}{\rpoly{\rvar}{\typeb}}
                     & \doteq & \cswith{\uf{\kvar}}{\tfun{\params{\type}}{\tbool}}{\cstr}  \\
\quad \mbox{where}   &        &                              \\
\quad \quad \cstr    & =      & \chk{\tcenv'}{\rvinst{\expr}{\rvar}{\aref}}{\rvinst{\typeb}{\rvar}{\aref}} \\
\quad \quad \tcenv'  & =      & \tcenvext{\uf{\kvar}}{\tfun{\params{\type}}{\tbool}} \\
\quad \quad \aref    & =      & \elam{\params{\evar}}{\uf{\kvar}(\params{\evar})} \\ 
\quad \quad \tb{\kvar}{\tfun{\params{\type}}{\tbool}} & = &  \rvar \\[0.05in]


\toprule
\synsym & : & (\tcenv \times \Expr) \rightarrow (\Cstr \times \Type) \\
\midrule
\syn{\tcenv}{\rapp{\expr}}
                    & \doteq & (\cstr, \rvinst{\typeb}{\rvar}{{\lambda \params{\tb{x}{\base}},\tb{x}{\base}.{\pred}}}) \\[0.05in]
\quad \mbox{where}  &        & \\
\quad \quad (\cstr, \rpoly{\rvar}{\typeb}) & = & \syn{\tcenv}{\expr} \\ 
\quad \quad \tb{\kvar}{\tfun{\params{\base}, \base}{\tbool}} & = & \rvar \\ 
\quad \quad \params{x} & = & \mbox{fresh variables of sort} \ $\params{\base}$ \\
\quad \quad \refb{\base}{\reft{x}{\pred}} & = & \fresh{\tcenv; \params{\tb{x}{\base}}}{\refb{\base}{\ureft}} \\[0.05in]  
\bottomrule
\end{array}$$
\caption{Algorithmic Checking for $\langsix$, extends cases of~\cref{fig:five:algo}}
\label{fig:six:algo}
\end{figure}

\section{Discussion}

In \langsix we saw how we often want to specify signatures 
that abstract over refinements, how these signatures can be 
checked using uninterpreted functions, and how we can extend 
the Horn-constraint based inference method from \S~\ref{sec:lang:three} 
to automatically instantiate abstract refinements at usage sites.
The method was introduced by \cite{Vazou13} who illustrated
a variety of applications in establishing invariants of 
data structures.
\cite{Gordon17} demonstrated that abstract refinements could 
be used to encode a form of concurrent rely-guarantee reasoning,
enabling the verification of implementations of lock-free data structures. 
Subsequent work by \cite{polikarpova16} showed how refinement 
polymorphism allows writing compact specifications from which 
implementations can be automatically \emph{synthesized}.

\mypara{Bounded Quantification}
The abstract refinements in $\langsix$ were completely unconstrained. 
However, we could imagine a form of \emph{bounded quantification} 
for refinement variables analogous to type variables \citep{fbounded}, 
which would restrict instantiating refinements to those satisfying 
particular conditions.
Such an extension was explored by \cite{Vazou15} who showed how 
the programmer can use Horn constraints to express bounds which
allow specifying expressive signatures whilst preserving decidable 
and automatic verification. 
These extensions enable, for example, encoding a monadic 
information flow control (IFC) mechanism, purely within 
refinement types \citep{liftyICFP20} enabling 
the construction of web applications adhering to expressive 
data-sensitive security policies \citep{binah}.

\chapter{Termination} \label{sec:lang:seven}

The problem of verifying that the execution of a certain 
piece of code always terminates is perhaps one of the 
oldest in computing, going back all the way to \cite{Turing36}'s 
work on the undecidability of the Halting Problem.
Of course, just because a problem is undecidable, 
doesn't mean that it goes away!
Indeed, as \cite{rice53} pointed out \emph{all} non-trivial 
semantic properties of programs are undecidable. That is, 
it is just as undecidable to guarantee that, \eg a program 
will not attempt to add \tint and \tbool values. 
Yet, we have developed syntactic disciplines that have turned 
verifying the absence of errors like adding \tint and \tbool 
values into into a routine part of compiling code.
Next, lets see how refinements allow us to develop a simple 
and practical discipline for verifying, at compile time, 
that functions always terminate. 

\section{Examples} \label{sec:seven:examples}

Lets see some examples that illustrate
how to verify termination.

\mypara{Well-founded Metrics}
Consider the function @sum@ which adds up the numbers from @0@ to @n@: 

\begin{code}
  val sum : n:nat => nat
  let rec sum = (n) => {
    if (n < 1) { 
      0 
    } else { 
      n + sum(n-1) 
    }
  }
\end{code}

Why \emph{does} @sum n@ terminate? First, notice that 
@sum@ will not terminate if invoked on a negative number, 
\eg @sum(-3)@ will diverge. We eliminate this possibility 
by requiring the precondition that @n:nat@ \ie the inputs 
be non-negative.
Now, when @n@ equals @zero@, the procedure simply returns 
the result @0@. Otherwise, it recurses on a \emph{strictly} 
smaller input, until, ultimately, it reaches @0@, at which 
point it terminates.

\mypara{Proving Termination by Induction}
Thus we can, somewhat more formally, prove termination 
\emph{by induction} on @n@.

\begin{itemize}
  \item \emph{Base case}
    @sum@ terminates for inputs $k =$ @0@. 

  \item \emph{Induction Hypothesis} 
    Assume @sum@ terminates on all $k \leq$ @n@.

  \item \emph{Inductive Step} 
    Check that @sum(n+1)@ only recursively invokes @sum(n)@
    which satisfies the induction hypothesis and hence terminates.
\end{itemize}

This reasoning suffices to convince ourselves that @sum(n)@ 
terminates for every non-negative @n@. That is, we have shown 
that @sum@ terminates because a \emph{well-founded} metric:
here the non-negative @n@ is \emph{strictly decreasing} 
at each recursive call.

\mypara{Proving Termination with Types}
We can capture the above reasoning via the type system as follows.
First we require that @sum@ only be called with non-negative @nat@
values, which were defined as 
\begin{code}
  type nat = int[v|0 <= v]
\end{code}
Second, we to ensure that the recursion is on \emph{strictly smaller} 
values, we need only typecheck the \emph{implementation} of @sum@ in 
an environment that requires @sum@ only be called with inputs smaller 
than @n@, \ie we check the body in a environment of the form
$$\tcenv_\mcode{sum} \ \doteq\ \tb{n}{\tnat},\ \tb{\mcode{sum}}{\trfun{n'}{\refb{\tint}{\rreft{0 \leq n' < n}}}{\tnat}}$$ 
The above ensures that any (recursive) call in the body 
only calls @sum@ with inputs smaller than the ``current'' 
parameter @n@. Notice that if we had not required @n@ to be 
non-negative, then the parameter @n-1@ passed in at the 
recursive call would be smaller than @n@ but would not 
be non-negative, and hence, would fail the strengthened 
precondition for @sum@, as indeed it should, as such a 
computation does not terminate!

\mypara{Recursion on Multiple Parameters}
The above method works even when there are multiple
parameters, as long as there is \emph{some} @nat@-valued 
parameter that is used to limit the recursion. 
Consider the following tail-recursive variant of @sum@
\begin{code}
  let rec sumT = (total, n) => {
    if (n == 0) {
      total
    } else {
      sumT(total + n, n - 1)
    }
  } 
\end{code}
The function @sumT(total, n)@ takes two parameters:
@n@ as before, and @total@ which holds the accumulated 
summation of the ``previously seen'' seen. 
That is, @sumT(0,3)@ evaluates as follows: 
\begin{code}
 sumT(0,3) $\dashrightarrow $ sumT(3,2) $\dashrightarrow$ sumT(5,1) $\dashrightarrow$ sumT(6,0) $\dashrightarrow$ 6
\end{code}

\mypara{Specifying Termination Metrics}
The accumulation parameter @total@ is not strictly decreasing.
However, the parameter @n@ \emph{is} decreasing and non-negative 
and serves to witness that @sumT@ always terminates. 
But how might the type-checker 
\emph{guess} that it should use @n@ instead of @total@? While one can imagine 
a variety of pragmatic and effective heuristics to make such guesses, for our 
purposes, we shall simply give the type checker an explicit \emph{termination metric}
\begin{code}
  val sumT: total:nat => n:nat => nat / n
\end{code}
In the above, we end the type signature for @sumT@ with @/ n@ to denote 
that the value @n@ should be used as the termination metric.
The typechecker will verify that the value of the metric 
@n@ is indeed well-founded: \ie non-negative and strictly 
decreasing at each recursive call, and if so, will deem 
the function terminating.

\mypara{Metric Expressions}
Metrics generalize to situations where no single parameter
is decreasing, but some \emph{expression} over the parameters 
is. For example, consider the function @range(i, j)@ which returns 
the list of @int@egers between @i@ and @j@
\begin{code}
  val range : i:int => j:int => list(int) / j - i
  let rec range = (i, j) => {
    if (i < j) {
      Cons(i, range(i+1, j))
    } else {
      Nil 
    }
  } 
\end{code}
In the above, neither argument is decreasing: @i@ \emph{increases}
at each call, and @j@ is unchanged. Nevertheless, the function terminates 
as the \emph{gap} between @i@ and @j@ diminishes at each recursive call,
and the function terminates when that gap reaches @0@. 
We can make this intuition precise via the termination metric @/ j-i@. 
Armed with this information, the type checker ensures 
that at each recursive call in the body, the value of 
@j-i@ is decreasing and non-negative. 
That is, we will check the implementation of @range@ 
in an environment
\begin{equation}
\tcenv_\mcode{range} \ \doteq\ 
    \tb{i,j}{\tint},\ 
    \tb{\mcode{range}}{\trfun{i'}{\tint}{\trfun{j'}{\refb{\tint}{\rreft{0 \leq j' - i' < j - i}}}{\tint}}}
    \label{eq:range:env}
\end{equation}
that stipulates that recursive calls to @range@ must have 
strictly smaller, non-negative gaps, to verify that @range@ 
indeed always terminates.


\mypara{Lexicographic Metrics}
Sometimes, it is convenient to split up the termination 
metric across \emph{multiple} smaller metrics. 
For example consider Ackermann's function 
\begin{code}
  val ack : m:nat => n:nat => nat / m, n
  let rec ack = (m, n) => {
    if (m == 0) { n + 1 } else {
      if (n == 0) { ack (m - 1, 1) } else {
        ack (m - 1, ack (m, n - 1))
      }
    }
  }
\end{code}
Why does @ack@ terminate? At each iteration either the 
\emph{first} parameter @m@ decreases, or @m@ remains the same 
and the \emph{second} parameter @n@ decreases.
Each time that @n@ reaches @0@, it cannot decrease further 
so @m@ must decrease. Hence, @m@ will eventually reach @0@ 
and @ack@ will terminate.
In other words, the \emph{pair} @(m, n)@ decreases in the 
\emph{lexicographic order} on pairs, which is a well-ordering 
that has no infinite descending chains \citep{NipkowRewriting}.

\mypara{Specifying Lexicographic Orders via Types}
We can extend our notion of metrics to account for lexicographic 
orders by allowing the user to write a \emph{sequence} of metrics. 
For example, we type @ack@ with the signature with the termination 
metric @/ m, n@ and then we will use the sequence to check the 
implementation of @ack@ in environment $\tcenv_\mcode{ack}$
$$
  \tb{m,n}{\tnat},\ \tb{\mcode{ack}}{\trfun{m'}{\tnat}{\trfun{n'}{\refb{\tnat}{\rreft{m' < m \vee (m' = m \wedge n' < n)}}}{\tnat}}}
$$
The signature for @ack@ limits recursive uses of @ack@ to parameters 
that satisfy the lexicographic ordering to ensure that @ack@ terminates.

\mypara{Structural Recursion}
Often the recursion is over the elements of a datatype 
like a list or a tree. For example, consider the 
function that appends two lists @xs@ and @ys@
\begin{code}
  val append : xs:list('a) => ys:list('a) => list('a) 
             / len(xs)
  let rec append = (xs, ys) => {
    switch (xs) {
      | Nil           => ys
      | Cons (x, xs') => Cons (x, append (xs', ys))
    }
  }
\end{code}
The function @append@ recurses on the \emph{tail} of the first list @xs@ 
and stops when that list is empty \ie equal to @Nil@.
This is a form of \emph{structural recursion} where each recursive call 
is over sub-structures (\eg the tail) of some input parameter. 
We can verify the termination of structurally recursive functions by 
using \emph{measures} to specify suitable metrics. 
For @append@ we specify the metric @/ len(xs)@ which tells the type 
checker to limit recursive calls in the implementation of @append@
to lists whose length is smaller than @xs@. That is, we 
check the implementation of @append@ in an environment where
$\tb{xs, ys}{\tlist{\tvar}}$, and \mcode{append} is limited to
$$
\trfun{xs'}{\refb{\tlist{\tvar}}{\rreft{\mcode{len}(xs') < \mcode{len}(xs)}}}{\trfun{ys'}{\tlist{\tvar}}{\tlist{\tvar}}}
$$

\mypara{Non-Structural Recursion}
Finally, the notion of metrics over measures scales up to account for 
more general scenarios where the recursion over the datatypes is not 
structural. For example, consider the function @braid@ which takes 
two lists $x_1,\ldots$ and $y_1,\ldots$ and returns the list $x_1, y_1,\ldots$
\begin{code}
  val braid : xs:list('a) => ys:list('a) => list('a) 
            / len(xs) + len(ys)
  let rec braid = (xs, ys) => {
    switch (xs) {
      | Nil => ys 
      | Cons (x, xs') => Cons(x, braid(ys, xs'))
    }
  }
\end{code}
The recursion in @braid@ is not structural: the recursive call flips the order 
of the lists to ensure the values alternate in the output. 
However, in this case, the \emph{sum} of the lengths of the two input lists 
shrinks. We specify this via the metric @/ len(xs)  + len(ys)@ which tells 
the type checker to check the body of @braid@ under the following environment 
$$
\tb{xs,\ ys}{\tlist{\tvar}},\ \tb{\mcode{braid}}{\trfun{xs'}{\tlist{\tvar}}{\trfun{ys'}{\refb{\tlist{\tvar}}{\rreft{p}}}{\tlist{\tvar}}}} 
$$
where the refinement
$$
p \ \doteq\ 0 \leq \mcode{len}(xs') + \mcode{len}(ys') < \mcode{len}(xs) + \mcode{len}(ys)
$$
limits recursive calls to parameters the sum of whose lengths 
are decreasing, to verify that @braid@ terminates.

\section{Types and Terms} \label{sec:seven:types}

\begin{figure}[t!]
\begin{tabular}{rrcll}
\emphbf{Metric}
  & \metric & $\bnfdef$ & \pred                                               & \emph{decreasing expression} \\
  &         & $\spmid$  & \pred, \metric                                      & \emph{lexicographic metric}  \\[0.05in]
\emphbf{Terms}
  & \expr   & $\bnfdef$ & $\ldots$                                            & \emph{from Fig.~\ref{fig:six:syntax}} \\
  &         & $\spmid$  & \eletr{\evar}{\expr_1}{ \type / \metric }{\expr_2}  & \emph{recursive binder}    \\[0.05in]
\end{tabular}
\caption{{$\langseven$: Syntax of Types and Terms}}
\label{fig:seven:syntax}
\end{figure}

As demonstrated by the examples, $\langseven$ requires two small extensions 
to its syntax: a way to specify termination metrics, and a means of specifying 
metrics in type signatures, as summarized in \cref{fig:seven:syntax}.

\mypara{Metrics}
A \emph{termination metric} (or just metric in brief) 
is either a single \emph{decreasing expression} $\pred$ 
which is an @int@-sorted term from the refinement logic 
(\cref{fig:smt:pred}), or a \emph{lexicographic metric} 
comprising a sequence of decreasing expressions.

\mypara{Recursive Signatures}
In $\langseven$ we require that recursive \mcode{rec} 
binders be annotated with signatures that also specify 
a termination metric $\metric$.
For example, we would type @ack@ as 
$$\eletr{\mcode{ack}}{\elam{m,n}{\expr_\mcode{ack}}}{\type_\mcode{ack}}{\ldots}$$
where the signature $\type_\mcode{ack}$ specifies 
the lexicographic metric with the sequence 
of decreasing expressions $m, n$
\begin{equation}
  \type_\mcode{ack} \ \doteq\ \trfun{m}{\tnat}{\trfun{n}{\tnat}{\tnat}}~/~m, n 
  \label{eq:ack:type}
\end{equation}

\section{Declarative Typing} \label{sec:seven:check}

\begin{figure}[t]
\judgementHead{Metric Well-formedness}{\wfm{\tcenv}{\metric}}
\begin{mathpar}
\inferrule
  {\tcenv \vdash \pred : \tint}
  {\wfm{\tcenv}{\pred}}
  {\ruleName{Wfm-Base}}

\inferrule
  {\wfm{\tcenv}{\pred} 
   \quad
   \wfm{\tcenv}{\metric}
  } 
  {\wfm{\tcenv}{\pred, \metric}}
  {\ruleName{Wfm-Lex}}
\end{mathpar}
\caption{Rules for Checking Metric Well-formedness}
\label{fig:seven:wfm}
\end{figure}

As we saw with the examples in \cref{sec:seven:examples} termination 
checking reduces quite directly to plain refinement checking after 
\emph{limiting} recursive applications within the implementation to 
types that are strengthened with special refinements that ensure 
that the recursion is \emph{well-founded}. 

\mypara{Metric Well-formedness}
The judgment $\wfm{\tcenv}{\metric}$ says that 
a termination metric $\metric$ is \emph{well-formed} 
in an environment $\tcenv$. 
The judgment is established by the rules \rulename{Wfm-Base} 
and \rulename{Wfm-Lex} which, in concert, check that each decreasing 
expression in the metric can be typed as an @int@-valued term under $\tcenv$.

\begin{figure}[t]
$$\begin{array}{lcl}
\toprule
\wfpsym & : & (\Metric \times \Metric) \rightarrow \Pred \\
\midrule
\wfp{\pred^*}{\pred'}
   & \doteq & 0 \leq \pred' \wedge \pred' < \pred^* \\[0.05in]

\wfp{\pred^*;\metric^*}{\pred';\metric'}
   & \doteq & 0 \leq \pred' \wedge (\pred' < \pred^* \vee r) \\
\quad \mbox{where} &   & \\
\quad \quad r      & = & \pred' = \pred^* \wedge \wfp{\metric^*}{\metric'} \\[0.05in]
\bottomrule
\end{array}$$
\caption{Computing Well-foundedness Refinements}
\label{fig:seven:wfr}
\end{figure}

\mypara{Well-foundedness Refinements}
The procedure $\wfp{\metric^*}{\metric'}$ shown in \cref{fig:seven:wfr} 
takes as input two termination metrics, and returns as output a predicate 
that guarantees the metric demonstrates well-founded (terminating) recursion.
The inputs $\metric^*$ and $\metric'$ respectively 
denote the values of a function's termination metric over the 
\emph{original} and \emph{recursive} call parameters.
The output is a \emph{well-foundedness refinement} corresponding 
to the precondition that must hold at each recursive call in order 
for the metric to demonstrate the function terminating.

For @range@, we would use the termination metric $/~j-i$
to compute the well-foundedness refinement
\begin{equation}
  \wfp{j' - i'}{j - i} \doteq 0 \leq j'-i' < j - i
  \label{eq:wfr:range}
\end{equation}
Similarly, for @ack@, we would use the termination metric $/~m, n$ 
to compute a well-foundedness refinement
\begin{equation}
  \wfp{(m',n')}{(m, n)} \doteq 0 \leq m' \wedge (m' < m \vee (m' = m \wedge 0 \leq n' < n))
  \label{eq:wfr:ack}
\end{equation}

\begin{figure}[t]
$$\begin{array}{lcl}
\toprule
\limtsym & : & (\tcenv \times \Metric \times \Type) \rightarrow \Type \\
\midrule
\limt{\tcenv}{\metric}{\type} & \doteq & \lim{\tcenv}{\metric}{\metric}{\type} \\[0.1in]

\midrule
\limsym & : & (\tcenv \times \Metric \times \Metric \times \Type) \rightarrow \Type \\[0.05in]
\lim{\tcenv}{\metric^*}{\metric}{\trfun{\evar}{\refb{\base}{\rreft{p}}}{\type}} & & \\
\quad \mid \wfm{\tcenvext{\evar}{\base}}{\metric}
                     & \doteq & \trfun{\evar'}{\refb{\base}{\rreft{p'}}}{\type'} \\
\quad \mbox{where}   &        & \\
\quad \quad \pred'   & =      & \SUBST{\pred}{\evar}{\evar'}   \wedge \wfp{\metric^*}{\metric'} \\
\quad \quad \metric' & =      & \SUBST{\metric}{\evar}{\evar'} \\
\quad \quad \type'   & =      & \SUBST{\type}{\evar}{\evar'}   \\[0.05in]

\lim{\tcenv}{\metric^*}{\metric}{\trfun{\evar}{\typeb}{\type}}
                     & \doteq & \trfun{\evar'}{\typeb'}{\type''} \\
\quad \mbox{where}   &        & \\
\quad \quad \type''  & =      & \lim{\tcenvext{\evar}{\typeb}}{\metric^*}{\metric'}{\type'}   \\
\quad \quad \metric' & =      & \SUBST{\metric}{\evar}{\evar'} \\
\quad \quad \typeb'  & =      & \SUBST{\typeb}{\evar}{\evar'}  \\
\quad \quad \type'   & =      & \SUBST{\type}{\evar}{\evar'}   \\[0.05in]

\lim{\tcenv}{\metric^*}{\metric}{\tpoly{\tvar}{\type}}
                     & \doteq & \tpoly{\tvar}{\lim{\tcenv}{\metric^*}{\metric}{\type}} \\

\bottomrule
\end{array}$$
\caption{Checking termination by type limiting}
\label{fig:seven:lim}
\end{figure}

\mypara{Type Limiting}
The procedure $\limt{\tcenv}{\metric}{\type}$ shown in \cref{fig:seven:lim} 
computes a type $\type'$ which strengthens the input types (preconditions) 
to require that all (recursive) calls be \emph{limited} to values that are 
allowed by the metric $\metric$.
The real work is done by the helper $\lim{\tcenv}{\metric^*}{\metric}{\type}$ 
which takes the \emph{original} metric $\metric^*$ and the \emph{recursive} 
metric $\metric$ where all the input binders ($x$) are replaced with ``primed'' 
versions ($x'$) and returns a version of $\type$ where 
(1)~the inputs are renamed with their primed variants and 
(2)~the well-foundedness refinement is used to strengthen 
    the first input refinement where it is well-formed, 
    \ie where all the binders appearing in the refinement 
    are in scope.
To this end, the procedure recurses over the structure of the 
(function) type $\type$, adding the binders to $\tcenv$ and 
renaming the inputs $\metric$ with their primed versions, 
\emph{until} it has added enough binders to $\tcenv$ for $\metric^*$
to be well-formed, at which point, it strengthens the current 
parameter's refinement $\pred$ with the well-foundedness 
refinement $\wfp{\metric^*}{\metric'}$.

For simplicity, the definition of \limsym and the rule 
\ruleName{Chk-Term} do not support refinement polymorphism. 
It is straightforward to remove this restriction by extending 
the definitions to allow for types abstracted over refinements.

\begin{myexample}{Type of \mcode{range}}
Recall that @range@ has type
\begin{equation}
\type_\mcode{range} \doteq \trfun{i}{\tint}{\trfun{j}{\tint}{\tlist{\tint}}}~/~j-i
\label{eq:range:type}
\end{equation}
We will check its body with the metric limited type
\begin{align*}
  &\ \limt{\emptyset}{\ j-i}{\ \trfun{i}{\tint}{\trfun{j}{\tint}{\tlist{\tint}}}}
\intertext{which is defined as}
= &\ \lim{\emptyset}{\ j-i}{\ j-i}{\ \trfun{i}{\tint}{\trfun{j}{\tint}{\tlist{\tint}}}}
\intertext{as the metric $j-i$ is not well-formed, we rename $i$ to $i'$ and recurse}
= &\ \trfun{i'}{\tint}{\lim{\tb{i}{\tint}}{\ j-i}{\ j-i'}{\ \trfun{j}{\tint}{\tlist{\tint}}}}
\intertext{now, since $\wfm{\tb{i}{\tint}, \tb{j}{\tint}}{j-i}$ the above is}
= &\ \trfun{i'}{\tint}{\trfun{j'}{\refb{\tint}{\rreft{ \wfp{j-i}{j'-i'} }}}{\tlist{\tint}}}
\intertext{which, after substituting the well-foundedness refinement (\cref{eq:wfr:range})}
= &\ \trfun{i'}{\tint}{\trfun{j'}{\refb{\tint}{\rreft{ 0 \leq j'-i' < j -i }}}{\tlist{\tint}}} 
\end{align*}
\end{myexample}

\begin{figure}[t]
\judgementHead{Termination Checking}{\mchk{\tcenv}{\evar}{\expr}{\typeb}{\metric}{\type}}
\begin{mathpar}
\inferrule
  {
    \type = \fresh{\tcenv}{\typeb} = \tpoly{\params{\tvar}}{\trfuns{\evarb}{\typeb}{\type^*}}
    \quad
    \expr = \tabs{\params{\tvar}}{\elam{\params{y}}{\expr^*}}
    \\
    \tchk{\tcenvextv{\params{\tb{\tvar}{\kind}}; \params{\tb{\evarb}{\typeb}}; \tb{\evar}{ \limt{\tcenv}{\metric}{\type} }}}
         {\expr^*}
         {\type^*}
  }
  {
    \mchk{\tcenv}{\evar}{\expr}{\typeb}{\metric}{\type}
  }
  {\ruleName{Chk-Term}}
\end{mathpar}

\judgementHead{Type Checking}{\tchk{\tcenv}{\expr}{\type}}
\begin{mathpar}
\inferrule
  {
    \mchk{\tcenv}{\evar}{\expr_1}{\typeb_1}{\metric}{\type_1}
    \and
    \tchk{\tcenvext{\evar}{\type_1}}
         {\expr_2}
         {\type_2}
  }
  {
    \tchk{\tcenv}
         {\eletr{\evar}{\expr_1}{\typeb_1 / \metric}{\expr_2}}
         {\type_2}
  }
  {\ruleName{Chk-Rec}}
\end{mathpar}
\caption{Bidirectional Checking: Other rules from $\langsix$ (Fig.~\ref{fig:six:check})}
\label{fig:seven:check}
\end{figure}

\mypara{Termination Checking}
We introduce a new \emph{termination checking} 
judgment $\mchk{\tcenv}{\evar}{\expr}{\typeb}{\metric}{\type}$
which \emph{guarantees} that in an environment $\tcenv$,
the (recursive) definition $\evar = \expr$ is terminating 
with the metric $\metric$, and that downstream definitions 
can \emph{assume} that $\evar$ behaves as $\type$.
The rule \ruleName{Chk-Term} shown in \cref{fig:seven:check} 
establishes this judgment by
\begin{enumerate}
\item \emphbf{Instantiating} the \emph{holes} in $\typeb$ to obtain 
      the complete signature $\type$, 
\item \emphbf{Splitting} the definition $\expr$ into its 
      input \emph{binders} $\params{\tb{\evarb}{\typeb}}$ 
      and  \emph{body} $\expr^*$,
\item \emphbf{Checking} the body $\expr^*$ in an environment containing
      the input binders $\params{\tb{\evarb}{\typeb}}$ which 
      name the \emph{current} input, \emph{and} where $\evar$ 
      is bound to its \emph{metric limited type} $\limt{\tcenv}{\metric}{\type}$.
\end{enumerate}
That is, the key change is to check the body of the recursive 
binder in an environment that limits the recursion using the 
specified metric.

\mypara{Checking Recursive Definitions}
We can now use the termination checking judgment 
in the updated rule \rulename{Chk-Rec} shown in 
\cref{fig:seven:check}.
To check that $\eletr{\evar}{\expr_1}{\typeb_1 / \metric}{\expr_2}$ 
has type $\type_2$, the new rule requires that the binder 
$\evar$ terminates with $\metric$, and then, (as before), 
checks $\expr_2$ against $\type_2$ in the environment 
extended with $\evar$.

\begin{myexample}{Checking \mcode{range}}
Lets see how \rulename{Chk-Term} checks the term 
\begin{align}
  \expr_\mcode{r} &\ \doteq\ 
    \eletr{\mcode{range}}{(\elam{i, j}{\expr_\mcode{range}})}{\ \type_\mcode{range}}{\ldots}
    \label{eq:range:full}
\intertext{where}
  \expr_\mcode{range} &\ \doteq\ \eif{i < j}{\tapp{\mcode{Cons}}{\tint}\ i\ (\mcode{range}\ (i+i)\ j)}{\tapp{\mcode{Nil}}{\tint}} 
  \notag \\ 
  \type_\mcode{range} &\ \doteq \trfun{i}{\tint}{\trfun{j}{\tint}{\tlist{\tint}}}~/~j-i
  \notag
\end{align}
\rulename{Chk-Term} stipulates that we use the metric 
to limit the type and check the body $\expr_\mcode{range}$
against the specified output $\tlist{\tint}$ in environment 
$\tcenv_\mcode{range}$ from \cref{eq:range:env}
\begin{align*}
  \tcenv_\mcode{range} & \ \vdash\ \expr_\mcode{range} \Leftarrow \tlist{\tint} 
\intertext{Rule \rulename{Chk-If} reduces the above to checks on each branch. Eliding the trivial \mcode{else} case, we get}
  \tcenv_\mcode{range}; i < j & \ \vdash\ {\tapp{\mcode{Cons}}{\tint}\ i\ (\mcode{range}\ (i+i)\ j)} \Leftarrow \tlist{\tint} 
\intertext{which \rulename{Syn-App} splits into a check that} 
  \tcenv_\mcode{range}; i < j & \ \vdash\ \mcode{range}\ (i+i)\ j \Leftarrow \tlist{\tint} 
\intertext{Again, \rulename{Syn-App} splits the above into checks that ensure each \emph{input} to \mcode{range} satisfies the 
environment's (limited) input type}
  \tcenv_\mcode{range}; i < j & \ \vdash\ \refb{\tint}{\reft{i'}{i' = i + 1}} \ \lqsubt\ \tint \\
  \tcenv_\mcode{range}; i < j; \tb{i'}{\rreft{i' = i + 1}} & \ \vdash\ \refb{\tint}{\reft{j'}{j' = j}} \ \lqsubt\ \refb{\tint}{\reft{j'}{0 \leq j' - i' < j - i}}
\intertext{the first of those is trivial; the second reduces to the entailment}  
  i < j; i' = i + 1; j' = j & \ \vdash\ 0 \leq j' - i' < j - i
\end{align*}
that is readily verified by the SMT solver.
\end{myexample}

\section{Verification Conditions} \label{sec:seven:algo}

\begin{figure}[t]
$$\begin{array}{lcl}
\toprule
\chktsym & : & (\tcenv \times \evar \times \Expr \times \Type \times \Metric) \rightarrow (\Type \times \Cstr) \\
\midrule
\chkt{\tcenv}{\evar}{\expr}{\typeb}{\metric}
    & \doteq & (\type,\ \cswiths{\evarb}{\typeb}{\cswith{\evar}{\type'}{\cstr}}) \\
\quad \mbox{where}   &    & \\
\quad \quad \cstr  & =  & \chk{\tcenvextv{\params{\tvar}; \params{\tb{\evarb}{\typeb}}; \tb{\evar}{\type'}}}
                                {\expr^*}
                                {\type^*} \\
\quad \quad \type' & =  & \limt{\tcenv}{\metric}{\type}             \\
\quad \quad \tpoly{\params{\tvar}}{\trfuns{\evarb}{\typeb}{\type^*}} \ \mbox{as}\ \type
                     & =  & \fresh{\tcenv}{\typeb}                            \\
\quad \quad \tabs{\params{\tvar}}{\elam{\params{y}}{\expr^*}}
                     & =  & \expr                                           \\[0.05in]
\bottomrule
\end{array}$$
\caption{Algorithmic Termination Checking} 
\label{fig:seven:algo:term}
\end{figure}

The algorithmic VC generation procedure for $\langseven$
extends how $\chksym$ generates Horn constraints for 
$\eletr{\evar}{\expr_1}{\typeb_1}{\expr_2}$, 
as summarized in \cref{fig:seven:algo}.
The implementation mirrors the declarative formulation 
\rulename{Chk-Rec}, where we first check that the recursive 
definition $\expr_1$ is terminating, and then use its type $\type_1$ 
to check $\expr_2$.

\mypara{Algorithmic Termination Checking}
We implement the termination checking judgment with a procedure 
\chktsym summarized in \cref{fig:seven:algo:term}, which takes 
as input a recursive definition $\evar = \expr$ and the type 
$\typeb$ and metric $\metric$ ascribed to the definition and 
returns as output a pair comprising the Horn VC $\cstr$ 
whose satisfiability indicates that the definition is 
well-typed and terminating, and the type $\type$ that 
subsequent binders can assume for $\evar$. 
First, (as in \cref{fig:three:algo}) we use $\freshsym$ 
to instantiate the holes in the specification $\typeb$ 
with new Horn variables.
Second, we obtain the body $\expr^*$ and input binders 
$\params{\tb{\evarb}{\typeb}}$.
Finally, we invoke $\chksym$ to compute the VC $\cstr$ 
for the body in an environment containing the input 
binders and the metric-limited type $\type'$.
The following states the correspondence between the algorithmic 
and declarative versions of termination checking:

\begin{proposition}
If $\chkt{\tcenv}{\evar}{\expr}{\typeb}{\metric} = (\type, \cstr)$ 
and $\cstr$ is Horn satisfiable then $\mchk{\tcenv}{\evar}{\expr}{\typeb}{\metric}{\type}$.
\end{proposition}

Following the same reasoning as the declarative checking,
\ie generating a VC for the body in the metric-limited environment, 
$\chk{\emptyset}{\expr_\mcode{r}}{\ldots}$ generates the 
following VC for the term from \cref{eq:range:full}:
$$
  \forall \tb{i, j, i', j'}{\tint}.\ i < j \Rightarrow i' = i + 1 \Rightarrow j' = j \Rightarrow 0 \leq j' - i' < j - i
$$
The above VC is valid, which proves that \mcode{range} terminates.

\begin{figure}[t]
$$\begin{array}{lcl}
\toprule
\chksym & : & (\tcenv \times \Expr \times \Type) \rightarrow \Cstr \\
\midrule
\chk{\tcenv}{\eletr{\evar}{\expr_1}{\typeb_1 / \metric}{\expr_2}}{\type_2}
    & \doteq & \cstr_1 \wedge \cswith{\evar}{\type_1}{\cstr_2} \\
\quad \mbox{where}   &    & \\
\quad \quad (\type_1, \cstr_1)  & =  & \chkt{\tcenv}{\evar}{\expr_1}{\typeb_1}{\metric} \\ 
\quad \quad \cstr_2  & =  & \chk{\tcenvext{\evar}{\type_1}}{\expr_2}{\type_2}   \\[0.05in]
\bottomrule
\end{array}$$
\caption{Algorithmic Checking for $\langseven$, extends cases of~\cref{fig:six:algo}}
\label{fig:seven:algo}
\end{figure}

\section{Discussion}

To recap, in $\langseven$ we saw how to extend the type system
to also check that (recursive) functions terminate.
We started with functions like @sum@ and @sumT@ that recurse 
on some natural number @n@ that directly demonstrates termination. 
Then, we saw how to generalize the above idea to decreasing expressions 
like the one that we used to demonstrate @range@ terminates.
We saw how the idea of expressions can be generalized to sequences 
to yield termination metrics which demonstrate termination via 
well-founded lexicographic ordering.
Finally, we saw how the notion of measures generalizes the above 
to functions that work on datatypes.
The key idea in all of the above, is simply to check the body of 
the recursive call with a strengthened type that limits (recursive)
inputs to be well-founded, thereby enforcing termination.

\mypara{Incompleteness}
Of course, thanks to the Halting problem there are terminating functions 
that cannot be proven terminating via the approach shown above, because 
there is no algorithmic procedure to find termination metrics, and because 
the metrics themselves may, in general, fall outside the SMT solver's 
decidable theories.
For example, it would be a major result to find a suitable terminating 
metric for the @collatz@ function \citep{collatzWiki}.

\mypara{Termination in Practice}
Nevertheless, there is evidence that the mechanisms shown here 
with simple extensions, \eg to account for mutually recursive
functions, suffice to verify termination in the vast majority 
of programs that arise in practice. 
For example, the metric-based approach extended with simple heuristics 
to generate default metrics associated with particular datatypes, suffices 
to verify 96\% of recursive functions on a corpus of more than 10,000 lines 
of widely used Haskell libraries, whilst requiring only about 1.7 metrics per 
100 lines of code \citep{Vazou14}.
Other SMT-based verifiers like \fstar \citep{fstar} and \dafny \citep{dafny} 
use similar strategies to check termination very effectively.

\mypara{Other Strategies for Proving Termination}
There is a vast literature on techniques for proving termination 
all of which ultimately find their roots in the notion of well-founded 
metrics, introduced by \cite{Turing49}.
\cite{JonesB04} and \cite{Sereni05} embody this idea via the 
``size-change principle'' that they use to verify 
termination of recursive functions, and which, can 
be rephrased as a \emph{contract} to enable \emph{dynamic} 
termination checking \cite{Nguyen19}.
Proof assistants like \textsc{Coq} \citep{coq-book} 
and \textsc{Isabelle} \citep{IsabelleManual} employ 
\emph{structural} termination checks wherein 
recursive calls can only be made on strict 
sub-structures of the inputs (\eg the tail of 
an input list.)
\cite{HughesParetoSabry96} and \cite{BartheTermination} 
show how to generalize this idea via \emph{sized types} 
wherein the bodies of recursive functions are checked 
under metric limited environments.
An alternative approach formulated by \cite{Giesl11} is to 
reduce termination for functional programs 
to termination of term rewriting systems.
\cite{Rybalchenko04} show how to generalize and unify 
the notion of termination metrics and program invariants, via 
the notion of \emph{transition invariants}, which also allow 
the use of abstract interpretation based methods to automatically 
synthesize suitable metrics (or ``ranking functions''), which 
was the basis of the \textsc{Terminator} tool \citep{CookPR11} 
which verifies the termination of device drivers written in \textsc{C}.

Our formulation for $\langseven$ is inspired by the method 
introduced by \cite{XiTerminationLICS01} to encode 
sized types in a refinement setting.
Refinements provide the advantage of \emph{unifying} 
reasoning about invariants of data with reasoning about 
termination. 
This unification is crucial for large real-world code bases, 
where termination requires functions only be called under 
certain pre-conditions (\eg @int@ valued inputs are non-negative), 
or require specific post-conditions (\eg the \mcode{split} function 
in a merge-sort routine returns output lists that are strictly smaller 
than the input), or let us specify arithmetic metrics like those in 
\mcode{range} where termination depends crucially on the 
path-sensitive reasoning performed by the rest of the type checking.

\chapter{Programs as Proofs}
\label{sec:lang:eight}

We have been rather timid in what we allow 
specifications to \emph{say}, limiting them
mostly to facts about integers or sets or 
ordering extended with uninterpreted measures 
that describe properties of algebraic data, 
to ensure that the VCs are SMT decidable.
Next, lets see how to break out of this shell, 
to allow users to write specifications over
user defined functions and then prove theorems 
about those functions by supplying proofs 
structured as programs.
We will do so by \emph{reflecting} the implementation 
of the user-defined function into its output type, 
thereby converting its type signature into a precise 
description of the function's behavior.
Reflection has a profound consequence: at \emph{uses} 
of the function, the standard rule \rulename{Syn-App} for function 
application turns into a means of explicating how the function 
behaves at the given input, which lets us encode the function's 
behavior at the (refinement) type level. 
The above idea, coupled with a small set of combinators, 
lets us write sophisticated proofs simply as programs.

\section{Examples}

Lets start with an overview of how refinement reflection works 
by seeing how it lets us write paper-and-pencil-style proofs 
as programs.

\subsection{Propositions as Types} 

Refinements let us encode propositions as types.
For example, 
%
%
a unit type can be refined with a logical proposition
that encodes that @1 + 1 = 2@
%
\begin{code}
  type one_plus_one_eq_two = ()[v|1 + 1 = 2]
\end{code}
As the @v@ and @()@ are unimportant, we will elide them and just write
\begin{code}
  type one_plus_one_eq_two = [1 + 1 = 2]
\end{code}
As another example, here is the proposition 
that $\tint$ addition is commutative, \ie 
$\forall \tb{x, y}{\tint}. x + y = y + x$:
\begin{code}
  type plus_comm = x:int => y:int => [x + y = y + x]
\end{code}

\mypara{Programs as Proofs}
Notice that we can represent universal quantification as a function type,
following the Curry-Howard correspondence \citep{CHIso,Wadler15}.
Thus, following the correspondence, any term $e$ whose type corresponds 
to a proposition $P$ can be viewed as a \emph{proof} of that proposition.
Here is a trivial ``proof'' of the proposition @one_plus_one_eq_two@
\begin{code}
  val thm_one_plus_one_eq_two : one_plus_one_eq_two
  let thm_one_plus_one_eq_two = ()
\end{code}
Note that the VC generation procedure we outlined in \cref{sec:lang:one}
would verify the above program by checking the validity of the VC
$$1 + 1 = 2$$ 
which is validated by the SMT solver. Here is a proof for @plus_comm@
\begin{code}
  val thm_plus_comm : plus_comm
  let thm_plus_comm = (x, y) => ()
\end{code}
The VC generation procedure from \cref{sec:lang:one} generates the VC
$$ \forall x, y.\ x + y = y + x $$ 
which the SMT solver validates via the theory of linear 
arithmetic, giving us our ``theorem''. 
These two propositions fell squarely within decidable theories 
and hence had trivial proofs, with the SMT solver doing all 
the work. 

\subsection{Refinement Reflection} \label{sec:eight:refl:ex}

Next, lets extend the language of refinements 
with user-defined functions, and write propositions 
and proofs over those functions.

\mypara{Step 1: Propositions over User-defined Functions} 
First, $\langeight$ introduces a way to define functions, 
\eg to @sum@ numbers from @0@ to @n@
\begin{code}
  val sum : n:nat => nat / n
  def sum = (n) => { 
    if (n == 0) { 
      0 
    } else {
      n + sum(n-1)
    } 
  }
\end{code}
The @def@-bound functions are just like the usual @rec@ binders 
--- the metric @/ n@ ensures they terminate --- except 
that we can refer to them in refinements as the 
\emph{uninterpreted} function $\rcode{sum}$.
Consequently, we can now specify the proposition 
\begin{equation}
\forall i, j.\ i=j \Rightarrow \rcode{sum}(i) = \rcode{sum}(j) 
\label{eq:sum_eq_vc}
\end{equation}
and verify it via a trivial proof
\begin{code}
  val sum_eq : i:nat => j:nat[i=j] => [sum(i) = sum(j)]
  let sum_eq = (i, j) => ()
\end{code}
The above program generates the VC corresponding directly 
to the proposition \cref{eq:sum_eq_vc} which the SMT solver 
automatically validates via congruence closure \citep{Nelson81}.

\mypara{Step 2: Refinement Reflection}
\langeight imbues @def@ binders with a second 
crucial property: we strengthen their user-specified 
types with a refinement that exactly 
\emph{reflects the function's implementation}.
That is, let $\defn{sum}(n)$ be an abbreviation for the refinement
$$\defn{sum}(n) \ \doteq\ \ite{n = 0}{0}{n + \rcode{sum}(n-1)}$$ 
where $\rcode{sum}$ is the \emph{uninterpreted} function 
representing @sum@ in the refinement logic. Now, we assign 
@sum@ the type
$$ 
  \mcode{sum} : \trfun{n}{\tnat}{ \refb{\tnat}{\reft{\vvar}{\vvar = \rcode{sum}(n) \wedge \vvar = \defn{sum}(n)}}}
$$
which says that the output value $\vvar$ equals 
to the logical representation $\rcode{sum}(n)$ 
which itself equals the value of the reflected 
body $\defn{sum}(n)$.

\mypara{Step 3: Proofs using Function Applications}
As we have reflected the function's definition into 
its output type, each \emph{application} of @sum@ 
\emph{instantiates} its definition at the given 
input. 
For example, here is a proof that $\rcode{sum}(2) = 3$
\begin{code}
  val sum_2_eq_3 : () => [sum(2) = 3]
  let sum_2_eq_3 = () => { 
    let $t_0$ = sum(0);
    let $t_1$ = sum(1);
    let $t_2$ = sum(2);
    ()                 
  }
\end{code}
The usual rules \rulename{Chk-Let} 
and \rulename{Syn-App} yield the VC 
$$\begin{array}{l}
\forall t_0.\ t_0 = \rcode{sum}(0) = (\ite{0 = 0}{0}{0 + \rcode{sum}(0-1)}) \Rightarrow \\ 
\quad \forall t_1.\ t_1 = \rcode{sum}(1) = (\ite{1 = 0}{0}{1 + \rcode{sum}(1-1)}) \Rightarrow \\ 
\quad \quad \forall t_2.\ t_2 = \rcode{sum}(2) = (\ite{2 = 0}{0}{2 + \rcode{sum}(2-1)}) \Rightarrow \\ 
\quad \quad \quad \rcode{sum}(2) = 3 
\end{array}$$
Intuitively, each application $\mcode{sum}(i)$ \emph{instantiates} the definition
of $\mcode{sum}$ at the input $i$, after which the SMT solver's theories for 
equality, congruence and arithmetic kick in to internally simplify the above 
VC to the following valid formula
$$\begin{array}{l}
\rcode{sum}(0) = 0 \Rightarrow \\ 
\quad \rcode{sum}(1) = 1 + \rcode{sum}(0) \Rightarrow \\
\quad \quad \rcode{sum}(2) = 2 + \rcode{sum}(1) \Rightarrow \\
\quad \quad \quad \rcode{sum}(2) = 3 
\end{array}$$
We need \emph{all} the instances for $0$, $1$ and $2$ 
to prove the goal. A proof term that omitted the 
binding $t_1$ would be \emph{rejected} as it
yields the VC
$$\begin{array}{l}
\rcode{sum}(0) = 0 \Rightarrow \\ 
\quad \rcode{sum}(2) = 2 + \rcode{sum}(1) \Rightarrow \\
\quad \quad \rcode{sum}(2) = 3 
\end{array}$$
which is \emph{invalid} as $\rcode{sum}(1)$ is unconstrained.

\subsection{Structuring Proofs via Combinators}

Writing proofs like @sum_2_eq_3@ can be 
tedious: how are we to divine \emph{which}
terms to instantiate the definition of @sum@ at? 

\mypara{Equational Proofs}
We can solve this problem by structuring 
proofs to follow the style of calculational 
or equational reasoning \citep{Bird89, Dijkstra76} 
and implemented in \textsc{Agda}\citep{agdaequational}, 
and \textsc{Dafny} \citep{LeinoPolikarpova16},
via an \emph{equality-chaining} combinator
\begin{code}
  val (===) : x:'a => y:'a[y == x] => [v=x && v=y]
  let (===) = (x, y) => y
\end{code}
The combinator's type specification says that 
$\expr_1 \ \mcode{===}\ \expr_2$ is 
a \emph{proof} that $\expr_1$ and $\expr_2$ 
are equal, and further, a term that equals 
$\expr_1$ and $\expr_2$.
We can use equality-chaining to 
rewrite @sum_2_eq_3@ in a way that 
might mirror a pencil-and-paper proof
\begin{code}
  val sum_2_eq_3 : _ -> [sum 3 = 6] @-}
  let sum_2_eq_3 = () => { 
    sum(2)
      === 2 + sum(1)
      === 2 + 1 + sum(0)
      === 3 
  }
\end{code}
The precondition of @(===)@ checks that each 
\emph{intermediate} equality holds, simply by 
using the \emph{applied} instance of @sum@ at 
the respective call.
The postcondition of @(===)@ lets us chain the
equalities together to prove the goal.
Thus, if we skip a step, \eg if we write 
\begin{code}
  let sum_2_eq_3' = () => { 
    sum(2)
      === 2 + 1 + sum(0)
      === 3 
  }
\end{code}
the precondition for the first equality-chain 
will fail, and so type checking pinpoints where 
information is needed to complete the proof.

\mypara{Functions as Lemmas}
Suppose we want to verify the proposition @sum(3) == 6@.
We could, of course, repeat all the calculations we did 
to prove @sum_2_eq_3@ but instead, it would be nice to 
\emph{reuse} the proposition that we have already proved 
as a \emph{lemma} to prove the new goal.
We do so by introducing a \emph{because} combinator that 
conjoins propositions
\begin{code}
  val (?) : x:'a[p x] => y:'b[q y] => 'a[v| p v && q y] 
  let (?) = (x,_) => x 
\end{code}
We can now reuse @sum_2_eq_3@ as a lemma, simply by 
applying it as
\begin{code}
  val sum_3_eq_6 : _ => [sum(3) = 6]
  let sum_3_eq_6 = () => { 
    sum(3)
      === 3 + sum(2) 
        ? sum_2_eq_3 () 
      === 6 
  }
\end{code}
The types of our combinators ensure that the above yields a VC like
$$\rcode{sum}(3) = \defn{sum}(3) \Rightarrow \rcode{sum}(2) = 3 \Rightarrow \rcode{sum}(3) = 6$$ 
where the fact establishing the value of $\rcode{sum}(2)$ is established
by applying, and hence, obtaining the output (post-condition) 
of the function @sum_2_eq_3@.

\subsection{Proofs as Programs} \label{sec:eight:proofs:ex}

The above proofs are quite unremarkable: they merely
confirm what a computation evaluates to. 
However, they introduce the building blocks of more
interesting examples that illustrate the 
correspondence between proofs and programs summarized 
in \cref{fig:proofs_as_programs}.

\begin{figure}[t]
\begin{center}
\begin{tabular}{*{2}{c}}
\toprule
    \textbf{Proof} & \textbf{Program} \\
\midrule
    Theorem        & Function \\
    Apply Theorem  & Call Function \\
    Case Split     & Branch \\ 
    Induction      & Recursion \\
\bottomrule
\end{tabular}
\end{center}
\caption{Correspondence between Proofs and Programs.}
\label{fig:proofs_as_programs}
\end{figure}

\mypara{Induction on Numbers}
Lets write and prove the proposition 
$$\forall n.\ \sum\limits_{i=0}^n i = \frac{n \times (n+1)}{2}$$
We can specify the proposition as a type and then
provide a proof as:
\begin{code}
  val thm_sum : n:nat => [2 * sum(n) = n * (n+1)] 
  let thm_sum = (n) => { 
    switch (n) { 
    | 0 => {   2 * sum(0) 
           === 0 * (0+1) 
           }
    | n => {   2 * sum(n)
           === 2 * (n + sum(n-1))
           === 2 * n + 2 * sum(n-1)
             ? thm_sum(n-1)
           === n * (n+1) 
           }
    }
  }
\end{code}
The above proof mirrors the classic proof \emph{by induction}.
We split cases on @n@ via a @switch@ that branches on the value of @n@. 
In the \emph{base} case we prove the equality via a calculation on @sum(0)@.
In the \emph{inductive} case we prove the equality by \emph{recursively} 
applying @thm_sum@ at @(n-1)@ which effectively allows us to use the 
\emph{induction hypothesis} for a smaller value of @n@. 
The \emph{termination} check --- made possible by the metric @/ n@ --- 
crucially ensures that the recursion (\ie induction) is well-founded, 
and hence, that the proof is not circular.

\mypara{Induction on Data}
We will see how \emph{measures} let us define \emph{selectors} that 
let us reflect functions on data types like lists into the refinement 
logic.
%
%
%
%
%
This lets us write and prove theorems over datatypes, such as 
this example that involves the usual function for concatenating lists
\begin{code}
  val app : list('a) => list('a) => list('a)
  def app = (xs, ys) => { 
    switch (xs) { 
    | Nil          => ys
    | Cons(x, xs') => Cons(x, app(xs', ys))
    }
  }
\end{code}
Lets verify that @app@ is \emph{associative}, \ie
$$ 
  \forall {xs}, {ys}, {zs}.\ \mcode{app}(\mcode{app}({xs},\ {ys}),\ {zs}) \ =\ \mcode{app}({xs},\ \mcode{app}({ys},\ {zs}))
$$
by writing a recursive proof:
\begin{code}
  val app_assoc : xs:list('a) => ys:list('a) => 
    [app(app(xs, ys), zs) = app(xs, app(ys, zs))] 
    / len(xs)
  
  let app_assoc = (xs, ys, zs) => { 
    switch (xs) { 
    | Nil => { 
            app (app(Nil, ys), zs)
        === app(ys, zs)
        === app(Nil, app(ys, zs)) 
      }
    | Cons(x, xs') => { 
            app(app(Cons(x,xs'), ys), zs)
        === app(Cons(x, app(xs', ys)), zs)
        === Cons(x, app(app(xs', ys), zs))
          ? app_assoc(xs', ys, zs)
        === Cons(x, app(xs', app(ys, zs)))
        === app(Cons(x, xs'), app(ys, zs))
      } 
    }
  }
\end{code}
This time, the induction is on @xs@ and is 
shown well-founded due to the metric @len(xs)@.
As before, we split on the base case where @xs@ is @Nil@, 
and the inductive case where @xs@ is @Cons(x, xs')@. 
In either case, we prove the respective goal via 
a calculation. In the inductive case, we get to 
invoke the induction hypothesis by \emph{calling} 
the theorem @app_assoc@ on the smaller input @xs'@. 

\section{Types and Terms}

\begin{figure}[t!]
\begin{tabular}{rrcll}
\emphbf{Terms}
  & \expr   & $\bnfdef$ & $\ldots$                                            & \emph{from Fig.~\ref{fig:seven:syntax}} \\
  &         & $\spmid$  & \erefl{\evar}{\expr_1}{ \type / \metric }{\expr_2}  & \emph{reflected binder}    \\ [0.05in]
\end{tabular}
\caption{{$\langeight$: Syntax of Types and Terms}}
\label{fig:eight:syntax}
\end{figure}

\Cref{fig:eight:syntax} summarizes the extensions in $\langeight$
needed to support reflection and proofs.
The syntax of types remains unchanged.
The terms are extended with a form $\erefl{\evar}{\expr_1}{\type/\metric}{\expr_2}$ 
which are just like @rec@-binders except that we will strengthen their 
output types using reflection.

\begin{figure}[t]
$$\begin{array}{lcl}
\toprule
\reflectsym & : & (\Evar \times \Expr \times \Type) \rightarrow \Type \\
\midrule
\reflect{f}{\expr}{\typeb} 
  & \doteq & \tpoly{\params{\tvar}}{\trfuns{\evarb}{\typeb}{\refb{\base}{\reft{\vvar}{\pred \wedge \pred'}}}} \\
\quad \mbox{where} & & \\
\quad \quad \tpoly{\params{\tvar}}{\trfuns{\evarb}{\typeb}{\refb{\base}{\reft{\vvar}{\pred}}}}
  & =  & \typeb                            \\
\quad \quad \tabs{\params{\tvar}}{\elam{\params{y}}{\expr^*}}
  & =  & \expr                             \\
\quad \quad \pred' 
  & =  & (\vvar = \uif{f}{\params{\evarb}} \wedge \vvar = \embed{\expr^*}) \\ [0.05in]
\bottomrule 
\end{array}$$
\caption{Reflecting Terms into Types}
\label{fig:eight:reflect}
\end{figure}

\mypara{Reflection}
The workhorse of \langeight is the procedure $\reflect{f}{\expr}{\typeb}$ 
shown in \cref{fig:eight:reflect}, which takes as input a binder $f$, 
the binder's definition $\expr$ and the binder's type $\typeb$, and returns 
a variant of $\typeb$ where the \emph{output} type is strengthened with a 
refinement that says that the returned value \emph{equals} the implementation 
of the function at the given inputs.
The procedure works by first obtaining the parameters $\params{\evarb}$ 
and \emph{body} $\expr^*$ of the definition $\expr$. Next, it translates 
the body $\expr^*$ into the refinement logic via the procedure $\embedsym$.
Finally, it strengthens the output type with the postcondition $\pred'$ that 
says that the output value $\vvar$ equals the function body.
For example, $\reflect{\mw{add}}{\expr}{\typeb}$ where 
$\expr \doteq \elam{x_1, x_2}{x_1 + x_2}$ and 
$\typeb \doteq \tfun{\tint}{\tfun{\tint}{\tint}}$ 
returns as output the type
$$\trfun{x_1}{\tint}
    {\trfun{x_2}{\tint}
      {\refb{\tint}{\reft{\vvar}{\vvar = \uif{\mw{add}}{x_1, x2} \wedge \vvar = x_1 + x_2}}}}$$

\begin{figure}[t]
$$\begin{array}{lcl}
\toprule
\embedsym & : & \Expr \rightarrow \Pred                     \\[0.05in]
\midrule 
\embed{n}
  & \doteq & n                                              \\[0.05in]
\embed{\ttrue}
  & \doteq & \ttrue                                         \\[0.05in]
\embed{\tfalse}
  & \doteq & \tfalse                                        \\[0.05in]
\embed{\vconst\ \evar_1\ \evar_2}
  & \doteq & \evar_1 \bowtie_{\vconst} \evar_2              \\[0.05in]
\embed{ \eapp{(\eapp{f}{\expr_1}) \ldots}{\expr_n} }
  & \doteq & \uif{f}{\embed{\expr_1},\ldots,\embed{\expr_n}} \\[0.05in]
\embed{\elet{\evar}{\expr_1}{\expr_2}}
  & \doteq & \SUBST{\embed{\expr_2}}{\evar}{\embed{\expr_1}} \\[0.05in]
\embed{\eif{\evar}{\evar_1}{\evar_2}}
  & \doteq & \ite{\evar}{\embed{\expr_1}}{\embed{\expr_2}}  \\[0.05in]
\embed{\ecase{\evar}{\params{\alt}}} 
  & \doteq & \embedAs{\evar}{\params{\alt}}                 \\[0.10in]
\bottomrule
\end{array}$$
\caption{Embedding Terms into the Refinement Logic}
\label{fig:eight:embed}
\end{figure}

\mypara{Embedding}
The hard work in $\reflectsym$ is done by the procedure 
$\embedsym$, summarized in \cref{fig:eight:embed} which 
recursively translates \emph{implementation} terms into 
\emph{logical} expressions.
The procedure translates literals like @2@ and @false@
into the corresponding values in the logic
(\ie $2$ and $\tfalse$ respectively); 
primitive function applications like @x+y@ or @a <= b@ 
into the corresponding terms or relations in the logic 
(\ie $x + y$ and $a \leq b$ respectively); and 
translates all other function calls into uninterpreted 
function applications.
Let-binders can be translated by substitution as type 
checking ensures that all reflected terms are terminating, 
and hence, well defined.
Branches are translated into ternary choices.

\begin{myexample}{Reflection of \mcode{sum}}
Suppose that $\expr_\mw{sum}$ is the implementation 
of the @sum@ function from \cref{sec:eight:refl:ex}. Then
\begin{equation}
\embed{\expr_\mw{sum}} \doteq \ite{n = 0}{0}{n + \uif{\mw{sum}}{n-1}}
\label{eq:embed:sum}
\end{equation}
where the recursive call is translated into an uninterpreted 
function application $\uif{\mw{sum}}{n-1}$.
\end{myexample}

\mypara{Embedding Case Alternatives}
Pattern match terms $\ecase{\evar}{\alt_1;\ldots}$ 
are embedded using $\embedAs{\evar}{\alt_1;\ldots}$ 
which translates them as nested ternary branches of 
the form $\ite{c_1}{e_1}{\ldots}$ where $c_1$ is 
a logical predicate that is true when $\evar$ matches 
the first constructor and $\expr_1$ the embedding of 
the corresponding result.
We translate case alternatives using 
two operators from the SMT decidable
theory of Algebraic Datatypes \citep{Nelson81}.
First, the \emph{test} predicate
$\iscon{\dcon}{\evar}$ determines 
whether $\evar$ equals to a term 
$\dcon(c_1, \ldots, c_k)$.
Second, if $\evar$ equals $\dcon(c_1,\ldots,c_k)$ 
then the \emph{projection} function 
$\selcon{\dcon}{i}{\evar}$ equals $c_i$.
Thus, given the scrutinee $\evar$, the procedure $\embedAssym$ 
translates each alternative, by using the test predicate 
to determine if that alternative matches: if so, translating 
its body with the binders replaced by the respective projections, 
and otherwise, recursing on the remaining alternatives.

\begin{myexample}{Reflecting \mcode{app}}
Recall the definition of the list @app@end 
function from \cref{sec:eight:proofs:ex}. Let $\expr_\mw{app}$
denote the body of the implementation, \ie @switch (ys) { ...}@.  
Then
$$
\embed{\expr_\mw{app}} \doteq 
  \ite{\iscon{\mw{Nil}}{\mw{ys}}}{\mw{Nil}}{\mw{Cons}( \uif{head}{\mw{xs}}, \mw{app}( \uif{tail}{\mw{xs}}, \mw{ys}))}
$$
where $\mw{head}$ and $\mw{tail}$ are the projections for the @Cons@
constructor.
\end{myexample}

\begin{figure}[t]
$$
\begin{array}{lcl}
\toprule
\embedAssym & : & (\Evar \times \params{\Alt}) \rightarrow \Pred \\[0.05in]
\midrule 
\embedAs{\evar}{\alt;\params{\alt}} 
  & \doteq & \ite{\iscon{\dcon}{\evar}}
                 {\embedA{\evar}{\alt}}
                 {\embedAs{\evar}{\params{\alt}}} \\[0.1in]
\embedAs{\evar}{\alt}
  & \doteq & \embedA{\evar}{\alt}                 \\[0.05in]
\midrule
\embedAsym  & : & (\Evar \times \Alt)  \rightarrow \Pred \\[0.05in]
\embedA{\evar}{\dcalt{\dcon}{\evarb}{\expr}}
  & \doteq & \SUBST{\embed{\expr}}{\params{\evarb_i}}{\params{\selcon{\dcon}{i}{\evar}}} \\[0.1in]
\bottomrule
\end{array}$$
\caption{Embedding Switch Alternatives into the Refinement Logic}
\label{fig:eight:embed:alt}
\end{figure}
 
\section{Declarative Checking}

\begin{figure}[t]
\judgementHead{Type Checking}{\tchk{\tcenv}{\expr}{\type}}
\begin{mathpar}
\inferrule
  {
    \mchk{\tcenv}{\evar}{\expr_1}{\typeb_1}{\metric}{\type_1}
    \quad
    \type_1' = \reflect{\evar}{\expr_1}{\type_1}
    \quad
    \tchk{\tcenvext{\evar}{\type_1'}}{\expr_2}{\type_2}
  }
  {
    \tchk{\tcenv}{\erefl{\evar}{\expr_1}{\typeb_1 / \metric}{\expr_2}}{\type_2}
  }
  {\ruleName{Chk-Refl}}
\end{mathpar}
	\caption{Bidirectional Checking: Other rules from $\langseven$ (Fig.~\ref{fig:seven:check})}
\label{fig:eight:check}
\end{figure}

The rule \ruleName{Chk-Refl} shown in \cref{fig:eight:check} shows how we use 
$\reflectsym$ to check reflect-binders $\erefl{\evar}{\expr_1}{\typeb_1 / \metric}{\expr_2}$.
First, we check that the definition $\evar = \expr_1$ is terminating with metric $\metric$\footnote{
  Non-recursive binders can be accomodated using the trivial metric $0$.}.
Next, we reflect the definition of $\expr_1$ to strengthen the type $\type_1$ to $\type_1'$ 
which is bound to $\evar$ in the environment used to check $\expr_2$.

\begin{myexample}{Using reflected \mcode{sum}}
Suppose that we wanted to check 
\begin{align*}
  \tcenv & \ \vdash\ \tchkgoal{\erefl{\mw{sum}}{\elam{n}{\expr_\mw{sum}}}{\tfun{\tint}{\tint}/ n}{\expr}}{\type} 
\intertext{Rule \ruleName{Chk-Refl} says we should first establish that \mw{sum} terminates}
  \tcenv & \ \vdash\ \mchkgoal{\mw{sum}}{\elam{n}{\expr_\mw{sum}}}{\tfun{\tint}{\tint}}{n}{\tfun{\tint}{\tint}} 
\intertext{Next, we reflect the definition of \mw{sum} into its type} 
  \type_\mw{sum} & \ \doteq\ \trfun{n}{\tint}{ \refb{\tint}{\reft{\vvar}{ \vvar = \uif{\mw{sum}}{n} \wedge \vvar = \embed{\expr_\mw{sum}} }}}
\end{align*}
where $\embed{\expr_\mw{sum}}$ is the embedding of $\mw{sum}$'s body \cref{eq:embed:sum}.
We can then check $\expr$ in the environment extended by binding $\mw{sum}$ to its reflected type:
$\tcenvext{\mw{sum}}{\type_\mw{sum}} \ \vdash \tchkgoal{\expr}{\type}$.
\end{myexample}

\section{Verification Conditions} \label{sec:eight:algo}

\begin{figure}[t]
$$\begin{array}{lcl}
\toprule
\chksym & : & (\tcenv \times \Expr \times \Type) \rightarrow \Cstr \\
\midrule
\chk{\tcenv}{\erefl{\evar}{\expr_1}{\typeb_1 / \metric}{\expr_2}}{\type_2}
                      & \doteq & \cstr_1 \wedge \cswithref{\evar}{\type_1'}{\cstr_2} \\
\quad \mbox{where}    &        & \\
\quad \quad (\type_1, \cstr_1)  & =  & \chkt{\tcenv}{\evar}{\expr_1}{\typeb_1}{\metric} \\ 
\quad \quad \type_1'            & =  & \reflect{\evar}{\expr_1}{\type_1} \\[0.05in]
\quad \quad \cstr_2             & =  & \chk{\tcenvext{\evar}{\type_1'}}{\expr_2}{\type_2}   \\[0.05in]

\bottomrule
\end{array}$$
\caption{Algorithmic Checking for $\langeight$, extends cases of~\cref{fig:seven:algo}}
\label{fig:eight:algo}
\end{figure}

Finally, we extend algorithmic VC generation function to account for reflected-binders,
as summarized in \cref{fig:eight:algo}. Following the declarative rule, we first invoke 
$\chktsym$ to obtain a constraint $\cstr_1$ that checks that the reflected definition 
$\evar_1 = \expr_1$ terminates with metric $\metric$. We then embed the definition  
into the returned type to obtain the reflected type $\type_1'$ that is used to compute 
the VC $\cstr_2$ for $\expr_2$.

\begin{myexample}{Checking use of \mcode{add}} 
Lets look at the VC generated by $\chk{\emptyset}{\expr}{\type}$ where
\begin{align*}
 \expr          & \doteq\ \erefl{\mw{add}}{\expr_\mw{add}}{\type_\mw{add} / 0}{\eapp{\eapp{\mw{add}}{4}}{5}} \\
 \type          & \doteq\ \refb{\tint}{\reft{\vvar}{\vvar = 9}} 
\intertext{where the definition and type of $\mw{add}$ are respectively}
 \expr_\mw{add} & \doteq\ \elam{x, y}{x + y} \\ 
 \type_\mw{add} & \doteq\ \tfun{\tint}{\tfun{\tint}{\tint}} 
\intertext{First, the termination check $\chkt{\emptyset}{\mw{add}}{\expr_\mw{add}}{\type_\mw{add}}{0}$ yields 
  the trivial type and constraint} 
  (\type_1, \cstr_1) & \doteq\ (\type_\mw{add}, \ttrue)
\intertext{as there are no holes or refinements in the specified signature. 
  Next, we reflect the definition into the output type to obtain} 
  \type_1' & \doteq \ \trfun{x}{\tint}{\trfun{y}{\tint}{ \refb{\tint}{\reft{\vvar}{\vvar = x + y}}}} 
\intertext{and then invoke $\chk{\tb{\mw{add}}{\type_1'}}{ \eapp{\eapp{\mw{add}}{4}}{5} } { \type }$ 
  to get the constraint}
  \cstr_2  & \doteq \ \forall \vvar. \vvar = 4 + 5 \Rightarrow \vvar = 9 
\end{align*}
which is proved valid by SMT.
\end{myexample}

\section{Discussion}

With \langeight we saw how to extend specifications with arbitrary 
user-defined functions whose definitions are reflected into the 
function's output type. This lets us prove propositions over those 
functions, simply by writing programs where each \emph{use} of a
function instantiates, via the reflected output, the defintion 
of the function at the given input.
Reflection dramatically expands the range of what refinements can 
be used for, from enforcing invariants of values or data types, 
to proving the functional correctness of \eg various parallelism 
constructs \citep{Vazou17}, dynamic information flow enforcement 
\citep{lweb}, and laws governing replicated data types \citep{vazou20}.

\mypara{Axioms}
Other SMT based verifiers, notably \textsc{Dafny} and \fstar support 
specifications over user defined functions by encoding their semantics 
with universally-quantified \emph{axioms}.
Modern SMT solvers have sophisticated heuristics for instantiating these 
axioms automatically using user specified \emph{triggers} \citep{Simplify} 
yielding proofs where the user need not spell out all the intermediate 
computations. 
One drawback of the axiomatic approach is that reckless triggering
can cause the SMT engine to diverge \citep{Leino16}. 
\textsc{Dafny} and \fstar use a notion of \emph{fuel} \citep{Amin2014ComputingWA} 
to limit the instantiation to some fixed depth. While fuel can be quite effective 
in practice, it lacks semantic completeness guarantees which are useful to characterize 
what kinds of proofs can be successfully automated.
\cite{Suter2011}, show completeness guarantees for a class 
of \emph{sufficiently surjective} recursive functions, which, 
informally, correspond to catamorphisms over algebraic 
datatypes, \eg functions like the \emph{measures} from \langfive 
that compute the length of a list or height of a tree.
Unfortunately, this result does not extend to arbitrary (terminating) 
user-defined functions. 

\mypara{Proof by Logical Evaluation}
\cite{Vazou18} observe that much of the verbosity in proofs 
arises from spelling out long \emph{chains} of computations, 
for example, the intermediate equalities in the proofs @sum_2_eq_3@  
(\S~\ref{sec:eight:refl:ex}) and @thm_sum@ and @app_assoc@ 
(\S~\ref{sec:eight:proofs:ex}).
Based on this observation, the paper introduces an the notion
of \emph{proof by logical evaluation} (PLE), an algorithm for 
strengthening the antecedents in the VCs by automatically unfolding 
the definition of recursive functions in way that is both terminating 
and complete for equational chains, effectively enabling a form 
of \emph{refinement-level computation}. 
With PLE, we can prove @sum_3_eq_6@ simply as
\begin{code}
  val sum_3_eq_6 : () => [sum 3 = 6]
  let sum_3_eq_6 = () => () 
\end{code}
as the PLE algorithm unfolds the definitions of @sum@ three times.
PLE greatly simplifies inductive proofs by eliminating the tedious 
internal steps, allowing the proof to focus on the important 
case-splitting and induction (recursion). 
That, is the proof of @thm_sum@ is reduced to
\begin{code}
  val thm_sum : n:nat => [2 * sum(n) = n * (n+1)]
  let thm_sum = (n) => { 
    switch (n) { 
    | 0 => () 
    | n => thm_sum(n-1)
    }
  }
\end{code}
which simply spells out the inductive skeleton, yielding a VC: 
%
%
$$\begin{array}{llll}
\multicolumn{4}{l}{\forall n. 0 \leq n \Rightarrow} \\
       & \multicolumn{3}{l}{n = 0 \Rightarrow} \\ 
       & \multicolumn{2}{l}{\quad 2 \times \uif{sum}{n} = n \times (n + 1)}  & \quad \quad \mbox{(base case)}\\
\wedge & \multicolumn{3}{l}{n \not = 0 \Rightarrow} \\     
       &        & 0 \leq n - 1 \wedge n - 1 < n                              & \quad \quad \mbox{(metric)}\\ 
       & \wedge & 2 \times \uif{sum}{n - 1} = (n - 1) \times n \Rightarrow   & \quad \quad \mbox{(ind hyp)}\\
       &        & \quad 2 \times \uif{sum}{n - 1} = n \times (n + 1)         & \quad \quad \mbox{(ind case)} 
\end{array}$$

Similarly, @app_assoc@ reduces to the below, where the proof need only 
include the recursive skeleton, and the recursive call that establishes 
the induction hypothesis for @xs'@: 
\begin{code}
  val app_assoc : xs:list('a) => ys:list('a) => 
    [app(app(xs, ys), zs) = app(xs, app(ys, zs))] 
    / len(xs)
  let app_assoc = (xs, ys, zs) => { 
    switch (xs) { 
    | Nil          => ()
    | Cons(x, xs') => app_assoc(xs', ys, zs)
    }
  }
\end{code}
Note that it is essential that the proof function 
\emph{terminates}: otherwise we could simply write 
circular ``proofs'' for \emph{any} proposition. 
While the increased automation of PLE or axioms makes the 
proofs much more concise, much work remains in devising 
interfaces that can help the programmer structure their 
proofs, and understand why particular proofs are rejected 
by the type checker.

\chapter{Conclusion}\label{sec:outro}

In this article we saw a progression of languages that 
incrementally implement a refinement type checker capable 
of enforcing a full spectrum of correctness requirements 
at compile time. 
We conclude with some remarks on our experience developing 
and using refinement type checkers over the past decade, and 
point the way to some interesting and important directions 
for future work.

\section{The Good: Types Enable Compositional Reasoning}

The great advantage of refinement types is that they 
\emph{align the abstractions} that the analysis uses 
with those that the programmer uses.
Consequently, they provide a simple syntax-directed 
way to \emph{decompose} reasoning about complex values 
like collections and higher-order functions or collections 
into VCs over simple values like integers.

For example, consider the goal of verifying array-access 
safety in the following function that sums the squares of 
an array @x@
\begin{code}
  val sumSquares : array(int) => int
  let sumSquares = (x) => { 
    sum [get(x, i)^2 for i in 0 .. length(x) - 1]
  }
\end{code}
The programmer might write the function using for-comprehension 
syntax, but internally, the function would be translated into 
\begin{code}
  val sumSquares : array(int) => int
  let sumSquares = (x) => { 
    let is   = range(0, length(x) - 1);
    let body = (i) => { get(x, i) ^ 2 };
    let ys   = map(body, is);
    sum(ys)
  }
\end{code}
which makes use collections, higher-order functions and polymorphism 
and collections. Each of these features are problematic for classical 
program logics or program analysis, but are decomposed away by types.
\begin{itemize}
  \item \emphbf{Collections} 
  First, we need a way to represent and establish 
  the fact that \emph{every} element of the collection @is@ was 
  a valid index for @x@. 
  With program logics, this would require universally quantified 
  invariants which make for brittle SMT solving.
  With program analysis, this would require tailoring sophisticated 
  abstract domains capable of performing shape analysis \citep{Gopan05}. 
  In contrast, refinements represent this fact 
  as $$\tlist{\refb{\tint}{\reft{\vvar}{0 \leq \vvar < \uif{length}{\evar}}}}$$ 
  the \emph{refinement} expressing the 
  constraint on a single @int@ and the 
  \emph{type constructor} @list@ 
  generalizing the constraint over 
  the collection.

  \item \emphbf{Closures}
  Next, we need to represent the fact that the closure 
  @body@ accesses the array @x@ at various indices 
  \emph{supplied by} @map@.
  Classical (\eg Floyd-Hoare) program logics do not 
  account for closures.
  Modern logics like \eg Hoare-Type Theory \citep{Nanevski} 
  or \textsc{Iris} \citep{iris} do handle them, and tools like 
  \textsc{Dafny} permit reasoning about closures, but with 
  significantly more overhead. 
  In the program analysis world, this problem is a variant
  of \emph{Control-flow Analysis} \citep{Shivers88} which is
  complicated by reasoning about properties of free variables
  (\eg @x@) \citep{Might07}, which has resisted a robust solution
  for several decades.
  Types make the problem practically disappear: we ascribe
  the type @i:nat[i < length(x)] => int@ to @body@ and then
  (contra-variant) function subtyping naturally ensures that
  only valid indices are in fact passed into the closure.

  \item \emphbf{Polymorphism}
  Finally, functions like @map@ are ubiquitous: they are reused
  in many different sites, and verification requires a way to
  specify a contract that is both \emph{general} enough to account
  for all the use-cases, yet \emph{precise} enough to facilitate
  verification at each site.
  In the program logic setting, this requires quantification
  over the possible invariants \citep{Nanevski} which 
  makes verification less ergonomic, as the programmer must 
  spell out where the quantifier is added (\ie generalized) and 
  removed (\ie instantiated).
  In the program analysis setting this the problem of 
  \emph{context sensitivity} which remains a notorious 
  source of imprecision \citep{Smaragdakis20}.
  In contrast, type- and refinement- polymorphism provides 
  a natural solution: we need only (automatically) instantiate 
  the type variables $\tvar$ and $\tvarb$ in @map@'s type
  $$
  \tfun{(\tfun{\tvar}{\tvarb}) }{\tfun{ \tlist{\tvar} }{ \tlist{\tvarb}  }}  
  $$
  with suitable refined types to enable context-sensitive verification. 
\end{itemize}

\section{The Bad: Reasoning about State}

A reader who has made it this far is clearly aware of the 
elephant in the room: this article has hewed closely to 
\emph{pure} programs, and entirely shied away from
discussing the topic of \emph{state}.
This is partly for exposition, partly because there 
is already a substantial literature on the topic that 
merits its own separate survey and partly because 
\emph{precise} reasoning about imperative features 
remains quite difficult.

\mypara{Invariant References}
The simplest way to account for state is by introducing 
a type 
\begin{code}
  type ref('a)      /* 'a is co- and contra- variant */
\end{code}
denoting pointers to values of type @('a)@ and then 
use the standard API for accessing pointers:
\begin{code}
  val new : 'a => ref('a)                /* allocate */
  val get : ref('a) => 'a                /* read     */
  val set : 'a => ref('a) => ()          /* read     */
\end{code}
The machinery described in \S~\ref{sec:lang:three} scales up 
to account for such references, but suffers from two problems.
First, it is \emph{flow-insensitive}: we end up assigning 
a \emph{single} type to a reference throughout its lifetime 
(\eg @nat@) as opposed to different types at different points
as the reference is \emph{updated}.
Second, there is no way to use references in refinements, 
as the values referred to might change.

\mypara{Alias and Ownership Types}
\cite{AliasTypes} introduces a mechanism called \emph{Alias Types} 
for reasoning about references and aliasing within a type system, 
essentially by typing each pointer with a singleton \emph{location}, 
and tracking a separate \emph{store} that holds the values of each 
location.
\cite{LiquidPOPL10} describes a way to combine logical predicates 
with alias types to obtain a refinement type checker for C programs.
Similarly, \cite{Chugh2012} shows how to combine alias types 
with refinements to obtain a refinement system for JavaScript, 
and \cite{Bakst16} shows how to extend the approach to recursive 
alias types thereby allowing refinements to specify and verify 
complex invariants of linked data structures.
\cite{Vekris16} shows how \emph{ownership types} \citep{Noble}
used to enforce reference immutability for Java \citep{zibin}
can provide a more lightweight mechanism wherein immutable 
refinements can be embedded within an imperative language 
like TypeScript.

\mypara{Monads}
None of the above methods scale up to handle
the combination of higher-order functions 
\emph{and} state.
\cite{Filliatre98} introduced a method for verifying 
higher order programs with references, and \cite{ynot} 
introduce \emph{Hoare Type Theory} which encapsulates 
that reasoning within \emph{Hoare Monads} which are 
the usual state-transformers indexed by pre- and 
post-conditions. This approach, while expressive,
is tricky to use as it lacks a way to algorithmically generate 
verification conditions whose validity implies correctness.
\fstar elegantly solves this problem via the notion 
of \emph{Dijkstra Monads} where the monad is indexed 
by a single \emph{predicate transformer}: a function
that computes the (most general) heap pre-conditions 
under which some desired post-condition will hold.
Crucially, the composition of the transformers yields 
a mechanism for computing VCs. This method, combined 
with SMT solvers' native support for McCarthy's axioms 
for reasoning about arrays via the @select@ and @store@
operations \citep{Mccarthy}, yields a powerful way to 
verify higher-order stateful programs.

\mypara{Separation Logic}
Finally, over the last two decades, Separation Logic 
\citep{BI01, Reynolds02} has transformed how we think 
about the verification of pointer-manipulating programs,
and is the basis for modern mechanized program logics 
like \textsc{Iris} \citep{iris} which has been used to 
verify a range of sophisticated concurrent, pointer 
manipulating algorithms using the \textsc{Coq} proof 
assistant.
In future work it would be interesting to investigate 
how refinements can be combined with the monadic approach
perhaps in combination with separation logic \citep{kloos} 
to yield simpler and easier to use tools for verifying 
stateful programs.

\section{The Ugly: Explaining Verification Failures}

The more sophisticated a static type system,
analysis or program logic, the more difficult
it is to \emph{explain} failures.
In our experience, the most challenging aspect 
of using refinement types is that the high degree 
of automation makes it difficult for beginners to
understand verification \emph{failures}, which can 
arise in several modes.

\mypara{Problem: The Implementation is Wrong}
The most common case is when the implementation 
does not respect the specification. For example,
suppose @noDups@ is a measure (\cref{sec:lang:five}) 
such that @noDups(xs)@ holds when the list @xs@ 
contains no duplicates. Hence we can define a type 
of \emph{unique lists} \ie without any duplicates as 
\begin{code}
  type ulist('a) = list('a)[v| noDups(v)]
\end{code}
The following code fails to verify
\begin{code}
  val append : ulist('a) => ulist('a) => ulist('a) 
  let append = (xs, ys) => {
    switch (xs) {
    | Nil => ys 
    | Cons(x, xs') => Cons(x, append(xs', ys))
    }
  } 
\end{code}
Unfortunately the error message will simply say that 
the result of @Cons(x,...)@ is not a unique list and 
the programmer may be quite puzzled as to why.

These failures are the easiest to explain, as one can
augment the type checker with some form of symbolic execution
to produce \emph{counterexamples} that describe why the property 
does not hold \citep{hallahan19}. 
For example, in this case, the programmer could be given 
a counterexample of the form
\begin{code}
  xs             := Cons(1, Nil)
  ys             := Cons(1, Nil)
  append(xs, ys) := Cons(1, Cons(1, Nil)) 
\end{code}
which would demonstrate a situation where the 
output refinement fails to hold even though the 
input requirements are met, and hopefully this 
will provide a hint as to how to modify 
the specification or the code.

\mypara{Problem: The Specification is Weak}
A more vexing situation arises when the code 
\emph{does} satisfy the specification, in that 
there are no counterexamples, but where verification 
fails because the specifications where not enough 
to \emph{prove} the property.
Continuing with the @ulist@ example from 
above, suppose that from the counterexample, 
the programmer has realized that the output 
is unique only when the input lists have no 
common elements. They will specify this extra 
requirement as:
\begin{code}
  val append: xs:ulist('a) 
           => ys:ulist('a)[intersect(xs, ys) = empty]
           => ulist(a')
\end{code}
But now, imagine their dismay when the code is \emph{again}
rejected by the type checker. 
Unfortunately, this time, we cannot find a counterexample 
as indeed the function \emph{does} correctly implement 
the given specification: the concatatenation of two 
unique lists with no common elements always yields 
a unique list. 

In this case, verification fails because typechecking 
is \emph{modular}. The only information that the type 
checker has about the output of a function application, 
is whatever was specified in the function's type. 
Thus, in the above example consider the case 
\begin{code}
  | Cons(x, xs') => Cons(x, append(xs', ys))
\end{code}
The signature for @append@ says that the (recursive) 
call @append(xs',ys)@ can return \emph{any} unique list,
including one that may possibly contain @x@, and so 
the @Cons(x,...)@  need not be a duplicate-free list.
Of course, this cannot happen \emph{in reality} 
because the list \emph{output from} @append@ can 
only contain elements from the lists \emph{passed into} 
@append@, but this information is absent from 
the type signature, preventing verification.

This classic failure mode --- widely known 
in the verification community as the difference 
between invariants and \emph{inductive} invariants 
--- is a significant stumbling block for programmers
as it is difficult to pinpoint exactly where the extra
information is needed, and what that information should be.
\cite{hallahan19} demonstrate an algorithm for generating 
\emph{counterfactual counterexamples} that can pinpoint 
the functions whose types need to be strengthened.
In future work it would interesting to see if ideas 
from the synthesis literature \citep{gulwani_synthesis} 
can be used to suggest hints on how to strengthen 
the specifications or code to facilitate proof,
or more broadly, help the programmer rapidly 
build a robust mental model of the requirements 
for formal verification.
This would go a long way towards flattening the 
steep learning curve that remains the most daunting 
hurdle limiting the broader adoption of formal 
verification in software development.    

\backmatter  

\printbibliography

\end{document}